\documentclass{article}

    \PassOptionsToPackage{numbers, compress, sort}{natbib}

\usepackage[preprint]{neurips_2024}

\usepackage[utf8]{inputenc} 
\usepackage[T1]{fontenc}    
\usepackage[hidelinks]{hyperref}       
\usepackage{url}            
\usepackage{booktabs}       
\usepackage{amsfonts}       
\usepackage{nicefrac}       
\usepackage{microtype}      
\usepackage{xcolor}         

\usepackage[acronym,shortcuts]{glossaries}
\glsdisablehyper 
\usepackage[capitalize]{cleveref}
\usepackage{chemformula}
\usepackage{amsmath}
\usepackage{multirow}
\usepackage{colortbl}
\usepackage{tabularx}
\usepackage{subcaption}

\usepackage{float}
\usepackage{longtable}
\usepackage{rotating}

\usepackage{tikz}
\usetikzlibrary{positioning, shapes.geometric, calc}

\newcommand{\nfer}[0]{\ch{N}}
\newcommand{\pfer}[0]{\ch{P2O5}}
\newcommand{\kfer}[0]{\ch{K2O}}
\definecolor{ec}{HTML}{ED7D31}
\definecolor{env}{HTML}{1F4E79}
\definecolor{agr}{HTML}{548235}

\title{Global Crop-Specific Fertilization Dataset from 1961–2019}

\author{%
Fernando Coello\textsuperscript{\textnormal{1,2,3,*,$\dag$}},
  Thomas Decorte\textsuperscript{\textnormal{4,$\dag$}}, 
  \textbf{Iris Janssens}\textsuperscript{\textnormal{5}},
  \textbf{Steven Mortier}\textsuperscript{\textnormal{5}} \\
\textbf{Jordi Sardans}\textsuperscript{\textnormal{2,3}}, 
  \textbf{Josep Peñuelas}\textsuperscript{\textnormal{2,3}},
  \textbf{Tim Verdonck}\textsuperscript{\textnormal{4}} \\
  \textsuperscript{1}Universitat Autònoma de Barcelona, 08193, Bellaterra, Spain \\ 
  \textsuperscript{2}CREAF - Centro de Investigación Ecológica y Aplicaciones Forestales, Barcelona, 08913, Spain \\
  \textsuperscript{3}Global Ecology Unit, CSIC-CREAF-UAB, Barcelona, 08913, Spain \\
  \textsuperscript{4}University of Antwerp - imec - IDLab, Department of Mathematics, Antwerp, 2000, Belgium \\
  \textsuperscript{5}University of Antwerp - imec - IDLab, Department of Computer Science, Antwerp, 2000, Belgium \\
   \textsuperscript{*}corresponding author:  Fernando Coello (f.coello@creaf.uab.cat) \\
    \textsuperscript{$\dag$}these authors contributed equally to this work
    }

\author{%
Fernando Coello\textsuperscript{\textnormal{1,2,3,*,$\dag$}},
  Thomas Decorte\textsuperscript{\textnormal{4,$\dag$}}, 
  \textbf{Iris Janssens}\textsuperscript{\textnormal{5}},
  \textbf{Steven Mortier}\textsuperscript{\textnormal{5}} \\
\textbf{Jordi Sardans}\textsuperscript{\textnormal{2,3}}, 
  \textbf{Josep Peñuelas}\textsuperscript{\textnormal{2,3}},
  \textbf{Tim Verdonck}\textsuperscript{\textnormal{4}} \\
  \textsuperscript{1}Universitat Autònoma de Barcelona, 08193, Bellaterra, Spain \\ 
  \textsuperscript{2}CREAF - Centro de Investigación Ecológica y Aplicaciones Forestales, Barcelona, 08913, Spain \\
  \textsuperscript{3}Global Ecology Unit, CSIC-CREAF-UAB, Barcelona, 08913, Spain \\
  \textsuperscript{4}University of Antwerp - imec - IDLab, Department of Mathematics, Antwerp, 2000, Belgium \\
  \textsuperscript{5}University of Antwerp - imec - IDLab, Department of Computer Science, Antwerp, 2000, Belgium \\
   \textsuperscript{*}corresponding author:  Fernando Coello (f.coello@creaf.uab.cat) \\
    \textsuperscript{$\dag$}these authors contributed equally to this work
    }

\newacronym{ML}{ML}{machine learning} 
\newacronym{XGB}{XGB}{eXtreme Gradient Boosting or XGBoost}
\newacronym{HGB}{HGB}{HistGradientBoosting}
\newacronym{CV}{CV}{cross validation}
\newacronym{MAE}{MAE}{mean absolute error}
\newacronym{MSE}{MSE}{mean squared error}
\newacronym{RMSE}{RMSE}{root mean squared error}
\newacronym{SHAP}{SHAP}{SHapley Additive exPlanations}
\newacronym{MAPE}{MAPE}{mean absolute percentage error}
\newacronym{OHE}{OHE}{one-hot encoded}
\newacronym{xAI}{xAI}{eXplainable artificial intelligence}
\newacronym{ICC}{ICC}{Indicative Crop Classification for agricultural census}
\newacronym{IFDC}{IFDC}{International fertilizer Development Center}
\newacronym{IFA}{IFA}{International Fertilizer Association}
\newacronym{FUBC}{FUBC}{fertilizer use by crop}
\newacronym{FAO}{FAO}{Food and Agriculture Organization}
\newacronym{FE}{FE}{Fertilizer Europe}
\newacronym{MAP}{MAP}{mean annual precipitation}
\newacronym{SOC}{SOC}{soil organic carbon}
\newacronym{MAT}{MAT}{mean annual temperature}
\newacronym{PET}{PET}{potential evapotranspiration}
\newacronym{CEC}{CEC}{cation exchange capacity}
\newacronym{LCU}{LCU}{local currency units}
\newacronym{USD}{USD}{United States dollars}
\newacronym{GDP}{GDP}{gross domestic product}
\newacronym{OCS}{OCS}{organic carbon stock}
\newacronym{USDA}{USDA}{United States Department of Agriculture}
\newacronym{DEFRA}{DEFRA}{Department for Environment, Food \& Rural Affairs}
\newacronym{EU}{EU}{European Union}
\newacronym{UK}{UK}{United Kingdom}
\newacronym{USA}{USA}{United States of America}
\newacronym{SFR}{SFR}{Social Federal Republic}
\newacronym{USSR}{USSR}{Union of Soviet Socialist Republics}

\begin{document}

\maketitle

\begin{abstract}
As global fertilizer application rates increase, high-quality datasets are paramount for comprehensive analyses to support informed decision-making and policy formulation in crucial areas such as food security or climate change. This study aims to fill existing data gaps by employing two machine learning models, eXtreme Gradient Boosting and HistGradientBoosting algorithms to produce precise country-level predictions of nitrogen (\nfer), phosphorus pentoxide (\pfer), and potassium oxide (\kfer{}) application rates. Subsequently, we created a comprehensive dataset of 5-arcmin resolution maps depicting the application rates of each fertilizer for 13 major crop groups from 1961 to 2019. The predictions were validated by both comparing with existing databases and by assessing the drivers of fertilizer application rates using the model's  SHapley Additive exPlanations. This extensive dataset is poised to be a valuable resource for assessing fertilization trends, identifying the socioeconomic, agricultural, and environmental drivers of fertilizer application rates, and serving as an input for various applications, including environmental modeling, causal analysis, fertilizer price predictions, and forecasting. 

\end{abstract}


\thispagestyle{empty}

\section*{Background \& Summary}
Inorganic fertilizers are essential for replenishing the nutrients that are removed from soils during crop harvesting. The three main nutrients provided by fertilizers, nitrogen (\nfer{}), phosphorus (P) and potassium (K), play a key role in plant functions. While N and P, which are basic components of nucleotides, proteins and membrane lipids, are essential in energy metabolism \cite{Xu2012PlantNimportance, Shen2011PIntro}, 
K is essential for the transportation of water, metabolites, and nutrients across plant tissues, for defense against oxidative stresses, and for the maintenance of osmotic homeostasis \cite{Sardans2021KIntro1, Sardans2015KIntro2}. Although the first commercial inorganic fertilizers were developed in 1843, they were not the main anthropogenic inputs in the N, P, and K biochemical cycles until the second half of the 20th century \cite{Ludemann2023A19612020}. Today, inorganic fertilizers dominate as the primary nutrient input in croplands, surpassing the second human input, manure, by over double \cite{Ludemann2023A19612020}, and also serve as one of the main N input for grasslands \cite{Xu2019GrasslandsN}. This substantial surge during the 20th century not only facilitated the rapid growth in human population, but also had ecological and socioeconomic ramifications, such as water eutrophication, soil degradation, climate change, and mineral resource depletion \cite{Sutton2013OurNutrientWorld, Penuelas2020IntroConsequences}. In the remainder of this study, the term 'fertilizer' will refer to inorganic fertilizers, and all data and results regarding P and K will be presented in their oxidative forms (\pfer{} and \kfer{}, respectively), in accordance with common references in international standards and regulations.\\
Given their food security, socioeconomic and environmental implications, considerable research has been conducted to discern the temporal and regional trends in the use of \nfer{}, \pfer{}, and \kfer{} \cite{Adalibieke2023Global1961, Lu2017GlobalImbalance, Xu2019GrasslandsN, Nishina2017NH4NO3maps, MacDonald2011PImbalances}. Nevertheless, limited availability of temporal global spatial information regarding their application across various crops have restricted these analyses to a few global and regional studies that primarily focused on \nfer{} \cite{Adalibieke2023Global1961, Cao2018NitrogenCropUS, Yu2022NCrops_China}. These studies initially estimated consumption at the country- and state-level using simple equations, based on a few crop-specific fertilization features and changes in crop surface area \cite{Adalibieke2023Global1961, Cao2018NitrogenCropUS, Yu2022NCrops_China}, or using Bayesian Markov Chain Monte Carlo modeling \cite{conant2013patterns}. A global, crop-specific fertilization dataset is crucial for understanding crop nutrient management practices worldwide, identifying past trends and current gaps in fertilization, guiding agricultural policies to improve crop yields while minimizing environmental impacts, and providing input data for modeling \cite{Adalibieke2023Global1961}. Therefore, we aim to address this knowledge gap by providing insights into the application rates of \pfer{} and \kfer{} while also seeking to improve estimates for \nfer{}.\\
In order to accomplish this objective, we began by updating the panel datasets on cropland fertilization; enhancing the most comprehensive database developed by Ludemann \textit{et al.} (2022) \cite{Ludemann2022GlobalCountry} by incorporating global datasets covering data from the 1970s and 1980s \cite{Martinez1982FertilizerProduction, AdolfoMartinez1990FertilizerT-37}, country-specific data for European countries from 2001 to 2014 \cite{EFMA2002Fertilizer2001/02, EFMA2007Fertilizer2006/07, FertilizerEurope2012Fertilizer2011/12, FertilizerEurope2015Fertilizer2014/15}. This compilation process led to a 35\% expansion of the Ludemann \textit{et al.} (2022) database. Second, the dataset was expanded with data of various potential socioeconomic, environmental, and agricultural drivers of cropland fertilization. Third, two \ac{ML} regression models - \ac{XGB}\cite{Chen2016XGBoost:System} and \ac{HGB}\cite{Pedregosa2011Scikit-learn:Python}, both capable of handling the prevalent missing values within the dataset \cite{Pedregosa2011Scikit-learn:Python}- were applied to predict \nfer{}, \pfer{}, and \kfer{} fertilizer application rates for the different crop classes over 60 years. Since these models are considered black-box models, feature importance was incorporated using \ac{SHAP}\cite{Lundberg2017APredictions} values to identify the global socioeconomic, agricultural, and environmental drivers of cropland fertilization and to validate the \ac{ML} models. Fourth, the predictions were validated on national databases. However, since the \ac{ML} models were trained on global data, which show a discrepancy with the national data, the model predictions were first adjusted to match the total annual country-level \nfer{}, \pfer{}, and \kfer{} use in agricultural land, similar to previous studies \cite{Adalibieke2023Global1961, Lu2017GlobalImbalance, Zou2022, Nishina2017NH4NO3maps, MacDonald2011PImbalances}. 
Crucial in this adjustment was the fraction of total country-level fertilizer use allocated to grasslands and fodder crops, as an important portion of total fertilizer use in some countries is devoted to these areas, and little previous estimates existed \cite{Lassaletta2014, Einarsson2021, Zou2022, Ludemann2023A19612020}, especially for \kfer{}\cite{Ludemann2023A19612020}. Therefore, these fractions were estimated by reviewing scientific and technical information from 75 countries. The adjusted predictions were then validated using national databases of fertilizer application rates at the crop-level. Finally, the results were spatially allocated using crop maps of the year 2000, developed by Monfreda \textit{et al.} (2008) \cite{monfreda2008farming}; the annual harvested area of each crop class in each country; and the spatial changes in cropland surface based on the Hyde v3.3. project \cite{klein2017anthropogenic}.

\section*{Methods}

The following section outlines the comprehensive methodology that was adopted in this study. The methodology encompasses various stages, including the collection and aggregation of different datasets and the compilation into a unified dataset, as well as all preprocessing steps that were carried out. Additionally, we introduce the \ac{ML} models used in this study, as well as the respective training and evaluation procedures. Furthermore, we discuss the measures that were undertaken to explain the predictions made by the \ac{ML} models. Following this, we describe how we used the predictions to create detailed maps of global fertilizer application rates. Finally, we explain how we assessed the validity and plausibility of the dataset derived from our study.

\subsection*{Data collection and preprocessing}
\label{sec:fer_sd:data_collection_and_preprocessing}

\subsubsection*{Data collection}
\textbf{Fertilizer application rate by crops}
To compile a consistent and detailed dataset of fertilizer application rates for different crops, countries, and years, 14 global datasets \cite{Martinez1982FertilizerProduction, AdolfoMartinez1990FertilizerT-37, FAO1992Fertilizer1, FAO1994Fertilizer2, FAO1996Fertilizer3, FAO1999Fertilizer4, FAO2002Fertilizer5, EFMA2002Fertilizer2001/02, EFMA2007Fertilizer2006/07, PattrickHeffer2009Asessment2007/08, FertilizerEurope2012Fertilizer2011/12, Heffer2017Assessment2014-2014/15, FertilizerEurope2015Fertilizer2014/15, Ludemann2022GlobalCountry} were used. We discarded national databases, such as the \ac{USA} \cite{USDA2019} and India \cite{India1986Input, India1991Input, India1996Input, India2001Input, India2006Input, India2011Input, India2016Input}, to construct a homogeneous database. This approach avoids multiple year-nutrient-crop-country entries from both global and national databases, and allows us to retain external databases for validating the \ac{ML} model predictions
. To standardize all these datasets and minimize data loss, we classified all crop types into 13 crop groups (wheat, maize, rice, other cereals, soybean, palm fruit, other oilseeds, vegetables, fruits, roots and tubers, sugar crops, fiber crops and other crops) (\cref{tab:fer_sd:crop_classification}), in alignment with the \ac{ICC} Version 1.1 \cite{FAO2015ICC}. 

During the 80s, the \ac{IFDC} published two reports \cite{Martinez1982FertilizerProduction, AdolfoMartinez1990FertilizerT-37} regarding crop-specific data of \ac{FUBC} (hereinafter referred to as FUBC-\ac{IFDC}). After the crop grouping, these publications included data for 459 country-crop-years combinations (kg ha\textsuperscript{-1} of \nfer{}, \pfer{}, and \kfer{}) from 83 countries for 1973-1988. During the 90s, the \ac{FAO}, in collaboration with the fertilizer industry (\ac{IFDC} and \ac{IFA}), published five crop-specific datasets of fertilizer application rate (hereinafter referred to as \ac{FUBC}-\ac{FAO}). After grouping the data, these publications included data for 1693 fertilizer application rate specific to years and crops (kg ha\textsuperscript{-1} of \nfer{}, \pfer{}, and \kfer{}) from 108 countries for 1984–2002, although most of the data (98\%) covered 1988–2002. The data were collected using questionnaires from governmental agencies, members of industry companies, agronomists, and economic experts. In both datasets (\ac{FUBC}-\ac{IFDC} and \ac{FUBC}-\ac{FAO}), the use of fertilizer for each combination of nutrient, crop, country, and year was provided two ways: (a) as the average application rate of a fertilizer over total cropland area, and (b) as the percentage of fertilized cropland area and the application rate in that area. We transformed all data to the average application rate by multiplying the percentage of fertilized area by the application rate in that area. The data were either from a year (e.g., 1996) or a season (e.g., 1996/97). For seasonal data, we considered the starting year of the season as the year of the data in the analyses. Fore data for nutrient, crop, country, and year that were in more than one report, the data was selected from the most recent report. Data for crop, country, and year that were divided into crop varieties or management practices (e.g., irrigated or rain-fed rice, or soft or durum wheat) were aggregated and weighted by the area of the crop class included in the report. Data for sweet maize, or corn, were excluded, assuming that it referred to \textit{Zea mays } var. \textit{saccharata} and the data for silage maize, because FAOSTAT reports only the harvested area for maize grain. Values for the crop groups were derived from individual crops when either more than 90\% of the harvested area (based on FAOSTAT data\cite{FAOSTAT2023AreaDataset}) was dedicated to the production of a single crop, or when a combination of crops was available in the data, their weighted average was assigned to the entire group. 

Since the last \ac{FAO} publication, \ac{IFA} has released five reports detailing the total amount of \nfer{}, \pfer{}, and \kfer{} used for various crop classes, providing yearly or seasonal data spanning from 2006 to 2018 \cite{PattrickHeffer2009Asessment2007/08, PattrickHeffer2013Asessment2010/11, Heffer2017Assessment2014-2014/15, Ludemann2022GlobalCountry} (hereinafter referred to as \ac{FUBC}-\ac{IFA}). Initially covering 11 crop types, these reports expanded to 14 types in the fourth report. They encompassed information for the \ac{EU} together as well as 27 other countries. In 2022, Ludemann published a more comprehensive dataset covering data for 66 countries, featuring \ac{EU} data at the country scale, and information for 20 crop classes \cite{Ludemann2022GlobalCountry}. This report also included the \ac{FUBC}-\ac{FAO} data for the 1990s and prior data from IFA. However, small discrepancies between the \ac{FUBC}-\ac{FAO} original data and the one compiled by Ludemann \textit{et al.} (2022) prompted us to retain the original \ac{FUBC}-\ac{FAO} information. To estimate the average application rate for each combination of crop, country, and year, we divided the total used amount of each fertilizer by the harvested area provided by FAOSTAT \cite{FAOSTAT2023AreaDataset}. As previous research we assumed the harvested area as a proxy for the crop's annual surface on each country \cite{Lu2017GlobalImbalance, Adalibieke2023Global1961}. It is worth noting that the average application rate for maize was slightly overestimated because \ac{FUBC}-\ac{IFA} included the amount discharged to silage maize. According to Maiz’Europ’ \cite{MaizEurope2020}, the current area of forage maize crops is 17.3 million ha (approximately 1\% of the total area of maize crops in 2020) with the European Union as the most important producer of silage maize, with 6 million ha. We utilized the available raw data from Ludemann \textit{et al.} (2022)\cite{Ludemann2022GlobalCountry}, adopting \ac{FAO}-\ac{IFDC} datasets methods for grouping, and omitted certain countries where values were estimated based on the previous report and changes in crop surface. For the \ac{EU} countries, Norway and the \ac{UK}, four unpublished datasets from \ac{FE} spanning 2001–2015 (referred to as \ac{FUBC}-EFMA) \cite{EFMA2002Fertilizer2001/02, EFMA2007Fertilizer2006/07, FertilizerEurope2012Fertilizer2011/12, FertilizerEurope2015Fertilizer2014/15} were used. These datasets offered similar information to the \ac{FUBC}-\ac{FAO} publications for the \ac{EU} countries, the \ac{UK} and Norway and allowed us to exclude the fertilizer application to silage maize, which is important in the \ac{EU}\cite{MaizEurope2020}. However, \ac{FUBC}-EFMA datasets lacked individual crop classes for rice and soybeans, resulting in missing data at the country-level for these crops since 2000 in \ac{EU} countries. \\
The resulting dataset included data for the average fertilizer application for 3712 combinations of 13 crop classes, 114 countries, and years from 1973 to 2018. For most of the combinations of countries and crops, data were available for only a few years (on average, a country-crop combination had data for 4.1 $\pm$ 2.9 years, and 64\% of the combinations had five or fewer years with available data).\\

In order to later validate our estimations, we compiled a series of national databases. National data was quite limited, as only a few countries conduct surveys to study fertilizer management across different crops. The two countries with most available data were the \ac{USA} \cite{USDA2019}, and the \ac{UK} \cite{DEFRA2023FertiliserUKdataset}, which collected long time series on cropland fertilization for the three primary nutrients. The \ac{USA} dataset \cite{USDA2019} contains fertilization information for four crops -cotton, maize, soybean, and wheat- dating back to 1964. To compare with our predictions, we converted all data to average kg ha\textsuperscript{-1}. Additionally, based on the same surface threshold used for global datasets, we assumed that the application rate for cotton was equivalent to that of all fiber crop classes. The \ac{UK} dataset\cite{DEFRA2023FertiliserUKdataset} provides data for four crop classes  -roots and tubers, other oilseeds, sugar crops, and wheat-  starting from 1998 for the three nutrients across all Great Britain. We also compiled existing information from several Asian countries, including India, the Philippines, and Pakistan \cite{India1986Input, India1991Input, India1996Input, India2001Input, India2006Input, India2011Input, India2016Input, PAKISTAN, PSA2023}. The datasets from India \cite{India1986Input, India1991Input, India1996Input, India2001Input, India2006Input, India2011Input, India2016Input} and Pakistan \cite{PAKISTAN} did not require additional preprocessing, as they provided the data in average kg ha\textsuperscript{-1}. However, the dataset from Pakistan presented the information for all three nutrients combined\cite{PAKISTAN}. For the dataset from the Philippines, which covers rice and maize, we converted the raw data on the regional number of 50 kg bags per hectare of different fertilizers to \nfer{} and \pfer{} using the country-specific fertilizer nutrient information \cite{Briones2016Phillippines}. Finally, we also compiled existing data from Sweden\cite{SWEDEN2010, SWEDEN2012, SWEDEN2015, SWEDEN2018} and New Zealand\cite{NZ2021}. The data for \pfer{} and \kfer{} in the Sweden dataset, initially present in their pure nutrient form, were transformed to their oxidized forms by multiplying by the molecular weights of these elements.\\

\textbf{Fertilizer use in other agricultural lands} An important step in the methods involves adjusting \ac{ML} model predictions to national-level fertilizer use. We used the FAOSTAT database regarding fertilizer annual use at the country level for making this adjustment\cite{FAOSTAT2023FertilizerDataset}. This database includes data on all fertilizer use for agricultural lands, covering both croplands and grasslands \cite{FAOSTAT2023FertilizerDataset}. However, the crops included in the \ac{ML} models, as well as in the FAOSTAT harvested area data\cite{FAOSTAT2023AreaDataset} do not cover grasslands -whether permanent or temporary-  nor fodder crops such as silage maize or fodder beet. Therefore, the primary goal of this section is to estimate the fraction of total fertilizer used for these types of agricultural lands.

Data regarding fertilizer application rate for grasslands and fodder crops is even more scarce than fertilization for other croplands. Additionally, FAOSTAT lacks information about the surface of the majority of the fodder crops \cite{FAOSTAT2023AreaDataset}. Therefore, the methods used for estimation may not be as accurate as those used for other agricultural lands. Here, we reviewed technical information, such as the \ac{FUBC} compiled reports \cite{Martinez1982FertilizerProduction, AdolfoMartinez1990FertilizerT-37, FAO1992Fertilizer1, FAO1994Fertilizer2, FAO1996Fertilizer3, FAO1999Fertilizer4, FAO2002Fertilizer5, Heffer2017Assessment2014-2014/15, Ludemann2022GlobalCountry, FertilizerEurope2012Fertilizer2011/12, FertilizerEurope2015Fertilizer2014/15, EFMA2002Fertilizer2001/02, EFMA2007Fertilizer2006/07}, and scientific information from countries where the fertilization of grasslands was considered to be higher than 1\% of the total fertilizer consumption in previous research \cite{Lassaletta2014, Einarsson2021, Zou2022, Ludemann2023A19612020, Xu2019GrasslandsN}. Previous research typically focused only on permanent grassland fertilization, as their goal was to distinguish agricultural fertilizer usage between arable -croplands and temporary grasslands- and non-arable land -permanent grasslands- \cite{Lassaletta2014, Ludemann2023A19612020, Einarsson2021} . However, we included in the estimation the proportion of fertilizer used for temporary grasslands and fodder crops for two main reasons: 1) our main goal was to distinguish agricultural fertilizer usage between all croplands included in the thirteen crop classes defined in the previous section and the rest of the agricultural land, 2) the majority of data available in the compiled global reports give information about all grasslands and fodder crops together \cite{Martinez1982FertilizerProduction, AdolfoMartinez1990FertilizerT-37, FAO1992Fertilizer1, FAO1994Fertilizer2, FAO1996Fertilizer3, FAO1999Fertilizer4, FAO2002Fertilizer5, EFMA2002Fertilizer2001/02, EFMA2007Fertilizer2006/07, PattrickHeffer2009Asessment2007/08, FertilizerEurope2012Fertilizer2011/12, Heffer2017Assessment2014-2014/15, FertilizerEurope2015Fertilizer2014/15, Ludemann2022GlobalCountry}. The information estimated was the annual country proportion of \nfer{}, \pfer{}, \kfer{} fertilizers used for agriculture for grasslands and fodder crops. Depending on the available information, we have assessed at the country- or regional-level. In total, we reviewed scientific and technical reports for 75 countries. As in previous research\cite{Lassaletta2014, Einarsson2021, Zou2022, Ludemann2023A19612020}, the methods used for estimating the share of \nfer{}, \pfer{}, and \kfer{} usage for grasslands and fodder crops varied between countries and regions depending on the available information. Therefore, for every country, we argued the decisions taken based on the available data for providing at least as transparent as possible the estimations made. Moreover, we included a summary table (\cref{tab:fer_sd:grassland_share}) with the sources used for estimating the range of values used for each country.

\textit{Argentina}: In the 1960s, fertilizer application rate in Argentina was primarily directed towards sugar cane and citrus \cite{CEPAL1966}, with minimal application to grasslands, nearly zero in 1964 \cite{CEPAL1966}. Throughout the 1970s and 1980s, the fertilizer application rate remained low, although there was a notable increase in \pfer{} application to grasslands, reaching 28\% country consumption in 1979 \cite{Martinez1982FertilizerProduction}. The substantial expansion in \nfer{} and \pfer{} fertilizer occurred during the 90s, leading to a slight rise in the share of \nfer{} used for grasslands, and to a significant decrease in \pfer{} share for grasslands \cite{FAO1992Fertilizer1, FAO1994Fertilizer2, FAO1996Fertilizer3, FAO1999Fertilizer4, FAO2002Fertilizer5}. To fill data gaps, we adopted a methodology similar to Lassaletta \textit{et al.} (2014) \cite{Lassaletta2014}, utilizing linear interpolation of national \cite{Argentina2006Grasslands, Argentina2011Grasslands, Argentina2012Grasslands, Argentina2013Grasslands, Argentina2014Grasslands, Argentina2015Grasslands, Argentina2016Grasslands, Argentina2017Grasslands, Argentina2018Grasslands} and global datasets for the years lacking data, with grasslands' fertilizer share assumed as 0 in 1965\cite{CEPAL1966}. Despite potential limitations, setting the share to 0, as done in \ac{FAO} nutrient budgets \cite{Ludemann2023A19612020}, may underestimate fertilizer application rate, particularly for \pfer{}. \kfer{} fertilizer application rate in Argentina remains minimal due to soil composition, with all reports except one considering it as 0 in the use for grasslands and fodder crops \cite{Martinez1982FertilizerProduction, AdolfoMartinez1990FertilizerT-37, FAO1992Fertilizer1, FAO1994Fertilizer2, FAO1996Fertilizer3, FAO1999Fertilizer4, FAO2002Fertilizer5, Argentina2006Grasslands, Argentina2011Grasslands, Argentina2012Grasslands, Argentina2013Grasslands, Argentina2014Grasslands, Argentina2015Grasslands, Argentina2016Grasslands, Argentina2017Grasslands, Argentina2018Grasslands}.

\textit{Brazil}: According to several sources, the use of fertilizer in Brazil's grasslands has been very low \cite{FAO2004Brazil, Boddey2003}. The most recent values reported by \ac{IFA} in 2014 and 2018 indicate that less than 1\% of the fertilizer used in Brazil is used in grasslands \cite{Heffer2017Assessment2014-2014/15, Ludemann2022GlobalCountry}. However, Lassaleta \textit{et al.} (2014)  \cite{Lassaletta2014} and \ac{FAO} \cite{Ludemann2023A19612020} considered higher percentages for \nfer{} and \kfer{} based on regional averages \cite{Lassaletta2014} or previous research \cite{Ludemann2023A19612020}. For \pfer{} and \kfer{}, only \ac{FAO} includes an estimation, considering 0 for \pfer{}, while they estimate the \kfer{} consumption by calculating the average between \nfer{} and \pfer{} consumption \cite{Ludemann2023A19612020}. We have decided to consider 0 as the share used for grasslands and fodder crops due to the latest reported values and considering that no information is reported in previous reports \cite{Martinez1982FertilizerProduction, AdolfoMartinez1990FertilizerT-37, FAO1992Fertilizer1, FAO1994Fertilizer2, FAO1996Fertilizer3, FAO1999Fertilizer4, FAO2002Fertilizer5, Heffer2017Assessment2014-2014/15, Ludemann2022GlobalCountry}.

\textit{Canada}: Most of the compiled reports do not provide information about the use of fertilizers for fodder crops and grasslands \cite{Martinez1982FertilizerProduction, AdolfoMartinez1990FertilizerT-37, FAO1992Fertilizer1, FAO1994Fertilizer2, FAO1996Fertilizer3,FAO1999Fertilizer4, FAO2002Fertilizer5}. The latest report, with 2018 data, indicated that 0.5\% of \nfer{}, 0.9\% of \pfer{}, and 0.6\% of \kfer{} fertilizers were allocated to permanent grasslands, which increased to 12\%, 14.5\%, and 25\% respectively when considering tame hay and silage maize as well. Regarding \nfer{}, FAO \cite{Ludemann2023A19612020} and the 2014 estimation by Lassaleta \textit{et al.} (2014) \cite{Lassaletta2014} are consistent with the 2018 estimation for all forages. However, the values for \pfer{} and \kfer{} for all forages in the latest report differ significantly from those used by FAO \cite{Ludemann2023A19612020} (0\% for \pfer{} and 5\% for \kfer{}). This discrepancy in \pfer{} may be due to \ac{FAO}'s reliance on Heffer \textit{et al.} (2017) which does not consider nongrass perennial crops 0\% \cite{Pogue2023}, and the discrepancy for \kfer{} because \ac{FAO} considered the average value between \nfer{} and \pfer{}\cite{Ludemann2023A19612020}. We decided to utilize the percentage for all forages included in the last report \cite{Ludemann2022GlobalCountry} for the entire period. We maintained the same values throughout the period due to insufficient data to estimate any trends. Additionally, in 1974, Beaton and Berger noted that a significant share of fertilizer used in Canada was for forages, estimating 45\% of total use in 1970 was for hay and grazing grasslands \cite{Beaton1974}. They suggested that their estimation might be overestimated; however, it is unlikely that the fraction of fertilizers used for forages was 0 between 1960 and 1990.

\textit{Chile}: Based on the estimations of the \ac{FAO} and \ac{IFA} reports, Lassaleta \textit{et al.} (2014)  \cite{Lassaletta2014} and \ac{FAO} \cite{Ludemann2023A19612020} considered a significant share of fertilizer used for grasslands. For \nfer{} Lassaleta \textit{et al.} (2014) suggested an increasing percentage from 0\% in 1960 to 20\% in 2005, while \ac{FAO} maintained a constant percentage of 20\%. For \pfer{} and \kfer{}, the values used by \ac{FAO} were also high, at 30\% and 25\% respectively. However, for Chile, using a constant value for the period overestimated the early years as the share used for grasslands for \nfer{} and \pfer{} was only 1\% at the beginning of the 1960s \cite{CEPAL1966b}. We therefore decided to make a reconstruction similar to the one demonstrated by Lassaleta \textit{et al.} (2014), by considering 1\% as the starting share for each nutrient, and incorporating the reported values for all grasslands \cite{Martinez1982FertilizerProduction, AdolfoMartinez1990FertilizerT-37, FAO1992Fertilizer1, FAO1994Fertilizer2, FAO1996Fertilizer3, FAO1999Fertilizer4, FAO2002Fertilizer5, Heffer2017Assessment2014-2014/15, Ludemann2022GlobalCountry}.

\textit{Dominican Republic}: The values reported in global studies from the 1990s indicate that during this decade, the percentage of fertilizer application rate on grasslands and fodder crops ranged between 2\% and 4\% \cite{FAO1992Fertilizer1, FAO1996Fertilizer3, FAO1999Fertilizer4}. Considering these findings, Lassaleta \textit{et al.} (2014)  \cite{Lassaletta2014} allocated values ranging from 0\% to 2\% for \nfer{}. We have chosen to utilize the average values from the three reports \cite{FAO1992Fertilizer1, FAO1996Fertilizer3, FAO1999Fertilizer4} for the period 1990-2020. This decision was influenced by the lack of available data since 1997, and by the emergence of fertilizer application rate for pasture as a new and increasing practice during the 90s \cite{Nuñez1999FertilizerDomRepublic}.

\textit{Mexico}: The use of fertilizer for grasslands and fodder crops appears to be nearly zero, as indicated by previous research \cite{Lassaletta2014, Ludemann2023A19612020} and reported values \cite{FAO1992Fertilizer1, FAO2002Fertilizer5, Heffer2017Assessment2014-2014/15, Ludemann2022GlobalCountry}. The only relevant fertilizer used for grasslands and fodder crops in Mexico appears to be related with \pfer{} related with alfalfa production \cite{Michaud2015AlfalfaDistribution,FAO1992Fertilizer1, FAO2002Fertilizer5}. Due to limited available information, and the longstanding presence of alfalfa production in Mexico since the Spanish colonization, we opted to consider the average percentage (2.5\%) used in the two reports with data for alfalfa \cite{FAO1992Fertilizer1, FAO2002Fertilizer5}.

\textit{United States of America}: According to global and national estimates from previous research, the share of \nfer{} used for grasslands during the period ranged from 0\% to 20\% of the total \cite{Cao2018NitrogenCropUS, Lassaletta2014, Ludemann2023A19612020}. For \pfer{} and \kfer{}, the most recent estimation from \ac{FAO} indicated a constant share of 0\% for phosphorus and 10\% for potassium \cite{Ludemann2023A19612020}. To estimate the total fertilizer use for permanent and non-permanent grasslands from 1959 to 2014, we used all the available data \cite{Nelson1968USFertilizerUse, Beaton1974, FAO1992Fertilizer1, FAO1996Fertilizer3, Heffer2017Assessment2014-2014/15}. In many sources, the information for grasslands is combined, encompassing both permanent and non-permanent grasslands. We used linear interpolation to estimate the share used for all grasslands together, replicating the method from the most recent estimation \cite{Cao2018NitrogenCropUS}. However, we included data from three additional years (1974, 1992, 1996) \cite{Beaton1974, FAO1992Fertilizer1, FAO1996Fertilizer3}, and also extended the estimation to cover \pfer{}, and \kfer{}. 

\textit{Uruguay}: Grassland fertilization was actively promoted by the Uruguayan government during the 60s \cite{Russell1974ForagesFertilizationTropical}. As early as 1963, one-third of the fertilizer used in the country was applied to pastures, with a focus on \pfer{} due to the low P content of the Uruguayan soils \cite{Russell1974ForagesFertilizationTropical}. These trends are reflected in the first \ac{IFDC} report, which allocated 45\% of the \pfer{} used in the country for grasslands and fodder in the year 1986 \cite{AdolfoMartinez1990FertilizerT-37}. However, this share decreased to 22\% by 2018. In contrast, the percentage of \nfer{} used for grasslands has shown an increasing trend, from almost 0\% in 1986 \cite{AdolfoMartinez1990FertilizerT-37} to 12\% in recent years \cite{FAO2002Fertilizer5, Ludemann2022GlobalCountry}. \kfer{} is not used for these agricultural lands in the country \cite{AdolfoMartinez1990FertilizerT-37, FAO1992Fertilizer1, FAO2002Fertilizer5, Ludemann2022GlobalCountry}. Given the significant variation in percentages between decades and nutrients, we performed linear interpolation considering 33\% for \pfer{} in 1960, and 0\% for \nfer{} as starting points, and all the values included in the reports \cite{AdolfoMartinez1990FertilizerT-37, FAO1992Fertilizer1, FAO2002Fertilizer5, Ludemann2022GlobalCountry}.

\textit{Venezuela}: Information regarding grassland and fodder crop fertilization in Venezuela is limited. Due to the scarcity of data and discrepancies between reported values \cite{FAO1996Fertilizer3, FAO2002Fertilizer5}, \ac{FAO} has considered a fertilization rate of 0\% for grasslands during the specified period. Conversely, Lassaleta \textit{et al.} (2014)  \cite{Lassaletta2014} proposed different rates between 0\% and 9\% from 1960 to 2009 for \nfer{}. Given the challenge of determining the most appropriate criteria, we opted to adhere to the \ac{FAO} considerations. This decision is influenced by low government optimal use recommendations for grasslands compared to croplands \cite{Casanova2005VenezuelaFertilizer}, along with scientific evidence suggesting minimal fertilization for warm-climate grasslands \cite{Dubeux2007WarmGrasslands, Casanova2005VenezuelaFertilizer}.

\textit{Australia}: According to Lassaletta \textit{et al.} (2014), the share of \nfer{} used for grasslands never exceeded 8.5\% \cite{Lassaletta2014}, which is similar to the 10\% used by \ac{FAO} in their nutrient budgets assessments \cite{Ludemann2023A19612020}. Despite an intensification in the use of \nfer{} in Australian grasslands over the past three decades \cite{Rawnsley2019}, it is noted that these grasslands were already being fertilized in the late 1950s, primarily with \kfer{} \cite{Barrow1968}. For instance, in 1956, 15\% of the \kfer{} used in South Australia was directed towards pastures, a figure that rose to 42\% by 1966 \cite{Barrow1968}. Therefore, we have opted to consider a constant share of 6.4\% for \nfer{} use since 1960 derived from the mean value of the reports \cite{ AdolfoMartinez1990FertilizerT-37, FAO1992Fertilizer1, FAO1994Fertilizer2, FAO1996Fertilizer3, FAO1999Fertilizer4, FAO2002Fertilizer5, Heffer2017Assessment2014-2014/15,  Ludemann2022GlobalCountry}. Regarding \pfer{} and \kfer{} fertilizer, it appears that the \ac{FAO} estimations \cite{Ludemann2023A19612020} may have underestimated their use, particularly for \kfer{}. Thus, we decided to use the average value of all reports, because even with fluctuations, the variation in the reported values since 1985 is not too high, resulting in figures of 38.4\ $\pm$ 4.1\% for \pfer{} and 41.6 $\pm$ 6.9\% for \kfer{} \cite{ AdolfoMartinez1990FertilizerT-37, FAO1992Fertilizer1, FAO1994Fertilizer2, FAO1996Fertilizer3, FAO1999Fertilizer4, FAO2002Fertilizer5, Heffer2017Assessment2014-2014/15,  Ludemann2022GlobalCountry}.

\textit{New Zealand}: Previous global research presented contradictory estimates of fertilizer application rate for grasslands in New Zealand \cite{Ludemann2023A19612020, Lassaletta2014}, with figures ranging widely from 0\% to 90\%. However, both global and national reports consistently support the notion that the majority of the fertilizer application rate in the country is directed towards grasslands and fodder crops \cite{FAO1996Fertilizer3, FAO2002Fertilizer5, Heffer2017Assessment2014-2014/15, Ludemann2022GlobalCountry, NZ2021}. Therefore, we have adopted a constant percentage throughout the entire period as grasslands have been the primary type of agricultural land developed in the country since the British colonization, their fertilization has been relevant since the early 20th century \cite{Williams1990GrasslandsNZ}, and the fraction used for grassland has remained constant at least in the last 30 years \cite{FAO1996Fertilizer3, FAO2002Fertilizer5, Heffer2017Assessment2014-2014/15, Ludemann2022GlobalCountry}. The percentages selected were derived from the average of global reports \cite{FAO1996Fertilizer3, FAO2002Fertilizer5, Heffer2017Assessment2014-2014/15, Ludemann2022GlobalCountry}: 91.1 $\pm$ 1.4\% for \nfer{}, 93.0 $\pm$ 3.3\% for \pfer{}, and 88.8 $\pm$ 4.4\% for \kfer{}.

\underline{Europe}: Between 1980 and 2000, Europe accounted for at least half of the \nfer{} fertilizer used for grasslands and fodder crops, while consuming less than one-third of the total global fertilizer consumption \cite{Xu2019GrasslandsN}. Consequently, the available information was broader, and the methods applied could be more comprehensive. Einarsson \textit{et al.} (2021) provided the most comprehensive estimation for \nfer{} in most European countries \cite{Einarsson2021}. They compiled and estimated the surfaces of croplands, including fodder crops, as well as temporary and permanent grasslands for the \ac{EU} countries spanning from 1960 to 2019. Using their compiled data and the fertilizer application rate information from our study, we employed a similar methodology to estimate the fraction of \nfer{}, \pfer{}, and \kfer{} used in these areas.

However, we extended the analysis to include fodder crops and all types of grasslands together, while also estimating \pfer{}, and \kfer{}.  
First, we used \cref{eq:fer_sd:equation_EU_grasslands1} to estimate the ratio ($R_{f-a}$) between the fertilization intensity of grasslands and fodder combined, and the fertilization intensity of all agricultural land for the years with available data:

\begin{equation}
    \label{eq:fer_sd:equation_EU_grasslands1}
    \dfrac{Q_f}{Q_a} =\dfrac{R_f\times{A_f}}{R_a\times{A_a}} \rightarrow \dfrac{Q_f\times{A_a}}{Q_a\times{A_f}} = \dfrac{R_f}{R_a} = R_{f-a}
\end{equation}

where $Q_{f}$ is the amount of fertilizer (\nfer{}, \pfer{}, or \kfer{}) used for grasslands and fodder crops, $Q_{a}$ denotes all the fertilizer of the same nutrient used in the country, $A_{f}$ represent the grassland and fodder surface, and $A_{a}$ represents the total agricultural land, and $R_{f-a}$ the ratio of fertilizer application rate per area between fodder and grasslands ($R_{f}$), and all agricultural land ($R_{a}$). Therefore, $R_{f-a}$ represents the fertilizer application relationship between fodder and grassland in comparison to all agricultural lands. 

After estimating the annual $R_{f-a}$, we used two different procedures and equations depending on the years for which $R_{f-a}$ data was available. If scientific literature and the observed variation in $R_{f-a}$ indicated significant differences across the years, we performed a linear interpolation of the available values and then applied \cref{eq:fer_sd:equation_EU_grasslands2}. Otherwise, we applied \eqref{eq:fer_sd:equation_EU_grasslands3}. To assess the variation in $R_{f-a}$ we estimated the \ac{MAE} of the results derived from \cref{eq:fer_sd:equation_EU_grasslands3} compared with all the reported values. When the variation of $R_{f-a}$ occurred only in some decades within the period, we combined \cref{eq:fer_sd:equation_EU_grasslands2} and \cref{eq:fer_sd:equation_EU_grasslands3}. Detailed explanations were provided for each country individually. For non-\ac{EU} countries, we applied similar procedures as those used for the other continents. In \cref{eq:fer_sd:equation_EU_grasslands2,eq:fer_sd:equation_EU_grasslands3} presented below, $\overline{R_{f-a}}$ is the average $R_{f-a}$ of all reports with data, and $i$ is the year. 

\begin{equation}
    \label{eq:fer_sd:equation_EU_grasslands2}
    \dfrac{Q_{f_i}}{Q_{a_i}} ={R_{f-a}}_i\times{}\dfrac{A_{f_i}}{A_{a_i}}
\end{equation}

\begin{equation}
    \label{eq:fer_sd:equation_EU_grasslands3}
    \dfrac{Q_{f_i}}{Q_{a_i}} =\overline{R_{f-a}}\times{}\dfrac{A_{f_i}}{A_{a_i}}
\end{equation}

\textit{Austria}: The methodology used by Einarsson \textit{et al.} (2021) \cite{Einarsson2021}  results in an almost constant percentage of \nfer{} used for permanent grasslands of $\approx$10\% for the 1960-2019 period. This result led \ac{FAO} to consider that 10\% of fertilizer used in agricultural lands was used for permanent grasslands \cite{Ludemann2023A19612020}. The intensification of grassland management began in the 1970s and 1980s \cite{Buchgraber2011}, and the share used for grasslands was higher in the late 1970s than in the 1990s \cite{Martinez1982FertilizerProduction, FAO1992Fertilizer1}. For \pfer{} and \kfer{}, \ac{FAO} considered a constant 10\% allocation for permanent grasslands \cite{Ludemann2023A19612020}, based on previous estimations for \pfer{} \cite{Zou2022} and the average value between the fraction used for \nfer{} for \kfer{}. While historical data suggest fluctuations in the percentage of fertilizers used for grasslands and fodder crops over time \cite{Martinez1982FertilizerProduction, AdolfoMartinez1990FertilizerT-37, FAO1992Fertilizer1, FAO1994Fertilizer2, FAO1996Fertilizer3, FAO1999Fertilizer4, FAO2002Fertilizer5}, the application of \cref{eq:fer_sd:equation_EU_grasslands3} using constant $\overline{R_{f-a}}$ values of 0.33 for \nfer{}, 0.46 for \pfer{}, and 0.32 for \kfer{}, and surfaces changes \cite{Einarsson2021}, provided an \ac{MAE} of 2.33 $\pm$ 3.09\%, 3.87 $\pm$ 3.47\%, 3.31 $\pm$ 2.29\% respectively. Only two errors higher than 10\% occurred,  both underestimations, namely -11.8 \% for \nfer{} in 1977\cite{Martinez1982FertilizerProduction}, and -10.2\% for \pfer{} for 1990 \cite{FAO1992Fertilizer1}, suggesting higher $R_{f-a}$ during the 1970-1990 period. Based on these results, we decided to utilize the mentioned $\overline{R_{f-a}}$ values for the period from 1991 to 2020 as well as for the period from 1961 to 1969. For the years from 1970 to 1990, we calculated the average $R_{f-a}$ from 1977 and 1990 reports \cite{Martinez1982FertilizerProduction, FAO1992Fertilizer1} to minimize the errors during the period.

\textit{Belgium and Luxembourg}: Belgium and Luxembourg often share statistics as a single entity in historical statistics. Consequently, we adopted the same estimation for both countries. According to Einarsson \textit{et al.} (2021) \cite{Einarsson2021}, the percentage of fertilizer application rate for permanent grasslands ranged from ~53\% in the 1980s to  ~40\% in the last years. They deem the \nfer{} fertilization of permanent grasslands significant throughout the period based on the little available information they found \cite{Einarsson2021}. The same literature confirmed the use \pfer{}, \kfer{} for grasslands as early as 1955, although with slightly lower applications \cite{Einarsson2021} as in the actual reports. The use of constant $R_{f-a}$ values of 1.03 for \nfer{}, 0.91 for \pfer{}, and 0.81 for \kfer{} based on the technical reports values \cite{Martinez1982FertilizerProduction, FAO1992Fertilizer1, FAO1994Fertilizer2, FAO1996Fertilizer3, FAO1999Fertilizer4, FAO2002Fertilizer5, EFMA2002Fertilizer2001/02, EFMA2007Fertilizer2006/07, PattrickHeffer2009Asessment2007/08, FertilizerEurope2012Fertilizer2011/12, FertilizerEurope2015Fertilizer2014/15, Ludemann2022GlobalCountry} resulted in \ac{MAE} values of 2.18 $\pm$ 1.82\% for \nfer{}, 5.46 $\pm$ 4.04\% for \pfer{}, 3.62 $\pm$ 2.51\% for \kfer{}. Only two instances of overestimations exceeding 10\% were observed for \pfer{} in the two last reports \cite{FertilizerEurope2015Fertilizer2014/15, Ludemann2022GlobalCountry}. This may be linked with the enforcement of limits on \pfer{} application in the Flanders region since 2011 \cite{Vandermoere2020BelgiumP}. Therefore, for \pfer{} we decided to use  the average $R_{f-a}$ for the 1960-2010 period, and use a linear interpolation of the $R_{f-a}$ values since 2011.

\textit{Czech Republic, Slovakia, and Czechoslovakia}: Information regarding grasslands and fodder crops before the disintegration of the Czechoslovak Republic is very limited \cite{Einarsson2021}. Following assumptions made by Einarsson \textit{et al.} (2021) \cite{Einarsson2021}, we extended the average $\overline{R_{f-a}}$ reported for the Czech Republic and Slovakia since 1993 \cite{FAO2002Fertilizer5, FertilizerEurope2012Fertilizer2011/12, FertilizerEurope2015Fertilizer2014/15, Ludemann2022GlobalCountry} through the period 1960-1992, considering surfaces changes, and the agricultural land of each country \cite{Einarsson2021}. Potential overestimations could occur for the early years, as the fertilization of these areas compared to other croplands might have been lower than in the 1990s, like in neighboring countries such as Hungary or Germany \cite{Nemeth2000HungaryN, Raup1950Germany}. After 1993, there are four years with available data for both countries \cite{FAO2002Fertilizer5, FertilizerEurope2012Fertilizer2011/12, FertilizerEurope2015Fertilizer2014/15, Ludemann2022GlobalCountry}. The $R_{f-a}$ values for all years are similar for each nutrient in each country, so we used \cref{eq:fer_sd:equation_EU_grasslands3} to estimate the 1993-2020 period. This approach resulted in low deviations from the reported values for the Czech Republic (\ac{MAE} = 2.08 $\pm$ 1.58\% for \nfer{}, 2.57 $\pm$ 1.30\% for \pfer{}, 1.69 $\pm$ 1.47\% for \kfer{}) and Slovakia (\ac{MAE} = 1.49 $\pm$ 1.47\% for \nfer{}, 2.02$\pm$ 2.87\% for \pfer{}, 1.79 $\pm$ 2.13\% for \kfer{}).

\textit{Denmark}: Danish grasslands and fodder crop fertilization have a long history with \nfer{}, with average rates of 45 and 17 kg ha\textsuperscript{-1} for temporary and permanent grasslands respectively in the early 1960s \cite{Petersen1977DenmarkFertilizer}. The usage of \cref{eq:fer_sd:equation_EU_grasslands3} for the whole period for the three nutrients resulted in large deviations (\ac{MAE} = 8.89 $\pm$ 4.40\% for \nfer{}, 5.36 $\pm$ 3.71\% for \pfer{}, 8.42 $\pm$ 5.67\% for \kfer{}). Therefore, as the amount of available data was large in the compiled technical reports we used \cref{eq:fer_sd:equation_EU_grasslands2}, and linear interpolation of all $R_{f-a}$ values for the period 1980-2020 \cite{Martinez1982FertilizerProduction, FAO1992Fertilizer1, FAO1994Fertilizer2, FAO1996Fertilizer3, FAO1999Fertilizer4, FAO2002Fertilizer5, EFMA2002Fertilizer2001/02, EFMA2007Fertilizer2006/07, FertilizerEurope2012Fertilizer2011/12, FertilizerEurope2015Fertilizer2014/15, Ludemann2022GlobalCountry}. For the 1960-1980 period, we utilized \nfer{} data from 9 years within that timeframe \cite{Petersen1977DenmarkFertilizer}. Additionally, we considered the 1980-2020 relationship between \nfer{} $\overline{R_{f-a}}$ and \pfer{} or \kfer{} $\overline{R_{f-a}}$, and the available \nfer{} data for estimating the 1960-1980 timeframe regarding the \pfer{} or \kfer{} values. We regard this assumption as the only available information for the period spanning 1960-1980 for \pfer{} and \kfer{} \cite{Beaton1974} suggests a similar relationship in the application rates for all forages between \nfer{} and the other nutrients, at least in the reported values since 1980 \cite{Martinez1982FertilizerProduction, FAO1992Fertilizer1, FAO1994Fertilizer2, FAO1996Fertilizer3, FAO1999Fertilizer4, FAO2002Fertilizer5, EFMA2002Fertilizer2001/02, EFMA2007Fertilizer2006/07, FertilizerEurope2012Fertilizer2011/12, FertilizerEurope2015Fertilizer2014/15, Ludemann2022GlobalCountry}.

\textit{Finland}: Einarsson \textit{et al.} (2021) \cite{Einarsson2021} did not consider significant fertilization on permanent grasslands in Finland, as they mainly use arable land for forage production \cite{Virkajärvi2015GrasslandFinland}. However, fodder crops and temporary grasslands are key parts of the agricultural production in the country \cite{Virkajärvi2015GrasslandFinland}, and they are commonly fertilized \cite{Martinez1982FertilizerProduction, FAO1992Fertilizer1, FAO1994Fertilizer2, FAO1996Fertilizer3, FAO1999Fertilizer4, FAO2002Fertilizer5, EFMA2002Fertilizer2001/02, EFMA2007Fertilizer2006/07, FertilizerEurope2012Fertilizer2011/12, FertilizerEurope2015Fertilizer2014/15, Ludemann2022GlobalCountry}. Using \cref{eq:fer_sd:equation_EU_grasslands3} for the entire period across the three nutrients resulted in minimal deviations for \nfer{} and \pfer{} (\ac{MAE} = 1.57 $\pm$ 2.99\%, 2.10 $\pm$ 3.34\% respectively), but substantial deviations for \kfer{} (7.51 $\pm$ 7.38\% ). Given the substantial deviation for \kfer{}, and the large bias for ${R_{f-a}}$ in 1979 \cite{Martinez1982FertilizerProduction} for \nfer{} and \pfer{}, the first year with available data, we opted to use \cref{eq:fer_sd:equation_EU_grasslands2}, and the linear interpolation of the ${R_{f-a}}$. However, potential deviations may arise for the 1960s, as fertilizers were predominantly utilized for high-value crops during the early part of the decade \cite{Edwards1966FinlandAgriculture60}, yet no data are available for that period. 

\textit{France}: Data regarding grasslands and fodder crop fertilization is less limited than in the majority of \ac{EU} countries, although large differences exist between the available data. Two recent publications estimated the share of \nfer{} and \pfer{} used for permanent grasslands since 1960 \cite{Einarsson2021, LeNoe2018FranceNutrients} based on country surveys at the region-level \cite{SSP1984_EnqueteFrancePraire, SSP2001_EnqueteFrance, SSP2006_EnqueteFrance, SSP2020_EnqueteFrance}. However, the results obtained by them differ from the \ac{FUBC}-\ac{FAO} and \ac{FUBC}-\ac{FE} reports \cite{FAO1992Fertilizer1, FAO1994Fertilizer2, FAO1996Fertilizer3, FAO1999Fertilizer4, FAO2002Fertilizer5, EFMA2002Fertilizer2001/02, EFMA2007Fertilizer2006/07, FertilizerEurope2012Fertilizer2011/12, FertilizerEurope2015Fertilizer2014/15, Ludemann2022GlobalCountry}. For example, for 2006, Le Nöe  \textit{et al.} (2018) \cite{LeNoe2018FranceNutrients} report a share of \pfer{} used for permanent grasslands of 27\% whereas the \ac{FE} reports a value for all grasslands of 20\%. Considering other years with comparable data, such as 1990 or 2017,  Einarsson \textit{et al.} (2021) \cite{Einarsson2021} estimate a share of 16\% and 7\% respectively for \nfer{} used for permanent grasslands, while \ac{FAO} only reports 6\% for 1990, and the national survey reports 4.7\% for 2017 \cite{SSP2020_EnqueteFrance}. Therefore, as it is difficult to discern the more accurate value between the two estimations, we opted to use the average between the ${R_{f-a}}$ linear interpolated data from the global datasets \cite{FAO1992Fertilizer1, FAO1994Fertilizer2, FAO1996Fertilizer3, FAO1999Fertilizer4, FAO2002Fertilizer5, EFMA2002Fertilizer2001/02, EFMA2007Fertilizer2006/07, FertilizerEurope2012Fertilizer2011/12, FertilizerEurope2015Fertilizer2014/15, Ludemann2022GlobalCountry}, and from the national surveys \cite{SSP1984_EnqueteFrancePraire, SSP2001_EnqueteFrance, SSP2006_EnqueteFrance, SSP2020_EnqueteFrance}, considering for both as 0 the share in 1955 \cite{LeNoe2018FranceNutrients} and the single estimate for the 70s \cite{Beaton1974}.

\textit{Germany}:  The availability of data since the German reunification is substantial in global reports \cite{FAO1992Fertilizer1, FAO1994Fertilizer2, FAO1996Fertilizer3, FAO1999Fertilizer4, FAO2002Fertilizer5, EFMA2002Fertilizer2001/02, EFMA2007Fertilizer2006/07, FertilizerEurope2012Fertilizer2011/12, FertilizerEurope2015Fertilizer2014/15}. These reports suggest a decline since 1990 in fertilizer use for all forages compared to the rest of croplands, with the drop being particularly notable for \nfer{} and \pfer{}. As a result, we decided to use \cref{eq:fer_sd:equation_EU_grasslands2}, and interpolate the ${R_{f-a}}$ values, instead of $\overline{{R_{f-a}}}$ for the 1990-2020 period. For the 1960-1989 period, data on grassland and fodder fertilization is scarce and primarily pertains to West Germany \cite{Beaton1974}. Most of the data available for the period are relative to \nfer{}, except the 1982 \ac{IFDC}-\ac{FUBC} report. For the 1960-1989 period, We decided to use the linear interpolation assuming, similar to the case of France, zero fertilization of grasslands and fodder crops in 1955, as fertilization of these areas in Western Germany, where most of this agricultural land is located, was minimal before 1960 \cite{Raup1950Germany}, using the only report with available data for the three nutrients \cite{Martinez1982FertilizerProduction}. We extrapolate the data from Western Germany for the entire country due to data availability \cite{Martinez1982FertilizerProduction, Beaton1974, Einarsson2021}, the prevalence of these agricultural areas in Germany \cite{Einarsson2021}, and because grassland fertilization in East Germany was similar to that in West Germany, at least in the late 1970s \cite{Pearson1979_EastGermany}. Using these approaches, we deviate by approximately 3.9\% from the \nfer{} estimated data for the year 1974 \cite{Beaton1974}. Additionally, we deviated by about 10\% from the \nfer{} value for permanent grasslands reported by Einarsson \textit{et al.} (2021) for 1966 (based on real data) \cite{Einarsson2021}. This deviation is reasonable, considering that the average difference between $Q_{f}/Q_{a}$ only using information for permanent grasslands or all forages for \nfer{} is 7.9\% \cite{FAO1992Fertilizer1, FAO1994Fertilizer2, FAO1996Fertilizer3, FAO1999Fertilizer4, FAO2002Fertilizer5, EFMA2002Fertilizer2001/02, EFMA2007Fertilizer2006/07, FertilizerEurope2012Fertilizer2011/12, FertilizerEurope2015Fertilizer2014/15}. 

\textit{Greece}: Fertilization has not been considered for permanent grasslands in either previous research \cite{Lassaletta2014, Ludemann2023A19612020, Einarsson2021} or technical reports \cite{FAO1992Fertilizer1, FAO1994Fertilizer2, FAO1996Fertilizer3, FAO1999Fertilizer4, FAO2002Fertilizer5, EFMA2002Fertilizer2001/02, EFMA2007Fertilizer2006/07, FertilizerEurope2012Fertilizer2011/12, FertilizerEurope2015Fertilizer2014/15, Ludemann2022GlobalCountry}. However, since we are also considering fertilization for fodder crops, the technical reports have allocated fertilization for them, especially for alfalfa and sillage maize \cite{FAO1992Fertilizer1, FAO1994Fertilizer2, FAO1996Fertilizer3, FAO1999Fertilizer4, FAO2002Fertilizer5, EFMA2002Fertilizer2001/02, EFMA2007Fertilizer2006/07, FertilizerEurope2012Fertilizer2011/12, FertilizerEurope2015Fertilizer2014/15, Ludemann2022GlobalCountry}, which constitute the two main actual fodder crops in the country \cite{Einarsson2021}. Therefore, we used \cref{eq:fer_sd:equation_EU_grasslands2} and the linear interpolation of ${R_{f-a}}$ because the values of the 1990s are lower than the actual ones, and we have assumed a zero level of fodder fertilization in 1960, as it was only experimental in the country \cite{Panos1961GreeceGrasslands}.

\textit{Hungary}: Einarsson \textit{et al.} (2021) \cite{Einarsson2021} did not consider fertilization for permanent grasslands due to the scarcity of the data and because grassland fertilization is not a common practice nowadays  \cite{Einarsson2021}. Reported values suggest that a significant fraction, approximately 5\% of the fertilizer used since 1990 in the country was allocated to grasslands and fodder crops \cite{FAO1992Fertilizer1, EFMA2007Fertilizer2006/07, FertilizerEurope2012Fertilizer2011/12, FertilizerEurope2015Fertilizer2014/15, Ludemann2022GlobalCountry}, with an even higher proportion during the 1980s \cite{Martinez1982FertilizerProduction}. Scientific information confirms that the change in the political regime in 1989 was a key driver of fertilization practices in the country, reducing the fertilizer use by five-fold in the country, and limiting fertilization of these areas to managed grasslands \cite{Hou2015HungaryN}. Furthermore, fertilization in the country commenced in the 1960s and remained stagnant during the 1980s \cite{Nemeth2000HungaryN}. Therefore, for the period 1960-1989, we applied \cref{eq:fer_sd:equation_EU_grasslands2}, and the linear interpolation of ${R_{f-a}}$ from a 0 value in 1960, to the 1980 reported value \cite{Martinez1982FertilizerProduction}. For the 1990-2020 period, we used \cref{eq:fer_sd:equation_EU_grasslands3}, and the average $\overline{R_{f-a}}$, as there is no deviation larger than 10\% from the reported values using this method. 

\textit{Ireland}: Ireland is likely one of the countries that use a larger proportion of fertilizers for grasslands and fodder crops \cite{Ludemann2023A19612020, Einarsson2021, Lassaletta2014}, and also has more available information. Since 1972, six national surveys have been conducted, providing data for 22 years \cite{Murphy1978IrelandSurvey, Murphy1983IrelandSurvey, Murphy1987IrelandSurvey, Coulter2005IrelandSurvey, Lalor2010IrelandSurvey, Dillon2018IrelandSurvey}. Moreover, the global datasets also include information from ten different years since 1987 \cite{FAO1992Fertilizer1, FAO1994Fertilizer2, FAO1996Fertilizer3, FAO1999Fertilizer4, FAO2002Fertilizer5, EFMA2002Fertilizer2001/02, EFMA2007Fertilizer2006/07, FertilizerEurope2012Fertilizer2011/12, FertilizerEurope2015Fertilizer2014/15, Ludemann2022GlobalCountry}. For the 1986-2020 period, we used the average of the linear interpolation of the ${R_{f-a}}$ values based on national surveys \cite{Murphy1978IrelandSurvey, Murphy1983IrelandSurvey, Murphy1987IrelandSurvey, Coulter2005IrelandSurvey, Lalor2010IrelandSurvey, Dillon2018IrelandSurvey}, and surfaces data \cite{IrelandCSOSurface1980,IrelandCSOSurface1990, IrelandCSOSurface2000, IrelandCSOSurface2010, IrelandCSOSurface2020}, along with the ${R_{f-a}}$ values based on the global datasets \cite{FAO1992Fertilizer1, FAO1994Fertilizer2, FAO1996Fertilizer3, FAO1999Fertilizer4, FAO2002Fertilizer5, EFMA2002Fertilizer2001/02, EFMA2007Fertilizer2006/07, FertilizerEurope2012Fertilizer2011/12, FertilizerEurope2015Fertilizer2014/15, Ludemann2022GlobalCountry} and the Einarsson \textit{et al.} (2021) surface compilation \cite{Einarsson2021}. We excluded ${R_{f-a}}$ values based on the global datasets \cite{FAO1992Fertilizer1, FAO1994Fertilizer2, FAO1996Fertilizer3, FAO1999Fertilizer4, FAO2002Fertilizer5, EFMA2002Fertilizer2001/02, EFMA2007Fertilizer2006/07, FertilizerEurope2012Fertilizer2011/12, FertilizerEurope2015Fertilizer2014/15, Ludemann2022GlobalCountry} and the Einarsson \textit{et al.} (2021) surface compilation \cite{Einarsson2021} for the 2006-2010 period due to a change in the criteria for temporary grassland surface, which resulted in overestimations ($Q_{f}/Q_{a}$ > 1). For the 1960-1985 period, we only considered the linear interpolation of the available data, all from the national surveys ${R_{f-a}}$ \cite{Murphy1978IrelandSurvey, Murphy1983IrelandSurvey, Murphy1987IrelandSurvey}, and surfaces \cite{IrelandCSOSurface1980,IrelandCSOSurface1990}. In cases where there was no available surface data \cite{IrelandCSOSurface1980} in the national databases, like 1972, we used the closest year with available data (e.g., 1970). For 2008, which has two available national surveys \cite{ Lalor2010IrelandSurvey, Dillon2018IrelandSurvey}, we took the average of both. We considered the share of fertilizer used for grasslands and fodder crops as zero in 1955 because almost all fertilizer was used for tillage crops in that year \cite{Walsh1957Ireland}, with grassland fertilization increasing during the 1960s \cite{Heavy1969Ireland}. 

\textit{Italy}: Einarsson \textit{et al.} (2021) used a constant $\overline{R_{f-a}}$ for permanent grasslands for all years, as similar values are given in various reports and scientific information \cite{Einarsson2021}. When considering grasslands and fodder crops together, the ${R_{f-a}}$ were also consistent for each nutrient over all years \cite{Beaton1974, FAO1994Fertilizer2, FAO1996Fertilizer3, FAO1999Fertilizer4, FAO2002Fertilizer5, EFMA2002Fertilizer2001/02, EFMA2007Fertilizer2006/07, FertilizerEurope2012Fertilizer2011/12, FertilizerEurope2015Fertilizer2014/15, Ludemann2022GlobalCountry}, even including the 1974 data \cite{Beaton1974}. The \ac{MAE} using \cref{eq:fer_sd:equation_EU_grasslands3} for the entire period across the three nutrients resulted in minimal deviations comparing with the reported values \cite{Beaton1974, FAO1994Fertilizer2, FAO1996Fertilizer3, FAO1999Fertilizer4, FAO2002Fertilizer5, EFMA2002Fertilizer2001/02, EFMA2007Fertilizer2006/07, FertilizerEurope2012Fertilizer2011/12, FertilizerEurope2015Fertilizer2014/15, Ludemann2022GlobalCountry} (\ac{MAE} = 2.24 $\pm$ 1.55\% for \nfer{}, 2.00 $\pm$ 1.37\% for \pfer{}, and 3.21 $\pm$ 1.21\% for \kfer{}). Therefore, we used the the $\overline{R_{f-a}}$ for the three nutrients. However, there could be potential overestimations for the 1960s decade because nearby countries like France or Germany did not use fertilizers for these agricultural lands before 1955 \cite{LeNoe2018FranceNutrients}.

\textit{The Netherlands}: Information regarding grassland fertilization in the country is abundant \cite{Prins1983Netherlands, Einarsson2021}. However, before the development of global datasets, information regarding \pfer{} and \kfer{} is very limited. For the period 1979-2019, we used \cref{eq:fer_sd:equation_EU_grasslands2} considering the linear interpolation of the eleven ${R_{f-a}}$ data derived from the global datasets \cite{Martinez1982FertilizerProduction, FAO1996Fertilizer3, FAO1999Fertilizer4, FAO2002Fertilizer5, EFMA2002Fertilizer2001/02, EFMA2007Fertilizer2006/07, FertilizerEurope2012Fertilizer2011/12, FertilizerEurope2015Fertilizer2014/15, Ludemann2022GlobalCountry} and the agricultural surfaces changes \cite{Einarsson2021}. We used the global datasets instead of the national data available because they provide information regarding the three nutrients. For the years 1960 to 1979, we used the available compilation of \nfer{} application rates \cite{Prins1983Netherlands}, and the total \nfer{} fertilizer consumption \cite{FAOSTAT2023FertilizerDataset} to estimate the $Q_{f}/Q_{a}$ values for \nfer{}. For \pfer{} and \kfer{}, we used the ratio between the $Q_{f}/Q_{a}$ used for \nfer{} and these two nutrients for the most recent year with available data, 1979 \cite{Martinez1982FertilizerProduction}, to extrapolate the results for the 1960-1979 period. 

\textit{Poland}: The available data in reports from the period 1988-2018 \cite{FAO1992Fertilizer1, FAO1999Fertilizer4, FAO2002Fertilizer5, EFMA2002Fertilizer2001/02, EFMA2007Fertilizer2006/07, FertilizerEurope2012Fertilizer2011/12, FertilizerEurope2015Fertilizer2014/15, Ludemann2022GlobalCountry} did not show a constant ${R_{f-a}}$ for any nutrient \nfer{}, \pfer{} and \kfer{}. Data on fertilization before 1989, during the communist government, is sparse \cite{Einarsson2021, FAO1992Fertilizer1}. However, similar to other Eastern European countries like Hungary, it appears that fertilizer intensification in the country started during the 1960s \cite{Kurek1971PolandSurvey}, with a significant drop following the regime change \cite{FAOSTAT2023FertilizerDataset}. As a result, we adopted the same criteria used for other Eastern European countries, setting the 1960 value to zero, and applying two distinct linear interpolations of ${R_{f-a}}$: one for the 1960-1989 period, and another for the 1990-2020 period. For the 1990-2020 period, there are seven years with available data, whereas for the 1960-1989 only 1989 has data. Despite this limited data for the earlier period, survey estimates \cite{Kurek1971PolandSurvey} combined with FAOSTAT totals \cite{FAOSTAT2023FertilizerDataset} suggest that the combined share of the three nutrients was between 14\% and 15\% in the late 1960s, which aligns with the individual nutrient shares calculated by the linear interpolation which are between 10\% and 13\%. 

\textit{Portugal}: Einarsson \textit{et al.} (2021) did not consider fertilization of permanent grasslands, citing the relatively low surface area in the country \cite{Einarsson2021}. However, recent technical reports suggest that $Q_{f}/Q_{a}$ exceeds 20\% for the three major nutrients \cite{FAO1996Fertilizer3, FAO1999Fertilizer4, FAO2002Fertilizer5, EFMA2002Fertilizer2001/02, EFMA2007Fertilizer2006/07, FertilizerEurope2012Fertilizer2011/12, FertilizerEurope2015Fertilizer2014/15, Ludemann2022GlobalCountry}. We chose to apply \cref{eq:fer_sd:equation_EU_grasslands2} and to interpolate the 1977-2020 data \cite{Martinez1982FertilizerProduction, FAO1992Fertilizer1, FAO1994Fertilizer2, FAO1999Fertilizer4, FAO2002Fertilizer5, EFMA2002Fertilizer2001/02, EFMA2007Fertilizer2006/07, FertilizerEurope2012Fertilizer2011/12, FertilizerEurope2015Fertilizer2014/15, Ludemann2022GlobalCountry} because using \cref{eq:fer_sd:equation_EU_grasslands2} led to discrepancies greater than 10\% in some years. For the years before 1977, we retained the ${R_{f-a}}$ 1977 values \cite{Martinez1982FertilizerProduction} (which resulted in $Q_{f}/Q_{a}$ < 2\%) as there is no information for the earlier period.

\textit{Romania}: As with other Eastern European countries, there is no available information regarding grassland and fodder crop fertilization before the political regime change in 1989. However, between 1990 and 2020, data from five years suggest that about 5\% of fertilizer is used for grasslands and fodder crops \cite{FAO1992Fertilizer1, FAO1996Fertilizer3, FertilizerEurope2012Fertilizer2011/12, FertilizerEurope2015Fertilizer2014/15, EFMA2007Fertilizer2006/07}. For Romania, we applied \cref{eq:fer_sd:equation_EU_grasslands3}, using the average $\overline{R_{f-a}}$ value and the grassland and cropland surface data \cite{Einarsson2021}. Potential overestimations occurred during the first decades, although the estimated $Q_{f}/Q_{a}$ are less than 5\% for the first decades. 

\textit{Spain}: Previous research has not considered the fertilization of permanent grassland because this practice in the country is very uncommon \cite{Ludemann2023A19612020, Einarsson2021}. However, when considering temporary grasslands and fodder crops, this assumption changes, as forage crops occupy about 8\% of the arable land in the country and consume nearly the same percentage of fertilizers \cite{GarciaSerranoSpain2010}. To estimate the share of fertilizer use in these areas, we created a linear interpolation of the ${R_{f-a}}$ data from the ten years with available data, ranging from 1979 to 2014, and applied \cref{eq:fer_sd:equation_EU_grasslands2}. Using \cref{eq:fer_sd:equation_EU_grasslands3} resulted in estimations that were twice the reported values for the earlier years. Given the fraction used for these areas in 1979 was minimal ($Q_{f}/Q_{a}$ < of 2\%), potential overestimations for the first years are also likely minimal.

\textit{Sweden}: In the country, fertilization of forage production areas is closely linked to the transition from natural permanent grassland to temporary grassland production on arable land that occurred during the first part of the 20th century, especially during the 1940s and 1950s \cite{Aberg1955Sweden}. Moreover, based on the available data, fertilizer intensification of these areas compared to other croplands ${R_{f-a}}$ was lower during the 1970s than at the end of the century \cite{Beaton1974, FAO1999Fertilizer4, FAO2002Fertilizer5}. Therefore, we applied \cref{eq:fer_sd:equation_EU_grasslands2} and performed the linear interpolation of the ${R_{f-a}}$ of each nutrient of the 11 years with available data since 1974 \cite{Beaton1974, AdolfoMartinez1990FertilizerT-37, FAO1992Fertilizer1, FAO1994Fertilizer2, FAO1999Fertilizer4, FAO2002Fertilizer5, EFMA2002Fertilizer2001/02, EFMA2007Fertilizer2006/07, FertilizerEurope2012Fertilizer2011/12, FertilizerEurope2015Fertilizer2014/15, Ludemann2022GlobalCountry}. A slight overestimation might occur for the earlier years, as the intensification of these areas was increasing before the first year with available data \cite{Aberg1955Sweden}, but no data for the period was found.

\textit{United Kingdom and Northern Ireland (\ac{UK})}: The \ac{UK} has the world's longest and most complete dataset on the fertilization of grasslands and croplands \cite{DEFRA2023FertiliserUKdataset}. Annual time series data on fertilizer use for permanent and temporary grasslands are available for England and Wales since 1969 and for Great Britain since 1982 \cite{DEFRA2023FertiliserUKdataset}. Northern Ireland is not included in these surveys. Additionally, there are surveys for the years 1957, 1962, and 1966 for England and Wales \cite{Church1971EnglandSurvey}. Two problems arise for the estimation of $Q_{f}/Q_{a}$ from this data. The first one is that the surveys only include fertilization on permanent and temporary grassland, excluding rough grazing. The second challenge is that there is no information for Northern Ireland - which accounts for about 6\% of the country's fertilizer consumption \cite{DEFRA2023FertiliserUKdataset}-, and from 1960 to 1982, there is also no data for Scotland, who are responsible for about 14\% of the country's fertilizer consumption \cite{DEFRA2023FertiliserUKdataset}. For the period 1982-2019, we used the annual fertilizer application rates for Great Britain's tillage crops \cite{DEFRA2023FertiliserUKdataset} and the corresponding cropland surface area \cite{DEFRA2024AreasUK} (excluding temporary grasslands) to estimate the total fertilizer use for croplands. We considered grassland fertilization to be the complement of the value obtained, assuming the same application rates for Northern Ireland. To include these estimations in the fraction used for fodder crops, we add the average share used for them, which is less than the 3\% for all nutrients \cite{FAO1992Fertilizer1, FAO1996Fertilizer3, EFMA2002Fertilizer2001/02, EFMA2007Fertilizer2006/07, FertilizerEurope2012Fertilizer2011/12, FertilizerEurope2015Fertilizer2014/15, Ludemann2022GlobalCountry}. For the period 1960-1981, we applied the same methodology but using the application rates \cite{Church1971EnglandSurvey, DEFRA2023FertiliserUKdataset} and surfaces \cite{DEFRA2024AreasEngland} from England and Wales, adjusted by -2.5\% for \nfer{}, +2.8\% for \pfer{}, and +0.9\% for \kfer{}. These adjustments are based on the observed differences between the application rates in Great Britain and those in England and Wales during the 1980s decade. Moreover, for the 1960s decade for which there are no data available for all years, we applied the linear interpolation of the years with data. We used the national databases instead of the global datasets because they provide annual information covering almost the entire period for the three nutrients, and the values between them were quite similar.

\textit{Iceland}: Iceland's agriculture sector is primarily focused on livestock production, with about 90\% of its agricultural land being permanent grasslands \cite{Johannesson2010Iceland}. Additionally, most of the arable land is used for forage crops \cite{Johannesson2010Iceland}. While grassland fertilization is a common practice in Iceland \cite{Helgadottir2013Iceland}, there is limited information on application rates for different types of agricultural land, and no specific estimates on the proportion of fertilizer used for forage crops in the country. When we applied \cref{eq:fer_sd:equation_EU_grasslands3} using the average ${R_{f-a}}$ from other Nordic countries—Denmark, Sweden, and Finland, it resulted in a $Q_{f}/Q_{a}$ ratio greater than 100\%. To address this, we allocated a mid-value between 100\% and the proportion of agricultural land occupied by grasslands and fodder crops, ensuring it does not exceed 100\%.

\textit{Switzerland}: Data on fodder crop and grassland fertilization in the country from the period 1979-1999 suggest that between 30 and 50\% of the fertilizer used in the country is applied to these lands \cite{Martinez1982FertilizerProduction, FAO1992Fertilizer1, FAO1994Fertilizer2, FAO1996Fertilizer3, FAO1999Fertilizer4, FAO2002Fertilizer5}. However, whereas the data of the first two years indicate that almost 50\% of \nfer{} is used for grasslands and fodder crops \cite{Martinez1982FertilizerProduction, FAO1992Fertilizer1}, only about 30\% was used in 1999 \cite{FAO2002Fertilizer5}. Since 2000, the areas of artificial grasslands and silage maize (the two main forages that receive fertilizers \cite{FAO1992Fertilizer1}) have remained almost constant \cite{SwtizerlandOFSSurface}. As there is no information available regarding grassland fertilization before 1979 or after 2000, we used the 1979 data for the period 1960-1979 and the 2000 data for the period 2000-2020. For the period from 1979 to 2000, we applied linear interpolation to the six years with available data \cite{Martinez1982FertilizerProduction, FAO1992Fertilizer1, FAO1994Fertilizer2, FAO1996Fertilizer3, FAO1999Fertilizer4, FAO2002Fertilizer5}.

\textit{Norway}: Fodder crops and grasslands (both temporary and permanent) play a key role in the agricultural sector of the country \cite{Nordgård1975NorwayAgriculture,Steinshamn2016NorwayGrasslands}. Technical reports and scientific studies data indicate a nearly constant share of $Q_{f}/Q_{a}$ for \nfer{}, \pfer{}, and \kfer{} \cite{Beaton1974, FAO1994Fertilizer2, FAO1996Fertilizer3, FAO1999Fertilizer4, FAO2002Fertilizer5, EFMA2002Fertilizer2001/02, EFMA2007Fertilizer2006/07, FertilizerEurope2012Fertilizer2011/12, FertilizerEurope2015Fertilizer2014/15, Ludemann2022GlobalCountry}. Therefore, we used the average of all the available $Q_{f}/Q_{a}$ data \cite{Beaton1974, FAO1994Fertilizer2, FAO1996Fertilizer3, FAO1999Fertilizer4, FAO2002Fertilizer5, EFMA2002Fertilizer2001/02, EFMA2007Fertilizer2006/07, FertilizerEurope2012Fertilizer2011/12, FertilizerEurope2015Fertilizer2014/15, Ludemann2022GlobalCountry}, covering the period 1974-2018 for \nfer{}, and from the period 1990-2018 for \pfer{} and \kfer{}. The resulting values, with a share of 64.02\% $\pm$ 1.76\% for \nfer{}, 50.02\%  $\pm$ 2.25\% for \pfer{}, and 65.59\% $\pm$ 6.07\% for \kfer{}, were comparable to those estimated for other Scandinavian countries.

\textit{Yugoslav Socialist Federal Republic, and actual former countries}: Fodder crops and grasslands played a significant role in the agricultural production of the Yugoslav \ac{SFR} \cite{YugoslaviaReport1995}. Pastures and meadows occupied 33\% of the country's land, while fodder crops took up 20\% of the arable land \cite{YugoslaviaReport1995}. However, to the best of our knowledge, no information is available regarding fertilization for different agricultural lands before the dissolution of the country. After the dissolution, information became available in global reports for Croatia and Slovenia, but not in the other countries \cite{FAO1994Fertilizer2, FAO2002Fertilizer5, EFMA2002Fertilizer2001/02, EFMA2007Fertilizer2006/07, FertilizerEurope2012Fertilizer2011/12, FertilizerEurope2015Fertilizer2014/15, Ludemann2022GlobalCountry}. To estimate the $Q_{f}/Q_{a}$ values for Yugoslav \ac{SFR} during the period 1961-1991, we used the weighted average by agricultural land surface\cite{Einarsson2021} of the earliest ${R_{f-a}}$ values from Croatia and Slovenia \cite{FAO1994Fertilizer2, EFMA2002Fertilizer2001/02, Einarsson2021}, given that their ${R_{f-a}}$ values have changed significantly in recent years \cite{FAO1994Fertilizer2, FAO2002Fertilizer5, EFMA2002Fertilizer2001/02, EFMA2007Fertilizer2006/07, FertilizerEurope2012Fertilizer2011/12, FertilizerEurope2015Fertilizer2014/15, Ludemann2022GlobalCountry}. We also considered the cropland, grasslands, and fodder crop surfaces of Yugoslavia \ac{SFR} from the 1990s \cite{YugoslaviaReport1995} to estimate the $Q_{f}/Q_{a}$ used for the 1961-1991 period. For the period 1990-2019, for actual \ac{EU} former countries, we performed the linear interpolation of the ${R_{f-a}}$ values \cite{FAO1994Fertilizer2, FAO2002Fertilizer5, EFMA2002Fertilizer2001/02, EFMA2007Fertilizer2006/07, FertilizerEurope2012Fertilizer2011/12, FertilizerEurope2015Fertilizer2014/15, Ludemann2022GlobalCountry} to estimate $Q_{f}/Q_{a}$ considering the annual surfaces values \cite{Einarsson2021}. In Serbia, the largest country, forage production is a crucial component of its agricultural sector, with about two-fifths of the agricultural land dedicated to this purpose \cite{Lugic2010Serbia}. However, as no specific information on fertilization rates has been found. We considered the average weighted ${R_{f-a}}$ ratio of Croatia and Slovenia along with the 2004-2008 surfaces of agricultural lands, grasslands, and fodder crops \cite{Lugic2010Serbia}. For smaller countries like Montenegro of North Macedonia, we assumed the average annual $Q_{f}/Q_{a}$ values of Serbia and Croatia.

\textit{\ac{USSR} and Former \ac{USSR} Countries}: Quantitative and qualitative information about fertilization of grassland and fodder crops before the collapse of the \ac{USSR} is quite scarce \cite{Loza1977USSR1975, Klatt1976USSR, FAO1992Fertilizer1}. Some publications suggest that the use of fertilizers in these areas was minimal before 1975 \cite{Loza1977USSR1975, Klatt1976USSR}. However, data from 1990-1991, just before the collapse, from certain republics (Russia, Latvia, Estonia, or Belarus) indicate that a significant share of fertilizers was used for fodder crops and grasslands \cite{FAO1992Fertilizer1} (e.g., 40\% for \nfer{} in the Russian Federation \cite{FAO1992Fertilizer1}).
For the period 1960-1991, we estimated the ${R_{f-a}}$ for the entire \ac{USSR} in 1990, weighing the value of each republic ${R_{f-a}}$ \cite{FAO1992Fertilizer1, Shend1993USSRStats} in 1990-1991 by the total fertilizer use of each republic\cite{FAO1992Fertilizer1, Shend1993USSRStats}. The four republics with available data for this year (Russian Federation, Belarus, Latvia, and Estonia) account for 40\% of the agricultural land of the country and 62\% of its fertilizer consumption \cite{Shend1993USSRStats}. After estimating ${R_{f-a}}$ for each nutrient in 1990, we used linear interpolation to estimate the annual ${R_{f-a}}$ values, considering the value in 1975 as zero \cite{Loza1977USSR1975, Klatt1976USSR}. Finally, similar to the \ac{EU} countries, we considered the annual cropland, grassland, and fodder crop surfaces \cite{Shend1993USSRStats}, along with the calculated ${R_{f-a}}$, to estimate the annual $Q_{f}/Q_{a}$. For the period from 1992 to 2020, we considered individual country information where some data was available. However, for the following actual countries, there is no information in the global reports \cite{FAO1992Fertilizer1, FAO1994Fertilizer2, FAO1996Fertilizer3, FAO1999Fertilizer4, FAO2002Fertilizer5, EFMA2002Fertilizer2001/02, EFMA2007Fertilizer2006/07, PattrickHeffer2009Asessment2007/08, FertilizerEurope2012Fertilizer2011/12, FertilizerEurope2015Fertilizer2014/15, Ludemann2022GlobalCountry}: Armenia, Georgia, Kazakhstan, Kyrgyzstan, Tajikistan, and Turkmenistan. For all these countries, we considered a constant $Q_{f}/Q_{a}$ ratio during the 1992-2020 period due to the limited information. For Armenia and Georgia, we assumed the $Q_{f}/Q_{a}$ value in 1998 for Azerbaijan, the other Caucasian country \cite{FAO1992Fertilizer1}. For the Central Asian countries, we used the ratio for grasslands derived from Uzbekistan's 2014 data \cite{Heffer2017Assessment2014-2014/15}, which is significantly lower than the \ac{USSR}'s share in 1990. This reduction seems reasonable given the significant decrease in fertilizer use, temporary grasslands, and fodder crop surfaces in the region since the \ac{USSR} collapse \cite{Suleimenov2000CentralAsia}.

\textit{Estonia, Latvia, Lithuania}: The Baltic countries are the three former \ac{USSR} countries with the most available data in global datasets \cite{FAO2002Fertilizer5, EFMA2002Fertilizer2001/02, EFMA2007Fertilizer2006/07, FertilizerEurope2012Fertilizer2011/12, FertilizerEurope2015Fertilizer2014/15, Ludemann2022GlobalCountry}. Fertilizer intensification in these areas has changed significantly over the last three decades due to the abandonment of intensively managed areas \cite{Einarsson2021}. This trend is reflected in the changing ${R_{f-a}}$ values. Therefore, we used \cref{eq:fer_sd:equation_EU_grasslands2} and the linear interpolation with the six years with available data ${R_{f-a}}$ from the 1991-2018 period\cite{FAO2002Fertilizer5, EFMA2002Fertilizer2001/02, EFMA2007Fertilizer2006/07, FertilizerEurope2012Fertilizer2011/12, FertilizerEurope2015Fertilizer2014/15, Ludemann2022GlobalCountry} to estimate the $Q_{f}/Q_{a}$ values since the collapse of the \ac{USSR}. 

\textit{Belarus, Moldova, and Ukraine}: For these three countries, limited data is available regarding fodder crops and grasslands, but some information can be found in global reports \cite{Heffer2017Assessment2014-2014/15, Ludemann2022GlobalCountry, FAO1996Fertilizer3}. Thus, for each country, we used the average of the $Q_{f}/Q_{a}$ values from the 1992-2020 period. In the case of Belarus, where two sets of data were available for grasslands and one for fodder crops \cite{Heffer2017Assessment2014-2014/15, Ludemann2022GlobalCountry}, we took the average for grasslands from both reports and the ratio that accounts for the share of grasslands and the share including fodder crops.

\textit{Russian Federation}: There are three years with available data between 1992 and 2020 \cite{FAO1994Fertilizer2, FAO1996Fertilizer3, Heffer2017Assessment2014-2014/15}. In the first two years, the data showed that an average of approximately 25\% of the country's fertilizer was used on grasslands and fodder crops \cite{FAO1994Fertilizer2, FAO1996Fertilizer3}. However, in the latest report from 2014, only about 4\% was attributed to these areas (excluding fodder crops not used for hay or silage) \cite{Heffer2017Assessment2014-2014/15}. Therefore, we decide to use the linear interpolation of the $Q_{f}/Q_{a}$ values for the years with available data. For the late years, we likely underestimated the value because some fertilizer is used for fodder crops, like fodder beet, that are not intended for silage or hay.  However, these fodder crops only accounted for about the 8\% of the total fertilizer used for fodder crops and grasslands in 1990 \cite{FAO1992Fertilizer1}.

\textit{China}: Fertilization of China grasslands remains low at present \cite{Heffer2017Assessment2014-2014/15}. Among the compiled reports, only the latest one considers a proportion of the total fertilizer application rate in China, allocating 2\% for \nfer{}, 4\% for \pfer{}, and 3\% for \kfer{}. Other information on grassland fertilization in China is scarce, with the few authors that provided some information describing it as sparse \cite{Zhang2010}. \ac{FAO}, \cite{Ludemann2023A19612020} considers this proportion as 0\% for all three nutrients throughout the entire period, which differs from Lassaleta \textit{et al.} (2014), who, based on regional averages, estimated a percentage ranging between 0 and 4.7\% from 1960 to 2014. However, any global report or national more detailed information considers any fertilization. We have decided to adopt the same criteria as \ac{FAO} \cite{Ludemann2023A19612020}, albeit potentially underestimating values for the last decades. 

\textit{Iran}: Fertilization of Iran's grasslands and fodder crops appears to be minimal, with few reports providing data, and only since 1990, indicating values between 2\% and 6\% for all three nutrients \cite{FAO1992Fertilizer1, FAO1994Fertilizer2, Heffer2017Assessment2014-2014/15}. Other information is scarce and focused on experimental trials rather than broader country-wide applications. Considering that the first fertilization trials were developed during the 70s, and the first report with data is for 1990 \cite{FAO1992Fertilizer1}, which reported 2\% of \nfer{} and 6\% for \pfer{}, we considered as 0\% the share for the period 1960-1990, and the average of the reports for the period 1990-2020. 

\textit{Japan}: Since the first report with data, in 1979, almost all reports have underscored the importance of grassland and fodder fertilization in Japan. \ac{FAO} attributed a constant share of 20\% for \nfer{}, 0\% for \pfer{}, and 10\% for \kfer{} for the 1960-2020 period \cite{Ludemann2023A19612020}. Conversely, Lassaleta \textit{et al.} (2014)  \cite{Lassaletta2014} suggested a growing percentage of 20\% for \nfer{}, starting from 0\% in 1960, and increasing to 20\% in 2009. Although data before 1979 is unavailable, the reported data for \nfer{} use in 1979 was 15.7\%, higher than the 5.2\% estimated by Lassaleta \textit{et al.} (2014) \cite{Lassaletta2014}. Additionally, due to the lack of data, it is challenging to determine the inception of grassland fertilization in Japan, though it appears to coincide with the transition from semi-natural grasslands to more intensive pasture during the 60s \cite{Ushimaru2018_Japangrasslands}. Therefore, we opted to adhere to \ac{FAO}'s criteria, maintaining the same percentage throughout the period, despite the potential overestimated values for the initial years. We considered the average of all available reports with data  \cite{Martinez1982FertilizerProduction, FAO1992Fertilizer1, FAO1994Fertilizer2, FAO1996Fertilizer3, FAO1999Fertilizer4, FAO2002Fertilizer5, Heffer2017Assessment2014-2014/15, Ludemann2022GlobalCountry}, because \ac{FAO} criteria appears to underestimate the \pfer{}, and \kfer{} used for grasslands, resulting in percentages of 17.3\% for \nfer{}, 16.9\% for \pfer{}, 15.6\% for \kfer{}.

\textit{Korea Republic}: Grassland fertilization appears to be a common practice in the country nowadays \cite{Lee2019KoreaGrasslands}. However, there is no available data on the fertilization of these areas in global reports \cite{Martinez1982FertilizerProduction, FAO1992Fertilizer1, FAO1996Fertilizer3,  FAO2002Fertilizer5}, nor scientific publications. We used the same assumption as Lassaleta \textit{et al.} (2014), which is to consider the same proportion as in Japan, the geographically and socioeconomically closest country \cite{Lassaletta2014}. This assumption also aligns with the observation that the sum of this percentage, and the fertilizer used for the main crops \cite{Martinez1982FertilizerProduction, FAO1992Fertilizer1, FAO1996Fertilizer3,  FAO2002Fertilizer5} is less than the total for the country \cite{FAOSTAT2023FertilizerDataset}. 

\textit{Turkey}: Information about fertilization of grasslands and fodder crops in Turkey is scarce, suggesting that it is not a common practice. Lassaleta \textit{et al.} (2014) \cite{Lassaletta2014} considered percentages as high as 4.8\% for \nfer{} in 2009, whereas \ac{FAO} considered 0\% for all nutrients. All the available data since 1990 except for 2014 considered some amount of fertilizer used for grasslands, and forages \cite{Martinez1982FertilizerProduction, FAO1992Fertilizer1, FAO1994Fertilizer2, FAO1996Fertilizer3, FAO1999Fertilizer4, FAO2002Fertilizer5, Heffer2017Assessment2014-2014/15, Ludemann2022GlobalCountry}. Therefore, we used the average percentage of all reports for the period 1990 - 2020 \cite{Martinez1982FertilizerProduction, FAO1992Fertilizer1, FAO1994Fertilizer2, FAO1996Fertilizer3, FAO1999Fertilizer4, FAO2002Fertilizer5, Heffer2017Assessment2014-2014/15, Ludemann2022GlobalCountry}.

\textit{Other Asian Countries: Cambodia, Indonesia, Malaysia, The Philippines, Thailand, Vietnam, India, and Pakistan}: In Asian Southeast countries, only  Lassaleta \textit{et al.} (2014)  \cite{Lassaletta2014} considered that some fertilizer is used on grasslands, based on regional averages used for grasslands and other crops (including fruits, tea, vegetables, and forage and grasslands) \cite{Lassaletta2014}. However,  no global report  \cite{Martinez1982FertilizerProduction, AdolfoMartinez1990FertilizerT-37, FAO1992Fertilizer1, FAO1994Fertilizer2, FAO1996Fertilizer3, FAO1999Fertilizer4, FAO2002Fertilizer5, Heffer2017Assessment2014-2014/15, Ludemann2022GlobalCountry} or country-level sources \cite{FAO2005Indonesia} mentioned fertilizer application to grasslands as significant in these countries.  Therefore, we have chosen to align with \ac{FAO}'s criteria, which assumes no fertilizer application rate for grasslands in this region \cite{Ludemann2023A19612020}. We applied the same criteria for India and Pakistan, despite previous research considering a certain percentage used for grasslands \cite{Lassaletta2014, Xu2019GrasslandsN}. The data reports \cite{Martinez1982FertilizerProduction, AdolfoMartinez1990FertilizerT-37, FAO1992Fertilizer1, FAO1994Fertilizer2, FAO1996Fertilizer3, FAO1999Fertilizer4, FAO2002Fertilizer5, Heffer2017Assessment2014-2014/15, Ludemann2022GlobalCountry}, the scientific literature \cite{Ghosh2017GrasslandIndiaN,Irfan2022PakistanN}, and \ac{FAO} \cite{Ludemann2023A19612020} support the idea of non-fertilization of grassland in these two countries.

\textit{Egypt}: Data regarding grassland and fodder crop fertilization in Egypt are scarce \cite{AdolfoMartinez1990FertilizerT-37, FAO1999Fertilizer4}. As is common for many African countries, there is no fertilization of grasslands \cite{Elrys2019NutrientBudgetAfrica}. However, the few available data about the fertilization of Egyptian clover \cite{AdolfoMartinez1990FertilizerT-37, FAO1999Fertilizer4}, the main fodder crop in the country \cite{Michaud2015AlfalfaDistribution}, suggests that a significant portion of \nfer{} and \pfer{} is utilized for fodder production, aligning with country recommendations \cite{FAO2005Egypt}. Previous research, focused solely on grasslands, has either considered 0\% allocation for the three nutrients \cite{Ludemann2023A19612020} or a range between 0\% and 4\% for \nfer{} \cite{Lassaletta2014}. Here, we opted to consider the average of the two reports (1986, 1997) with data \cite{AdolfoMartinez1990FertilizerT-37, FAO1999Fertilizer4} for the entire period as Egyptian clover production has been significant since the beginning of the period \cite{Esfahani1988EgyptProduction}, and the available data is not sufficient to discern any trend.

\textit{Morocco}: Previous research has indicated various fractions of \nfer{} fertilizer used for grasslands in the country, ranging from 0\% to 11\% \cite{Lassaletta2014, Elrys2019NutrientBudgetAfrica, Ludemann2023A19612020}. With the available information, it is impossible to discern if any application for permanent grasslands occurred in the country, but not for forages such as alfalfa, Egyptian clover, or vetch \cite{Bounejmate1990ForagesMorocco, FAO2006Morocco}. Additionally, due to the scarce available data in the reports, discerning any trend is challenging \cite{FAO1992Fertilizer1, FAO1999Fertilizer4, FAO2006Morocco}, although the presence of improved pastures, usually linked to fertilizer application rate, doubled during the 80s decade \cite{Bounejmate1990ForagesMorocco}. Here, we have opted to use the same percentage, the average of all reports, to estimate the percentage of \nfer{}, \pfer{}, and \kfer{}, despite the potential overestimations in the first decades.

\textit{South Africa}: Fertilization of grasslands and fodder crops such as alfalfa appeared to be significant throughout the study period in South Africa. Both previous scientific research \cite{Lassaletta2014, Ludemann2023A19612020} and various technical reports \cite{FAO1992Fertilizer1, FAO1994Fertilizer2, FAO1996Fertilizer3, FAO1999Fertilizer4, FAO2002Fertilizer5, Heffer2017Assessment2014-2014/15, Ludemann2022GlobalCountry} indicated percentages ranging 0\% and 22.3\% for \nfer{}. For all three nutrients, the share used for grasslands and fodder crops during the 90s was higher than in the last decades \cite{FAO1992Fertilizer1, FAO1994Fertilizer2, FAO1996Fertilizer3, FAO1999Fertilizer4, FAO2002Fertilizer5, Heffer2017Assessment2014-2014/15, Ludemann2022GlobalCountry}. This percentage appears to be higher due to larger fertilizer application rates to croplands compared to grasslands and fodder crops \cite{FAO1992Fertilizer1, Ludemann2022GlobalCountry}, and not due to the relationship between cropland and grassland surface \cite{Niedertscheider2012SouthAfrica}. While information regarding grassland fertilization prior to 1990 is limited, several factors support the hypothesis of early fertilizer application rate for grassland and fodder production. These include the fraction used for grasslands and fodder in 1990 \cite{FAO1992Fertilizer1}, substantial research conducted on improved grasslands since 1920s \cite{Smith1984SouthAfrica}, and the early introduction of alfalfa in 1858 \cite{Michaud2015AlfalfaDistribution} which is a significant consumer of \pfer{} and \kfer{} in the country. Given the challenge of identifying any discernible trend and the likelihood of significant consumption at the beginning of the period, we have chosen to adopt the same percentage for the entire duration, aligning with \ac{FAO} assumptions \cite{Ludemann2023A19612020}, despite potential slight over- and underestimations throughout the period. The average of all reports \cite{FAO1992Fertilizer1, FAO1994Fertilizer2, FAO1996Fertilizer3, FAO1999Fertilizer4, FAO2002Fertilizer5, Heffer2017Assessment2014-2014/15, Ludemann2022GlobalCountry}, resulted in percentages of 12.4\% for \nfer{}, 13.3\% for \pfer{}, and 9.2\% for \kfer{}.\\

\textbf{Potential drivers} To develop our \ac{ML} models, we compiled a series of datasets that contain information on features that were identified in previous research as drivers or correlates of cropland fertilization. In this section and the next two, we clarify our rationale for the variable selection, the data sources and the methods used for estimating some of these variables. The list of all considered features can be found in \cref{tab:fer_sd:features} and further details about their estimation are provided below. \\
\textit{Environmental data} Environmental variables related to climate and soil characteristics have been identified as factors that influence fertilization management in farm-level studies \cite{Marenya2009SoilKenya} and regional panel data\cite{Bora2022FertilizerDriversIndia, Levers2016DriversEurope}. 
Therefore, we selected several potential factors, some of which have previously been shown to correlate with fertilization, such as \ac{MAP} \cite{Bora2022FertilizerDriversIndia}, or \ac{SOC} \cite{Levers2016DriversEurope}, as well as newer potential factors such as the aridity index. Data for these factors were sourced from two main databases: the CRU v.4. databases \cite{Harris2020VersionDataset}, for climatic factors, and the SoilGrids v.2. database \cite{Poggio2021SoilGridsUncertainty}, for soil characteristics. Obtaining values at the country-level while considering variations in climatic and soil conditions within a country can be imprecise. However, our fundamental unit of analysis is the country-level, as the \ac{FUBC} values are measured on this scale. To mitigate this limitation, we used spatial information for climatic and soil characteristics along with information about the location of crops \cite{monfreda2008farming}. All environmental variables were estimated using \cref{eq:equation_environemtal_variables1}, but preprocessing differed across variables. 

\begin{equation}
    \label{eq:equation_environemtal_variables1}
    Env_{jic} = \dfrac{\sum_{g\in G}(Env_{ig}\times HArea\_M2000_{gcj})}{HArea\_M2000_{cj}}
\end{equation}

Here, $Env_{jic}$ represents the mean value of the environmental variable for country \textit{j}, in year \textit{i}, where crop \textit{c} is located in the country; $Env_{ig}$ is the value of the environmental variable in year \textit{i}, for grid cell \textit{g}; $HArea\_M2000{gc}$ denotes the area of grid cell \textit{g} for crop \textit{c} in country \textit{j};  $HArea\_M2000_{cj}$ is the total surface of crop \textit{c} in country \textit{j} based on Monfreda \textit{et al.} (2008) crop maps\cite{monfreda2008farming}; and \textit{G} denotes the set of cells where the crop is located based on Monfreda \textit{et al.} (2008) crop maps\cite{monfreda2008farming}. 
For the \ac{MAP}, the $Env_{ig}$ values of \cref{eq:equation_environemtal_variables1} are calculated by summing the precipitation from all months in the CRU v.4. dataset \cite{Harris2020VersionDataset} for each grid cell \textit{g}, and year \textit{i}. For the \ac{MAT}, the $Env_{ig}$ values are calculated as the average of the monthly temperatures from the CRU v.4. dataset \cite{Harris2020VersionDataset}, weighed by the number of days of each month. The \ac{PET} values are derived by multiplying the daily month average from CRU v.4.\cite{Harris2020VersionDataset} by the number of days in each month and summing the results. For the aridity index, we used the United Nations (UN) definition \cite{Middleton1992DesertificationAtlas} of the ratio between \ac{MAP} and total \ac{PET} for each grid cell.  As soil variables do not have temporal resolution, we simplified \cref{eq:equation_environemtal_variables1} by removing the temporal factor. Additionally, for some soil variables like the soil \ac{CEC}, we aggregated the information by calculating the average for the first three depth layers from SoilGrid v.2. (0-5, 1-15 and 15-30 cm) \cite{Poggio2021SoilGridsUncertainty}.

\textit{Agrological data} We selected agrological features that were previously identified as factors potentially related to or driving fertilizer intensification, such as holding size \cite{Ju2016ReducingSize}, crop area \cite{Levers2016DriversEurope}, or irrigation implementation \cite{Bora2022FertilizerDriversIndia}, as well as features that should be connected to crop fertilization at the country-level, such as country fertilizer use per cropland area \cite{FAOSTAT2023FertilizerDataset}. Most of the agrological variables used are taken directly from the sources indicated in Table \ref{tab:fer_sd:features}. However, some required preprocessing. For holding size, we applied the methodology used by Zou \textit{et al.} (2022) \cite{Zou2022}, which involves standardizing the information based on the average holding size according to the total agricultural area. We used holding size data from the FAOSTAT agricultural censuses \cite{FAOSTAT2024HoldingCensus} and previous research \cite{Lowder2016HoldingSize}. To estimate the annual nutrient removal for each crop class based on annual production, we used the recent compilation by \ac{FAO} \cite{Ludemann2023A19612020} on nutrient removal in kilograms per tonne of crop produced, along with the annual country production data from FAOSTAT \cite{FAOSTAT2023AreaDataset}. Additionally, we used this compilation alone as a proxy for fertilizer recommendations, since these recommendations are generally based on the nutrient requirements of each crop \cite{JordanMeille2012Precommendations}.

\textit{Socioeconomic data} Economic factors, particularly those related to the profitability of fertilizer use, have been widely studied to understand fertilizer adoption at the farming-level \cite{Feder1985EconomicDrivers, Hossain2000AsianDrivers}. Both input prices (fertilizers) and output prices (crops) determine profitability and can be key factors influencing fertilization decisions. However, assessing inputs at the country-level is challenging, primarily due to a lack of standardized data. The only available dataset, FAOSTAT \cite{FAOSTAT2023FertilizersArchive}, does not cover all periods and lacks standardization. To address this, we used two variables as proxies of fertilizer prices: a) global real prices for Urea, phosphate rock, and muriate of potash, as compiled by the World Bank \cite{WorldBank2022WorldSheet}; and b) the distance from the production sites or mines, following the methodology proposed by McArthur \textit{et al.} (2017)  \cite{McArthur2017FertilizingDevelopment}. This methodology uses gravity models of trade, based on the premise that fertilizers are produced in a few specific countries \cite{McArthur2017FertilizingDevelopment}. The underlying hypothesis is that countries closer to fertilizer plants or mines are more sensitive to price variations because transport costs are a significant factor for farmers \cite{McArthur2017FertilizingDevelopment}. We applied a similar approach, estimating the minimum cost-adjusted distance by using the costDist function from \textit{terra} package \cite{hijmans2022package}, global friction maps \cite{Weiss2020FrictionMaps}, the locations and operational years of potash \cite{KleineKleffmann2023Potashmines} and phosphate mines, the locations of ammonia plants \cite{McArthur2017FertilizingDevelopment, Clarisse2019AmmoniaEmission} and the centroid of the cropland area on the country based on the Monfreda \textit{et al.} (2008) crop maps \cite{monfreda2008farming}. Assessing the output prices for crops faces a similar problem: there is no standardized dataset with national-level data for the entire period. To resolve this, we used two proxies for crop prices:  a) global real prices for specific commodities like wheat, maize, rice, palm oil, soybeans, sugar, and cotton, compiled by the World Bank \cite{WorldBank2022WorldSheet}, and b) standardized data from two FAOSTAT datasets \cite{FAOSTAT2023PodPrice, FAOSTAT2023PodPriceOld} that provide prices paid to producers at the country-level. The first dataset \cite{FAOSTAT2023PodPrice} contains information from 1990 onwards in \ac{USD}, and \ac{LCU}, while the second dataset \cite{FAOSTAT2023PodPriceOld} covers 1966 to 1990, only in \ac{LCU}. To standardize both datasets, we converted the older dataset into \ac{USD} using annual currency exchange rates \cite{WB2023ExchangeRate}. We then removed outliers independently for each crop by considering only values within 1.5 times the interquartile range. Before applying this method to the 1966-1990 dataset, we tested it on the \ac{LCU} data for maize, wheat, and rice from the 1990 onwards dataset. We compared the original \ac{USD} values with those obtained after converting the \ac{LCU} data using exchange rates. The outlier detection method retained more than 99\% of equivalent values (defined by a ratio between the original and calculated \ac{USD} values of 0.99 to 1.01), while removing over 90\% of non-equivalent values. Finally, the data was converted to real prices by applying the Consumer Price Index \cite{Ha2023InflationIndex}. 

Other socioeconomic factors, that are not directly related to the profitability of fertilizer use, have also been linked to country-level fertilizer use. These factors include the income level, reflected in the \ac{GDP} per capita \cite{Tilman2001ForecastingChange}; the population pressure, defined as the country's population divided by its agricultural land area \cite{Xiang2020PopuationPressure}; and the farmers' knowledge about fertilizer use, as well as general education levels \cite{Feder1985EconomicDrivers}, which we measured by the percentage of total \ac{GDP} spent on education. We used the sources listed in \cref{tab:fer_sd:features} to obtain data for these variables.

\subsubsection*{Data preprocessing}

Several preprocessing steps were performed to prepare the raw data for the \ac{ML} models. First, drawing from both expert domain knowledge and exploratory data analysis, the features relevant to \nfer{}, \pfer{} and \kfer{} fertilizer application rate were selected (\cref{tab:fer_sd:features}). Since not every feature was relevant for each of the three targets, we narrowed down the dataset to data points where the average fertilizer application rate is known for all three fertilizers. This restriction ensured that the dataset comprised only labeled data points, which is crucial for supervised \ac{ML} techniques.  Subsequently, anomalies in the data where the fertilizer application rate was unrealistically large, i.e., greater than 5000 kg ha$^{-1}$, were removed. Finally, categorical features were \ac{OHE}. 

\subsection*{Machine learning}

Previous studies within this domain typically propose linear equations to estimate the fertilizer application rate, and only consider a limited set of agricultural factors \cite{Adalibieke2023Global1961,Lu2017GlobalImbalance}. However, it is well-established that natural phenomena frequently exhibit nonlinear relationships \cite{Stenseth2002ClimateEnvironment}, rendering them unsuitable for modeling with linear methodologies. Similar studies have also employed Bayesian methods \cite{conant2013patterns}, with certain modeling assumptions that are not present in our study. \ac{ML} has the potential to overcome these limitations. The field of \ac{ML} has seen major increases in research \cite{Thiyagalingam2022ScientificBenchmarks} and industry \cite{Bertolini2021MachineReview}, and, more specifically, \ac{ML} has shown promising results in the field of ecology \cite{Thessen2016AdoptionScience,Christin2019ApplicationsEcology}, including agricultural research \cite{Bondre2019PredictionAlgorithms,Xiao2024SpatiotemporalProduction}, fertilizer consumption \cite{Grell2021Point-of-useNitrogen,Pacheco2022ExploringConsumption} and fertilizer management \cite{Xu2024fertilizerReduction}. For this reason, \ac{ML} was used in this study to estimate the annual fertilizer application rate at the crop- and country-level. The benefit of using \ac{ML} is threefold. First, \ac{ML} allows us to include a larger range of variables, for example also including socioeconomic factors. Second, nonlinear \ac{ML} techniques enable us to model nonlinear relationships between the variables in our dataset. Lastly, the model output can provide insights into the drivers associated with crop fertilization on a global scale, through the use of \ac{SHAP} values \cite{Lundberg2017APredictions} outlining the feature importance. The employed \ac{ML} methods to estimate fertilizer application rate for crops differ from previous research, which typically relied solely on changes in crop area, overall fertilizer consumption, and limited data regarding fertilizer application rate at the individual crop-level \cite{Adalibieke2023Global1961,Lu2017GlobalImbalance}. An advantage of our method is that it enables us to estimate values for countries where specific data is lacking by relying on other related variables. For example, the projected data for the \ac{USSR} aligns closely with national totals, even in the absence of country-specific data and without adjustments based on total consumption, as conventionally done \cite{Adalibieke2023Global1961,Lu2017GlobalImbalance}.

\subsubsection*{Models} In this study, two \ac{ML} models based on gradient boosted regression trees were selected to predict the average annual fertilizer application at the crop- and country-level. In gradient boosting \cite{Friedman2000AdditiveBoosting}, an ensemble of weak learners (in our case regression trees) is trained sequentially. First, a weak learner is fitted to the original data. In the next iteration, another weak learner is fitted to the residuals, i.e., the differences between the ground truth target values and the current predictions made by the ensemble. When fitting a new weak learner to the residuals, gradient boosting adjusts its parameters in the negative gradient direction, aiming to reduce the residual error of the ensemble. This sequential learning process enables gradient boosting models to create a strong learner by combining multiple weak learners. The specific gradient boosting models employed in this study are \ac{XGB} \cite{Chen2016XGBoost:System} and \ac{HGB} \cite{Pedregosa2011Scikit-learn:Python,Ke2017LightGBM:Tree}. \Ac{XGB} has been shown to be a powerful tool for predictive modeling in a wide range of applications in both industry and research, including agricultural research \cite{Xiao2024SpatiotemporalProduction} and fertilizer research \cite{Grell2021Point-of-useNitrogen}. It offers an optimized and scalable implementation of gradient boosting, and includes regularization techniques to prevent overfitting \cite{Chen2016XGBoost:System}. The \ac{HGB} model is primarily based on LightGBM \cite{Ke2017LightGBM:Tree}, which addresses one of the major bottlenecks in gradient boosting model training, namely the requirement to sort all samples at each node \cite{Pedregosa2011Scikit-learn:Python}. Indeed, in a traditional gradient boosting model, samples must be sorted at each node to determine the best split. This sorting process can become computationally expensive, especially when dealing with large datasets or deep trees. In \ac{HGB}, the samples are first collected into a histogram, which removes the need for sorting as samples in a histogram are implicitly ordered. This optimization results in a model that is much faster to train than traditional gradient boosting models, while still achieving similar or better performance \cite{Pedregosa2011Scikit-learn:Python}. The choice for these two methods over other conventional \ac{ML} approaches, such as neural networks, was primarily driven by the fact that both methods natively handle missing values. This constitutes a significant advantage, given that global fertilizer application rate data, along with the socioeconomic and agricultural variables used to predict the annual fertilizer application, are often incomplete. This also demonstrates another advantage of applying \ac{ML} to this problem over the conventional approach using linear equations. Indeed, the absence of just one variable in the equation renders it impossible to compute.

\subsubsection*{Model training and evaluation}

The selection of the optimal set of model hyperparameters is usually done using \ac{CV}, after which the \ac{CV} error is reported as the performance of a model \cite{Krstajic2014Cross-validationModels}. However, based on Stone (1974) \cite{Stone1974Cross-ValidatoryPredictions}, model assessment and model performance require different CV approaches. For this reason, we used nested CV, as it allowed us to find the optimal set of hyperparameters for a model and provide an unbiased estimate of the model’s performance \cite{Varma2006BiasSelection}. In nested \ac{CV}, two levels of \ac{CV} loops are used: an outer loop and an inner loop. In the outer loop, the dataset is split into training and testing sets, typically using k-fold CV. Each fold of the outer loop trains the model on the training set and evaluates the model on the testing set. Within each fold of the outer loop, the training data is provided to an inner \ac{CV} loop, in which the training data is further split into training and validation sets, also typically using k-fold \ac{CV}. The inner loop is responsible for selecting the set of hyperparameters that performs best on the validation set. Finally, the performance of the selected set of hyperparameters is evaluated on the corresponding test set in the outer loop. In our study, we used a $2 \times 5$ nested \ac{CV}, i.e., we had two outer loops and five inner loops. We employed a grid search that iteratively went over all possible combinations of hyperparameters, based on the explored hyperparameters as shown in \cref{tab:fer_sd:hyperparameters_models} for both the \ac{HGB} and \ac{XGB} models. The performance of the models was evaluated by averaging the performances of the two models in the outer \ac{CV} loop. The considered performance metrics included the determination coefficient ($R^2$), \ac{MAE}, \ac{MSE}, and \ac{RMSE}, all computed between the predicted and reported data points.

\subsubsection*{Model interpretability through SHAP value analysis}

Unfortunately, gradient boosting methods are so-called black-box models, i.e., it is not immediately clear how certain predictions are made. However, assessing the impact of the features on the predicted fertilizer application rate in the learned models could provide us with valuable insights into the drivers of fertilizer application rate. Therefore, we resorted to \ac{xAI} methods to understand the predictions made by our models. More specifically, we used \ac{SHAP} values \cite{Lundberg2017APredictions} as they are model-agnostic, can account for interactions between features and have an intuitive interpretation. Indeed, summing the \ac{SHAP} values for all features in one sample results in the prediction of the model. Additionally, like \ac{XGB} and \ac{HGB}, \ac{SHAP} values are robust with respect to missing data by design \cite{Lundberg2017APredictions}. Special attention was given to categorical values, as retrieving one \ac{SHAP} value for a categorical feature that is divided into \ac{OHE} features is non-trivial. However, as the \ac{SHAP} values are calculated using the preprocessed input data (i.e., containing the \ac{OHE} categorical features), the \ac{SHAP} values for one categorical variable were obtained by adding together all \ac{SHAP} values for its respective \ac{OHE} features.

\subsection*{Adjustment to country totals}

Previous research has always started with the same premise of allocating total fertilizer consumption at the country-level for estimating crop-level use \cite{Adalibieke2023Global1961, Lu2017GlobalImbalance}. However, here we adopt a different strategy, initiating the estimation of the fertilizer consumption at the crop-level directly. Despite this shift in strategy, we still consider country-level data to be more reliable than datasets compiled from various \ac{FUBC} sources. To reconcile our approach with the more dependable country-level data, we adjusted the \ac{ML} predictions to align with FAOSTAT's total fertilizer consumption at the country-level \cite{FAOSTAT2023FertilizerDataset}. As shown in \cref{eq:fer:equation_adjustment}, we distributed the differences between the predicted total fertilizer consumption and the FAOSTAT totals equally among crops, after excluding the fraction used for grasslands and fodder crops from FAOSTAT totals. 

\begin{equation}
    \label{eq:fer:equation_adjustment}
    Fert\_Pred_{icj} = FertML\_Pred_{icj} \times \dfrac{\sum_{d \in C}(FertML\_Pred_{idj}\times HArea\_FAO_{idj})}{FAOSTAT\_FERTng_{ij}}
\end{equation}

Where, $Fert\_Pred_{icj}$ represents the fertilizer application rate predictions after the adjustment for year \textit{i}, crop \textit{c}, and country \textit{j}. $FertML\_Pred_{icj}$ denotes the \ac{ML} model predictions, \textit{C} is the set of all crops classes included in the models,  $HArea\_FAO_{idj}$ the FAOSTAT harvested area\cite{FAOSTAT2023AreaDataset} of each crop class \textit{d}, and $FAOSTAT\_FERTng_{ij}$ is the total FAOSTAT fertilizer consumption for the country, after removing the fraction used for grasslands and fodder crops. 

\subsection*{Validation}
To validate the model predictions, we compared the model predictions with national databases containing information about the average use per hectare for different fertilizers and crops. This validation is quantified using the \ac{MAE} and \ac{MAPE} as well as with comparative plots if enough data was obtainable from the various national databases. The \ac{MAE} gives an idea about the actual deviation, whilst the \ac{MAPE} makes the comparison between prediction errors easier. The evaluated national databases include data obtained from for the \ac{USA} \cite{USDA2019}, \ac{UK} \cite{DEFRA2023FertiliserUKdataset}, India \cite{India1986Input, India1991Input, India1996Input, India2001Input, India2006Input, India2011Input, India2016Input}, Sweden \cite{SWEDEN2010, SWEDEN2012, SWEDEN2015, SWEDEN2018}, Philippines \cite{PSA2023}, and New Zealand \cite{NZ2021}. For Pakistan \cite{PAKISTAN}, only data for the sum of fertilizer application rate is available, hence the sum of \nfer{}, \pfer{}, and \kfer{} was used, expressed as NPK. 
This approach is restricted by available data in national databases for average fertilizer application rate across various crops and nutrients.

\subsection*{Gridded crop-specific application rate per fertilizer}
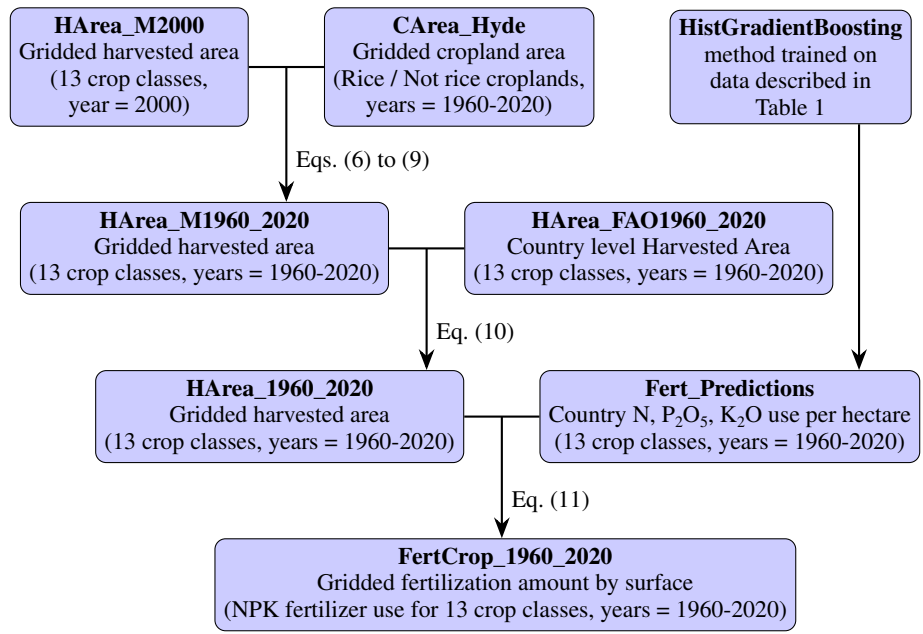
\begin{figure}[ht]
\centering
\begin{tikzpicture}[
    font=\small,
    node distance=1cm and 1cm,
    box/.style = {draw, rounded corners, align=center, minimum width=3cm, minimum height=1cm, fill=blue!20},
    arrow/.style = {-{Stealth[scale=1.2]}, thick},
    line/.style = {thick},  
]

\node[box] (HArea_M2000) {\textbf{HArea\_M2000} \\ Gridded harvested area \\ (13 crop classes,\\ year = 2000)};
\node[box, below=of HArea_M2000, xshift=1cm] (HArea_M1960_2020) {\textbf{HArea\_M1961\_2019} \\ Gridded harvested area \\ (13 crop classes, years = 1961-2019)};
\node[box, below=of HArea_M1960_2020, xshift=1cm] (HArea_1960_2020) {\textbf{HArea\_1961\_2019} \\ Gridded harvested area \\ (13 crop classes, years = 1961-2019)};
\node[box, below=of HArea_1960_2020, xshift=3cm] (FertCrop_1960_2020) {\textbf{FertCrop\_1961\_2019} \\ Gridded fertilization amount by surface \\ (NPK fertilizer use for 13 crop classes, years = 1961-2019)};

\node[box, right=of HArea_M1960_2020, xshift=0cm] (HArea_FAO1960_2020) {\textbf{HArea\_FAO1961\_2019} \\ Country level Harvested Area \\ (13 crop classes, years = 1961-2019)};
\node[box, right=of HArea_1960_2020] (Fert_Predictions) {\textbf{Fert\_Predictions} \\ Country \nfer{}, \pfer{}, \kfer{} use per hectare \\ (13 crop classes, years = 1961-2019)};

\node[box, right=of HArea_M2000, xshift=0cm] (CArea_Hyde) {\textbf{CArea\_Hyde} \\ Gridded cropland area \\ (Rice / Not rice croplands,\\ years = 1961-2019)};

\node[box, right=of CArea_Hyde, xshift=0cm] (HGB) {\textbf{HistGradientBoosting} \\ method trained on \\ data described in \\ \cref{tab:fer_sd:features}};

\node[coordinate, right=of HArea_M2000, xshift=-0.5cm] (firsteqs) {};
\node[coordinate] (end_of_first_eqs) at ($(firsteqs |- HArea_M1960_2020.north)$) {};

\node[coordinate, right=of HArea_M1960_2020, xshift=-0.5cm] (secondeq) {};
\node[coordinate] (end_of_second_eq) at ($(secondeq |- HArea_1960_2020.north)$) {};

\node[coordinate, right=of HArea_1960_2020, xshift=-0.5cm] (thirdeq) {};
\node[coordinate] (end_of_third_eq) at ($(thirdeq |- FertCrop_1960_2020.north)$) {};

\node[coordinate] (HGB_start) at ($(HGB.south east) - (0.7cm, 0)$) {};
\node[coordinate] (Fert_preds_end) at ($(HGB_start |- Fert_Predictions.north)$) {};

\draw[line] (CArea_Hyde) -- (HArea_M2000);
\draw[arrow] (firsteqs) -- (end_of_first_eqs) node[pos=0.7, right]{\cref{eq:fer_sd:equation_maps_carea_rice,eq:fer_sd:equation_maps_carea_notrice,eq:fer_sd:equation_maps_carea__harea_rice,eq:fer_sd:equation_maps_carea__harea_notrice}};

\draw[line] (HArea_M1960_2020) -- (HArea_FAO1960_2020);
\draw[arrow] (secondeq) -- (end_of_second_eq) node[pos=0.7, right]{\cref{eq:fer_sd:equation_maps_2}};

\draw[line] (HArea_1960_2020) -- (Fert_Predictions);
\draw[arrow] (thirdeq) -- (end_of_third_eq) node[pos=0.7, right]{\cref{eq:fer_sd:equation_maps_3}};

\draw[arrow] (HGB_start) --  (Fert_preds_end);

\end{tikzpicture}
\caption{Outline of the process for generating the gridded crop-specific fertilizer dataset.}
\label{fig:fer_sd:map_creation}
\end{figure}

Following the generated comprehensive dataset of global fertilizer application rate, we constructed detailed 5-arcmin resolution gridded maps for each fertilizer (\nfer{}, \pfer{}, and \kfer{}), crop class and year from 1961 to 2019.
The final gridded map dataset was compiled in a three-step process, as highlighted in \cref{fig:fer_sd:map_creation}. First, data of the gridded harvested area spanning from 1961 to 2019 for the 13 distinct crop classes (see \cref{tab:fer_sd:crop_classification}) were acquired by combining data from the EARTHSTAT project of the year 2000 ($HArea\_M2000$) \cite{monfreda2008farming}, supplemented with historical arable land and permanent crop areas per year ($CArea\_Hyde$) from the History Database of the Global Environment (HYDE version 3.3) \cite{klein2017anthropogenic}. The EARTHSTAT maps were created by combining national-, state-, and country-level census statistics with an up-to-date global dataset of croplands, organized on a 5-arcminute by 5-arcminute latitude-longitude grid. These datasets, reflecting land use around the year 2000, detail both the area harvested and the yield of 175 diverse crops worldwide \cite{monfreda2008farming}. Innovative maps outlining major crop groups were generated by consolidating these individual crop maps. The HYDE 3.3 project provides long time series estimates and maps for land use, including the cropland areas, based on an allocation algorithm with time-dependent weighting \cite{klein2017anthropogenic}. The elaborate information from the crop specific EARTHSTAT maps for the year 2000, in combination with the yearly changes in gridded cropland from HYDE 3.3, allowed us to make detailed gridded 5-arcmin resolution crop specific harvested areas for each of the evaluated years and crops using \cref{eq:fer_sd:equation_maps_carea_rice,eq:fer_sd:equation_maps_carea_notrice,eq:fer_sd:equation_maps_carea__harea_rice,eq:fer_sd:equation_maps_carea__harea_notrice}: 

For $CArea\_Hyde_{gi} > 0$ and crop is rice:

\begin{equation}\label{eq:fer_sd:equation_maps_carea_rice}
    HArea\_M_{gic} = CArea\_Hyde\_R_{gi}
\end{equation}

For $CArea\_Hyde_{gi} > 0$ and crop is not rice:

\begin{equation}\label{eq:fer_sd:equation_maps_carea_notrice}
    HArea\_M_{gic} = CArea\_Hyde\_NR_{gi}\times\dfrac{HArea\_M2000_{gc}}{CArea\_Hyde\_NR_{g2000}}
\end{equation}

For $CArea\_Hyde_{gi} > 0 \ \cup \ \sum_{c \in C} HArea\_M2000_{c} = 0$ and crop is rice:

\begin{equation}\label{eq:fer_sd:equation_maps_carea__harea_rice}
    HArea\_M_{gic} = CArea\_Hyde\_R_{gi}
\end{equation}

For $CArea\_Hyde_{gi} > 0 \ \cup \ \sum_{c \in C} HArea\_M2000_{c} = 0$ and crop is not rice:

\begin{equation}\label{eq:fer_sd:equation_maps_carea__harea_notrice}
    HArea\_M_{gic} = CArea\_Hyde\_NR_{gi}\times\dfrac{\sum_{k \in K} HArea\_M2000_{gc}/K}{CArea\_Hyde\_NR_{g2000}}
\end{equation}

Here, the indices denote the grid cell ($g$), the year ($i$), the crop ($c$). The harvested area ($HArea\_M_{gic}$) was generated through a series of conditional operations. These conditions stipulate that if the value of the HYDE3.3 cropland area map ($CArea\_Hyde_{gi}$) for that year $i$ and grid cell $g$ is larger than 0, and the crop is not rice, then the value of that grid cell for that specific crop and year is given by the HYDE3.3 cropland area ($CArea\_Hyde\_NR_{gi}$) for that grid cell/year combination. The value of the grid cell is then further adjusted by the ratio of the HYDE3.3 map of the year 2000 to the EARTHSTAT map of the year 2000 for the corresponding grid cell and crop ($\dfrac{HArea\_M2000_{gc}}{CArea\_Hyde\_NR_{g2000}}$). In the case of rice, the specific HYDE3.3 map for cropland area of rice was selected and not altered as this is readily available. Additionally, in instances where $CArea\_Hyde_{gi}$ was larger than 0 and the sum of all crops across the EARTHSTAT maps of the year 2000 is equal to 0 (e.g., when new lands are cultivated), a progressively expanding area $K$ was evaluated to find an appropriate ratio based on the average of the $k$ neighboring cells. The evaluated values for $k$ were 5, 10, 25, 50, 100, 150, 200 and 250, up until a value different from zero for the ratio is found. If no value different from zero was found, the ratio value was set equal to 1. This last step made the assumption that the crop distribution in neighboring cells adequately represents the distribution in the newly cultivated area, allowing for the calculation of adjusted harvested areas. Furthermore, as the $HArea\_M2000_{gc}$ is consistently used, we assumed that the changes in crop distribution over time remain constant. To ensure the accuracy of the generated maps, we capped the harvested area at the maximum feasible value in each cell.

To ensure consistency with FAOSTAT data used in the model predictions, the gridded harvested area ($HArea\_M1961\_2019$) was aligned with the country-specific harvested area reported by FAOSTAT ($HArea\_FAO2000$). Additionally, due to this alignment, some cells may initially have harvested area values that exceed the maximum possible for that cell. To correct this, we cap the harvested area at the maximum feasible value per cell and then redistribute any excess proportionally across other cells with harvested area values, ensuring overall consistency with FAOSTAT data. These adjustments, applied through \cref{eq:fer_sd:equation_maps_2}, provided a corrected gridded harvested area for the 13 crop classes over the 60-year period ($HArea\_1961\_2019$):

\begin{equation}
    \label{eq:fer_sd:equation_maps_2}
    HArea_{gic} = HArea\_M_{gic} \times \dfrac{\sum_{j \in J}HArea\_FAO_{icj}}{\sum_{j \in J}HArea\_M_{icj}}
\end{equation}

In this equation, $HArea\_FAO_{icj}$ represents the harvested area for year $i$, crop $c$, and country $j$ as reported by FAOSTAT, summed over all countries ($J$) in grid cell $g$ (to accommodate grid cells with multiple countries). Similarly, $HArea\_M_{icj}$ represents the estimated harvested area for the same combinations, also summed over all countries in grid cell $g$. The ratio of these sums adjusts the model gridded harvested area ($HArea\_M_{gic}$) to match FAOSTAT data, ensuring the resulting gridded harvested area on a country level is consistent with official statistics across the 60-year period.

Finally, the gridded harvested area ($HArea_{1961\_2019}$) was augmented with the average application rate of each predicted fertilizer (\nfer{}, \pfer{}, \kfer{}) as per \cref{eq:fer_sd:equation_maps_3}: 

\begin{equation}
    \label{eq:fer_sd:equation_maps_3}
    FertCrop_{gic} = HArea_{gic} \times \sum_{j \in J} (Fert\_Pred_{icj} \times PercCountry_{g})
\end{equation}

where $Fert\_Pred_{icj}$ is the country-level prediction resulting from the \ac{HGB} model after applying the adjustment, and $PercCountry_{g}$ refers to the percentage of grid cell $g$ that is occupied by the country $j$. This process was then applied to each fertilizer separately to obtain gridded maps for each fertilizer, year, and crop combination. 

\section*{Data Records}
The gridded fertilizer application data for \nfer{}, \pfer{}, and \kfer{} by crops from 1961 to 2019 are available in a Figshare repository \cite{Coello2024Dataset}. The dataset spans from 180ºE to 180ºW and 90ºS to 90ºN at a resolution of 5 arc-min in WGS84 (EPSG: 4326). It is provided in \textit{.tiff} format, which can be read by many tools, such as R and Python. The gridded application data by crops and fertilizers are stored in several files named ``${Crop\_Name}{Fertilizer}{Year}.tiff$''. Here, ``${Crop\_Name}$'' represents each crop class listed in \cref{tab:fer_sd:crop_classification}, ``${Fertilizer}$'' refers to \nfer{}, \pfer{}, or \kfer{}, and $'{Year}'$ indicates any year from 1961 to 2019.

\subsection*{Crop-specific \nfer{} application}
On a global scale, the \nfer{} application has grown for all crops (\cref{fig:fer_sd:map_nitrogen}). For example, the average \nfer{} use of the three main cereals has risen from 17.1 $\pm$ 6.1 kg ha\textsuperscript{-1}, 26.6 $\pm$ 7.2 kg ha\textsuperscript{-1}, 12.1 $\pm$ 3.9 kg ha\textsuperscript{-1} for wheat, maize and rice, respectively, in the 1960s, to 97.8 $\pm$ 4.2 kg ha\textsuperscript{-1}, 118.8 $\pm$ 4.2 kg ha\textsuperscript{-1}, 113.8 $\pm$ 1.9 kg ha\textsuperscript{-1} in the 2010s decade. Moreover, the largest increases in \nfer{} application occurred in vegetable crops, with a global growth of more than 120 kg ha\textsuperscript{-1} between these two decades (\cref{fig:fer_sd:map_nitrogen}). Conversely, the lowest increases occurred in soybean, where \nfer{} application rates grew by less than 20 kg ha\textsuperscript{-1}. At the regional scale, the intensification of \nfer{} fertilizer use has shifted from higher use at the beginning of the period in the \ac{USA} and Europe to being currently dominated by Asian countries such as China and India (\cref{fig:fer_sd:map_nitrogen}). This trend is particularly evident for some crops like vegetables and fruits, where China now has the areas with the highest \nfer{} use worldwide, whereas in the 1960s, these areas were primarily in Southern Europe and California.

\subsection*{Crop-specific \pfer{} application}
The application of \pfer{} also experienced global increases across all crops (\cref{fig:fer_sd:map_phosphorus}), but to a lesser extent than \nfer{}. The average \pfer{} used for the three main cereals and soybean rose from 13.8 $\pm$ 3.3 kg ha\textsuperscript{-1}, 13.1 $\pm$ 2.4 kg ha\textsuperscript{-1}, 6.3 $\pm$ 1.9 kg ha\textsuperscript{-1}, and 12.6 $\pm$ 2.4 kg ha\textsuperscript{-1} for wheat, maize, rice and soybean, respectively, in the 1960s to 35.5 $\pm$ 4.9 kg ha\textsuperscript{-1}, 43.0 $\pm$ 5.7 kg ha\textsuperscript{-1}, 39.9 $\pm$ 5.0 kg ha\textsuperscript{-1}, and 39.1 $\pm$ 6.6 kg ha\textsuperscript{-1} in the 2010s. Similar to \nfer{}, the largest increases occurred in vegetable crops, where \pfer{} application rates increased by more than 50 kg ha\textsuperscript{-1}. Conversely, the smallest increases were observed in the other cereal crop class, where the average \pfer{} application rate increased by only about 2.5 kg ha\textsuperscript{-1} between the two decades. Regionally, a similar pattern occurred with \pfer{} use, following the trend previously seen with \nfer{}, where the hotspot shifted from Europe to Asia. This shift is particularly notable for wheat, where the hotspot of \pfer{} intensification moved from Western Europe to northern India and northeastern China (\cref{fig:fer_sd:map_phosphorus}).  

\subsection*{Crop-specific \kfer{} application}
Globally, the use of \kfer{} has also increased across almost all crop classes (\cref{fig:fer_sd:map_potassium}). For wheat, maize, rice, and soybean, the average \kfer{} application rates have risen from 7.2 $\pm$ 1.6, 9.8 $\pm$ 2.0, 3.4 $\pm$ 0.5, and 11.6 $\pm$ 2.6 kg ha\textsuperscript{-1}, respectively, to 15.4 $\pm$ 4.1, 33.1 $\pm$ 4.8, 27.3 $\pm$ 3.9, and 9.8 $\pm$ 3.2 kg ha\textsuperscript{-1}. Unlike \nfer{} and \pfer{}, the largest difference in average \kfer{} application occurred for the oil palm crop, which increased from 3.7 $\pm$ 1.4 kg ha\textsuperscript{-1} during the 1960s to 87.6 $\pm$ 8.3 during the 2010s. Similar to \pfer{}, the other cereal class experienced the smallest change in \kfer{} use. In this case, the average \kfer{} application rate decreased from 11.7 $\pm$ 1.9 kg ha\textsuperscript{-1} during the 1960s to 9.8 $\pm$ 3.2 kg ha\textsuperscript{-1} during the 2010s. Regionally, a similar pattern emerged with \kfer{}, following the trend observed with \nfer{} and \pfer{}, with the hotspot of \kfer{} fertilization shifting from Europe and the \ac{USA} to Asia. However, this change was more pronounced in different crops, such as oil crops, where the use of \kfer{} has increased significantly in countries like Malaysia and Indonesia (\cref{fig:fer_sd:map_potassium}).

\section*{Technical Validation}

This section provides a detailed discussion of the validation efforts made to confirm the validity, consistency, and plausibility of our compiled dataset and predictions. First, the performance of the \ac{ML} models is evaluated. Subsequently, we use \ac{SHAP} values to confirm that our models used sensible features to make their predictions, based on literature. Finally, the predictions are validated by comparing them with reported data in both national and global databases.

\subsection*{ML Model performance}
The performance of the \ac{ML} models predicting the fertilizer application rates for the three fertilizers is shown in \cref{tab:fer_sd:metrics}. Both \Ac{XGB} and \Ac{HGB} significantly outperformed the naive prediction, which uses the mean fertilizer application as its prediction. \Ac{HGB} consistently outperformed (or matched) \ac{XGB} for all three fertilizers and performance metrics. For this reason, we will use the \ac{HGB} model in the remainder of this technical validation, as well as any subsequent analyses. 

\subsection*{SHAP value analysis}

\begin{figure}[ht]
\centering
\hspace*{-2cm}
\includegraphics[width=1.3\linewidth]{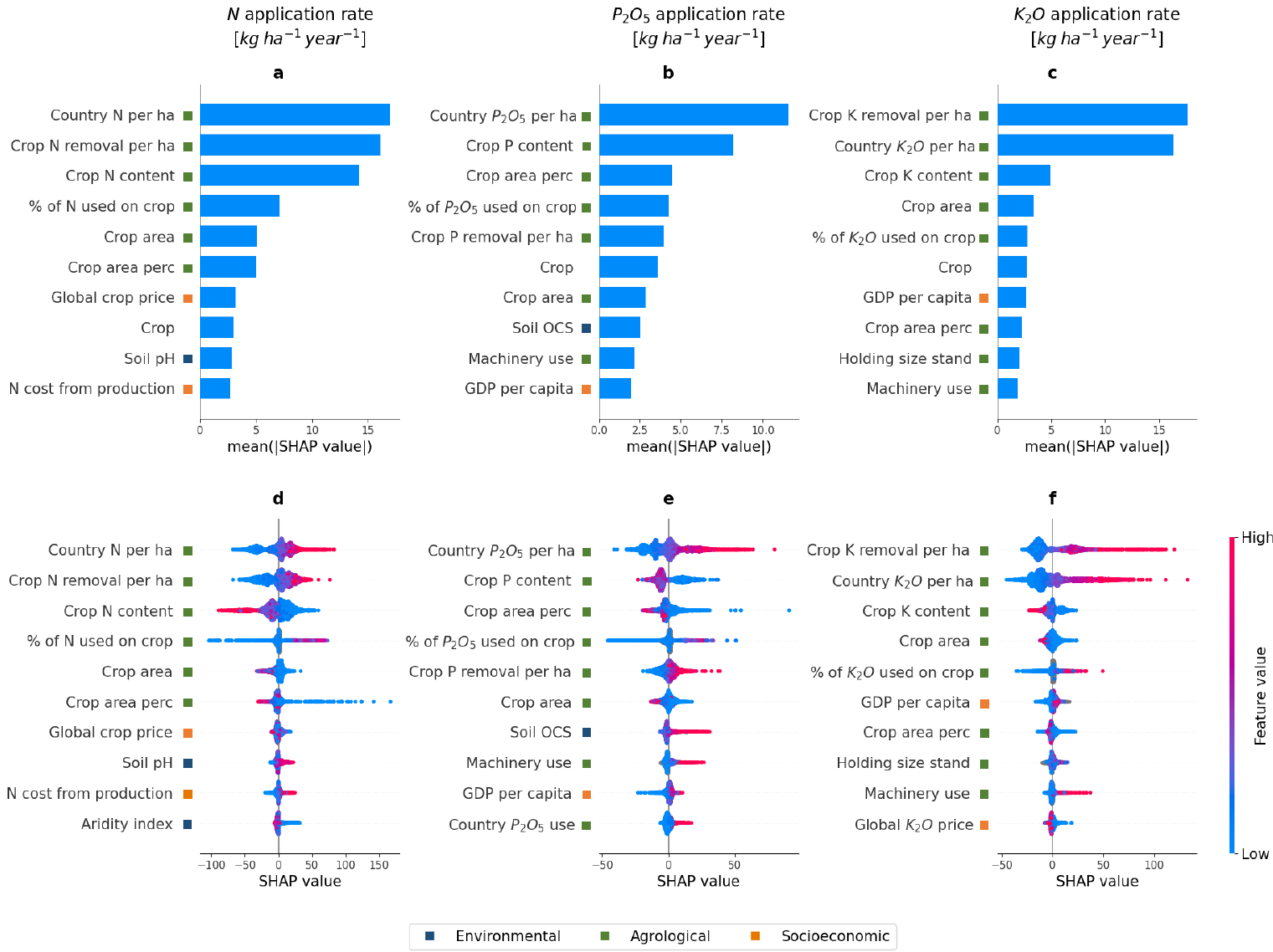}
\hspace*{2cm}
\caption{SHapley Additive eXplanation (SHAP) values of the top 10 most important features in the prediction of, respectively, the crop \nfer{} (a,d), \pfer{} (b,e) and \kfer{} (c,f) application rates using Histogram-based Gradient Boosted regression. (a,b,c) The top plots present the average feature importance, determined by the mean absolute SHAP value of each feature.  (d,e,f) The bottom plots depict a SHAP value for each prediction and show the local feature importance and the feature effect. The color of a dot represents the value of the feature in that instance - red indicating relatively high, blue indicating relatively low values. A dot with a high SHAP value for a feature suggests a positive contribution to the prediction, whereas a negative SHAP value leads to a lower prediction. The features are ranked in order of descending average importance and the blue, green and orange squares indicate whether the feature is an environmental, agrological or socioeconomic characteristic.}
\label{fig:fer_sd:shap}
\end{figure}

To examine the impact of the features on the prediction of the \nfer{}, \pfer{} and \kfer{} application rates, the \ac{SHAP} values of the ten most important features for the three corresponding \ac{HGB} models are illustrated in \cref{fig:fer_sd:shap}. Agrological drivers dominated the predictions, comprising six, seven, and eight of the ten highest ranked features, respectively. The impact of the features remained consistent across all fertilizers, albeit with varying magnitudes (\cref{fig:fer_sd:shap}-d,e,f). In particular, the predicted fertilizer application rates were consistently positively impacted by the country fertilizer per ha and the crop nutrient removal per ha (as red dots, i.e. high values of country fertilizer per ha and high nutrient removal per ha, corresponded with positive \ac{SHAP} values), while it was negatively impacted by the crop nutrient content (red dots corresponding with negative \ac{SHAP} values; \cref{fig:fer_sd:shap}-d,e,f). These relationships align with the expected influence of these features on fertilization at the crop-level \cite{Liu2015YieldFertilizer}. 
 Across the different fertilizers, the most important socioeconomic features varied. For instance, the GDP per capita was the most important socioeconomic feature in the prediction of the \pfer{} and \kfer{} application rates, while in the \nfer{} prediction, the global crop price was more important. Fertilization at the country-level is usually associated with the economic development of the country, measured by GDP \cite{Tilman2001ForecastingChange, Longo2008DriversEconomicFertilizer}. However, at the crop-level, this relationship only held true for the most expensive fertilizers, \pfer{} and \kfer{}. For \nfer{}, the most affordable nutrient \cite{WorldBank2022WorldSheet}, factors such as global crop price and \nfer{} cost from production appeared to be more significant (\cref{fig:fer_sd:shap}). Few environmental features seemed to be relevant for the predictions (\cref{fig:fer_sd:shap}); only the soil pH, soil \ac{OCS}, and aridity index appeared in the top ten for some nutrients. Although the influence of these variables appeared to be low, the direction of these relationships confirmed the findings of other authors at the farm- or regional-level for soil organic carbon content and soil pH. \cite{Marenya2009SoilKenya,Levers2016DriversEurope, Bora2022FertilizerDriversIndia}.

\subsection*{Validation}
\begin{figure}[ht]
\centering
\includegraphics[width=\linewidth]{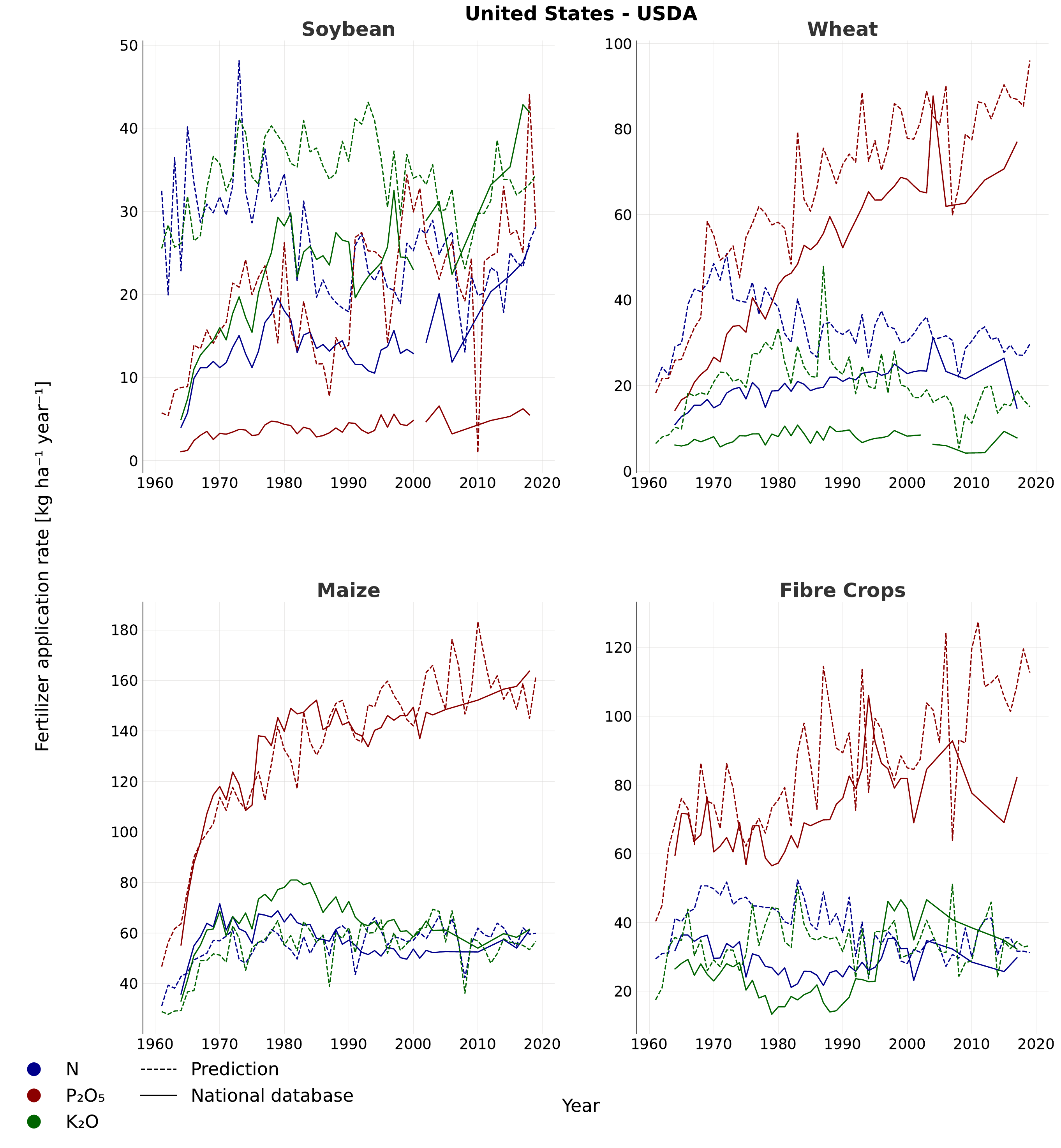}
\caption{Comparison of the application rates per ha per year for various crops between our predicted model output and the data reported by the United States Department of Agriculture (USDA) for the \ac{USA}.}
\label{fig:fer_sd:validation_NDB_PRED_USA}
\end{figure}

\begin{figure}[ht]
\centering
\includegraphics[width=\linewidth]{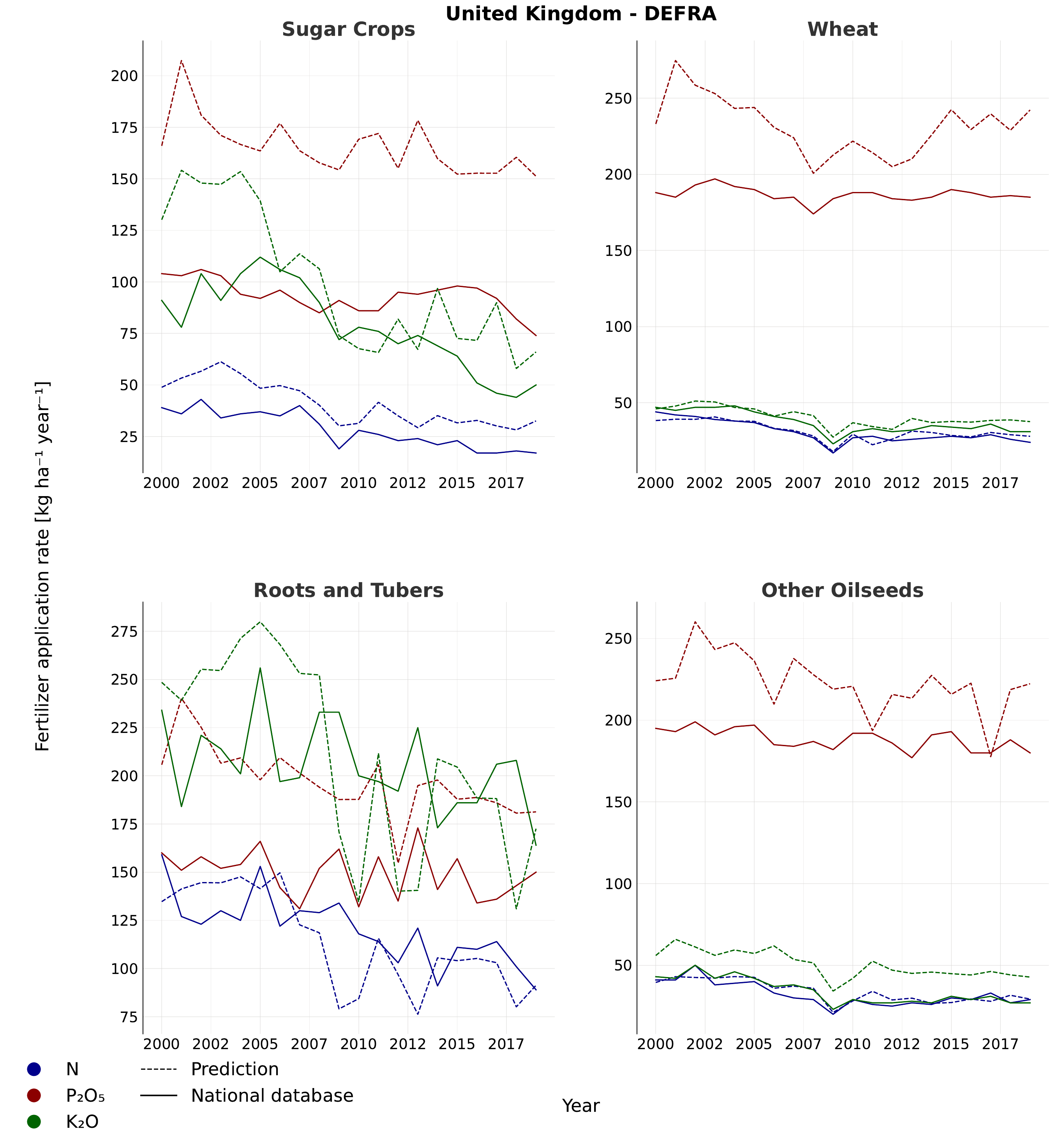}
\caption{Comparison of the application rates per ha per year for various crops between our predicted model output and the data reported by  the Department for Environment, Food \& Rural Affairs (DEFRA) for the \ac{UK}.}
\label{fig:fer_sd:validation_NDB_PRED_UK}
\end{figure}

\begin{figure}[ht]
\centering
\centerline{\includegraphics[width=1.4\textwidth]{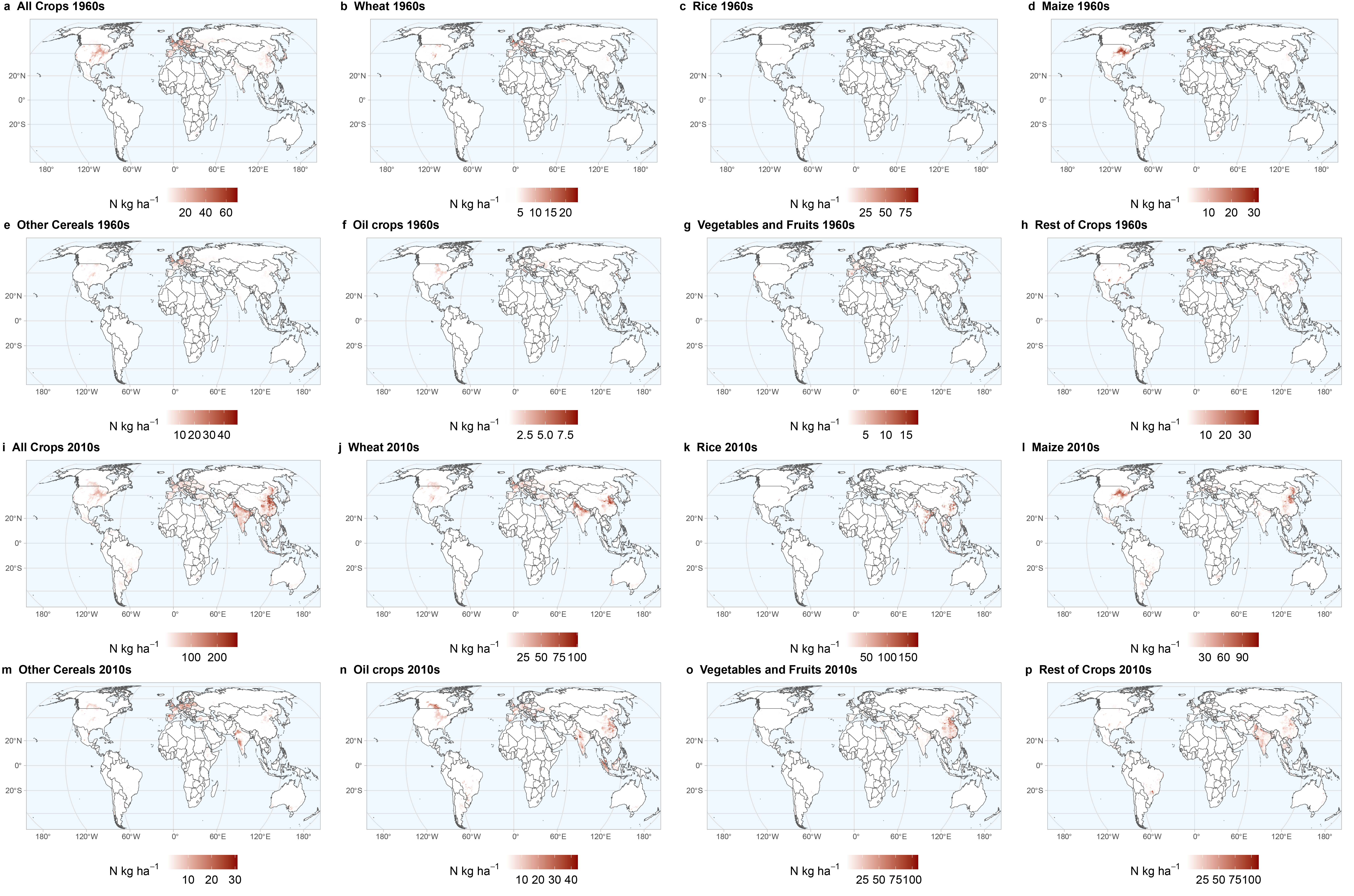}}
\caption{Spatial pattern of crop-specific fertilizer (\nfer{}) kg ha \textsuperscript{-1} consumed by each 0.05° grid cell for the following: a) average for the 1960s decade across all 13 crop classes, b) average for the 1960s decade for wheat, c) average for the 1960s decade for rice, d) average for the 1960s decade for maize, e) average for the 1960s decade for other cereals, f) average for the 1960s decade for all oil crops, g) average for the 1960s decade for vegetables and fruits, h) average for the 1960s decade for roots and tubers, sugar crops, fiber crops, and other crop classes, i) average for the 2010s decade across all 13 crop classes, j) average for the 2010s decade for wheat, k) average for the 2010s decade for rice, l) average for the 2010s decade for maize, m) average for the 2010s decade for other cereals, n) average for the 2010s decade for all oil crops, o) average for the 2010s decade for vegetables and fruits, p) average for the 2010s decade for roots and tubers, sugar crops, fiber crops, and other crop classes. The 1960s decade includes the years 1961-1969, and the 2010s decade includes the years 2010-2019}
\label{fig:fer_sd:map_nitrogen}
\end{figure}

\begin{figure}[ht]
\centering
\centerline{\includegraphics[width=1.4\textwidth]{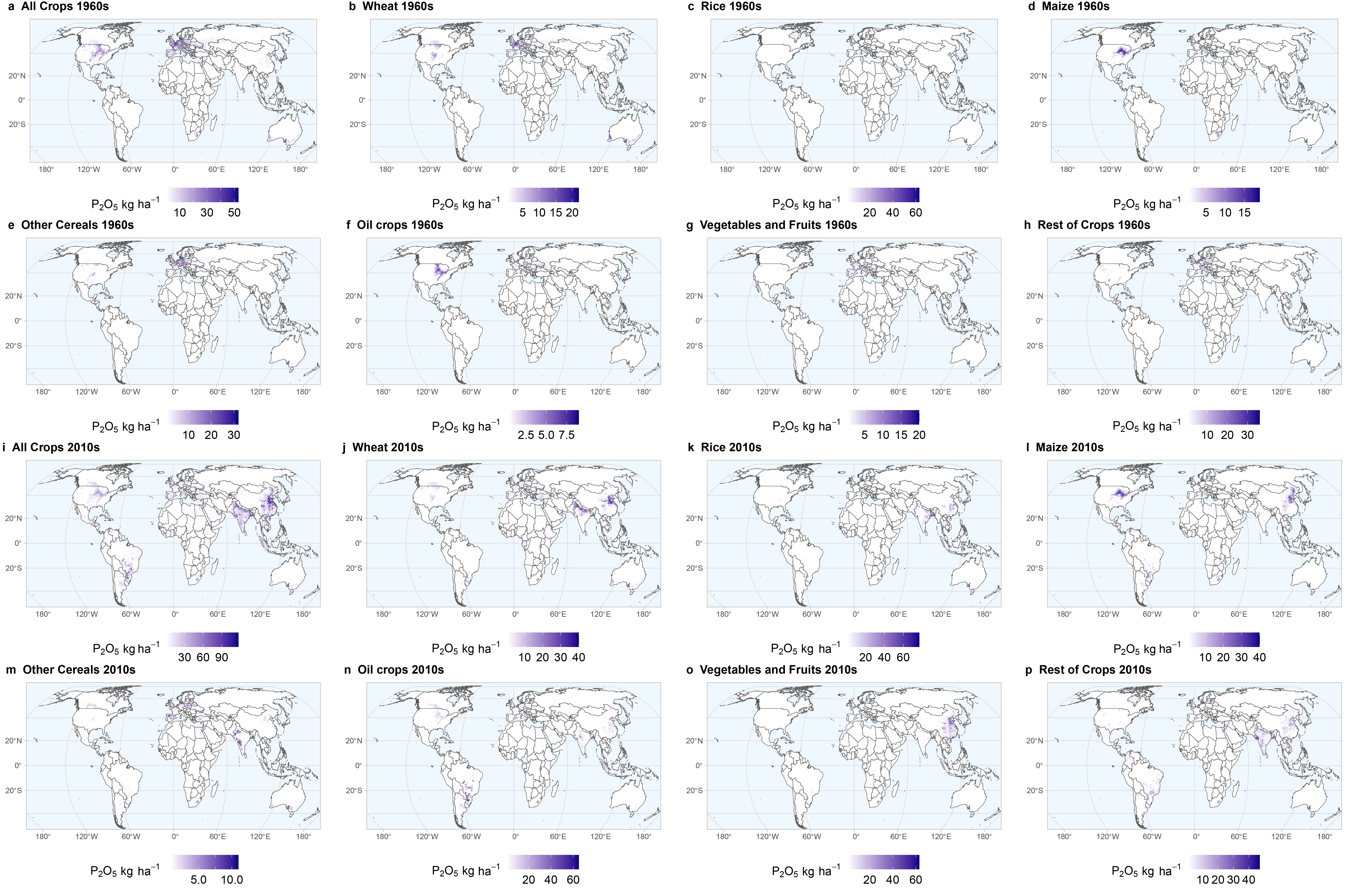}}
\caption{Spatial pattern of crop-specific fertilizer (\pfer{}) kg ha \textsuperscript{-1} consumed by each 0.05° grid cell for the following: a) average for the 1960s decade across all 13 crop classes, b) average for the 1960s decade for wheat, c) average for the 1960s decade for rice, d) average for the 1960s decade for maize, e) average for the 1960s decade for other cereals, f) average for the 1960s decade for all oil crops, g) average for the 1960s decade for vegetables and fruits, h) average for the 1960s decade for roots and tubers, sugar crops, fiber crops, and other crop classes, i) average for the 2010s decade across all 13 crop classes, j) average for the 2010s decade for wheat, k) average for the 2010s decade for rice, l) average for the 2010s decade for maize, m) average for the 2010s decade for other cereals, n) average for the 2010s decade for all oil crops, o) average for the 2010s decade for vegetables and fruits, p) average for the 2010s decade for roots and tubers, sugar crops, fiber crops, and other crop classes. The 1960s decade includes the years 1961-1969, and the 2010s decade includes the years 2010-2019}
\label{fig:fer_sd:map_phosphorus}
\end{figure}

\begin{figure}[ht]
\centering
\centerline{\includegraphics[width=1.4\textwidth]{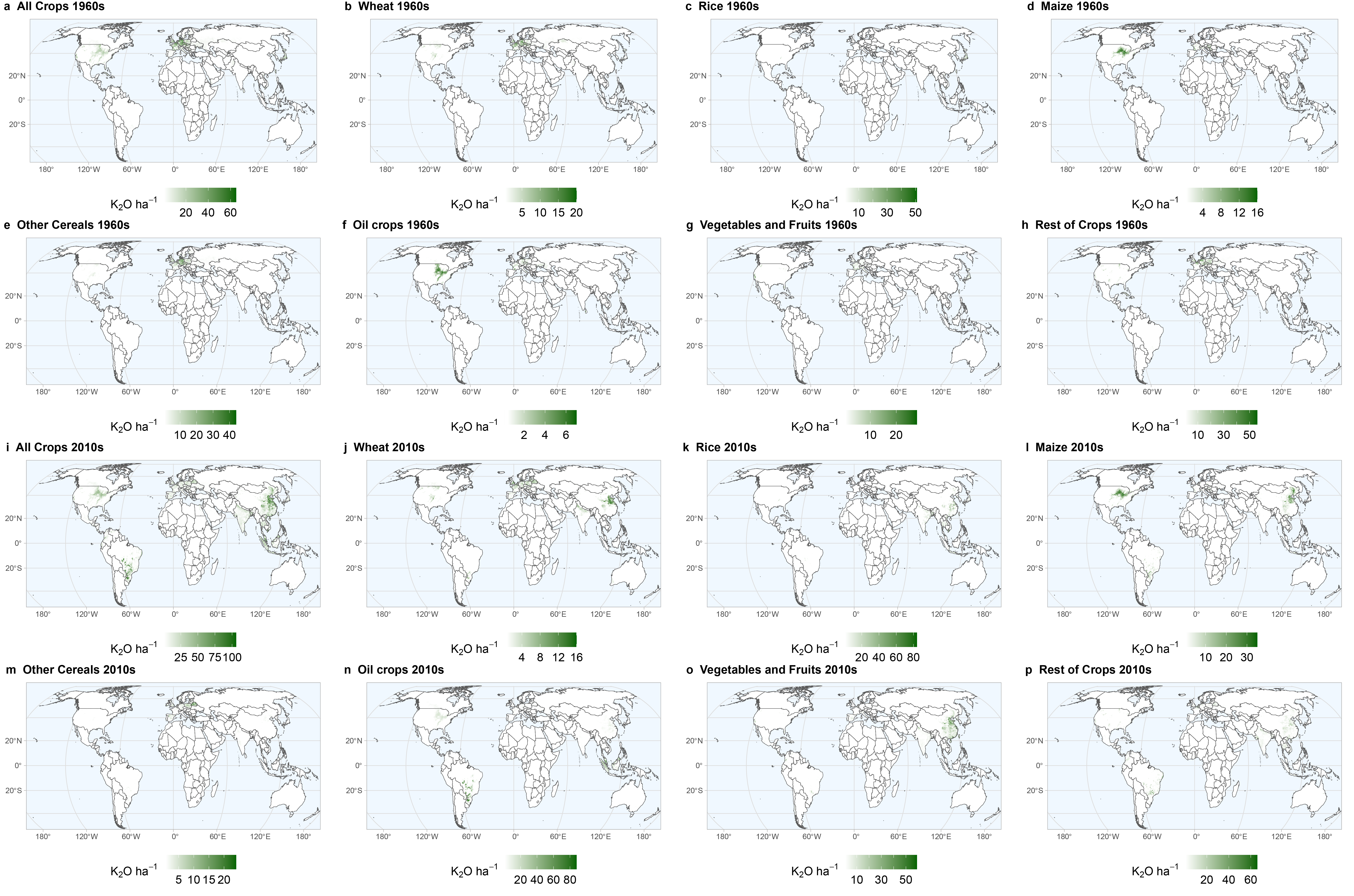}}
\caption{Spatial pattern of crop-specific fertilizer (\kfer{}) kg ha \textsuperscript{-1} consumed by each 0.05° grid cell for the following: a) average for the 1960s decade across all 13 crop classes, b) average for the 1960s decade for wheat, c) average for the 1960s decade for rice, d) average for the 1960s decade for maize, e) average for the 1960s decade for other cereals, f) average for the 1960s decade for all oil crops, g) average for the 1960s decade for vegetables and fruits, h) average for the 1960s decade for roots and tubers, sugar crops, fiber crops, and other crop classes, i) average for the 2010s decade across all 13 crop classes, j) average for the 2010s decade for wheat, k) average for the 2010s decade for rice, l) average for the 2010s decade for maize, m) average for the 2010s decade for other cereals, n) average for the 2010s decade for all oil crops, o) average for the 2010s decade for vegetables and fruits, p) average for the 2010s decade for roots and tubers, sugar crops, fiber crops, and other crop classes. The 1960s decade includes the years 1961-1969, and the 2010s decade includes the years 2010-2019}
\label{fig:fer_sd:map_potassium}
\end{figure}

To evaluate the validity of our results, we compare the compiled dataset based on the predictions against several national databases \cite{USDA2019,DEFRA2023FertiliserUKdataset, India1986Input, India1991Input, India1996Input, India2001Input, India2006Input, India2011Input, India2016Input, SWEDEN2010, SWEDEN2012, SWEDEN2015, SWEDEN2018,PSA2023,PAKISTAN,NZ2021} based on the \ac{MAE} and \ac{MAPE} errors between both, averaged over the available years as illustrated in \cref{tab:fer_sd:validation_model_mae}. For most country/crop combinations, the differences are within reasonable ranges, with \ac{MAE} values between 5-40 kg ha\textsuperscript{-1} and \ac{MAPE} values between 10\%-50\%. However, for some countries, the deviations are larger, suggesting that our models may not capture all the underlying intricacies in the data for each country or crop. This can be seen for Sweden where most results deviate between 20\%-100\%, or New Zealand where similar results can be found. However, it should be noted that these larger differences between our compiled dataset and the national databases cover only limited years as data was not always available for certain countries, as was the case for Sweden and New Zealand. Still, these discrepancies are slightly better than in earlier research \cite{Adalibieke2023Global1961}.  Additionally, more detailed plots to evaluate the results per year for the \ac{USA} and \ac{UK}, based on the \ac{USDA} and \ac{DEFRA} respectively, are included in \cref{fig:fer_sd:validation_NDB_PRED_USA,fig:fer_sd:validation_NDB_PRED_UK}. For the \ac{USDA} and \ac{DEFRA} crop nutrient data, the \ac{MAPE} values are less than 50\% and usually less than 25\%, except for \ac{USDA} soybean \nfer{} (\cref{fig:fer_sd:validation_NDB_PRED_USA}). \Cref{fig:fer_sd:validation_NDB_PRED_USA,fig:fer_sd:validation_NDB_PRED_UK} show that our predictions follow the real observed trend for the samples and thus form a reliable end source with only minimal differences. These discrepancies between the national databases and our compiled dataset can be attributed to occasional disparities between the application rates in the training data (data provided by the global dataset compilation) and the data in the national databases, e.g., the \ac{USA} data for soybean N in 1998 differed by 400\% between the two samples. These differences should be taken into account when comparing our results to the national databases, as our predictions are based on the global compiled dataset. As can be seen in \cref{tab:fer_sd:FAOSTAT_vs_NDB}, where the global databases data and the national databases are compared based on \ac{MAE} and \ac{MAPE}, most country/crop combination indicate an \ac{MAPE} values between 10\%-50\%, which is similar to our resulting error in \cref{tab:fer_sd:validation_model_mae}. Also, the lack of training samples for some country/crop combinations resulted in larger errors for these occurrences. 

To conclude, the model performances and logical feature importances, derived from the \ac{SHAP} values, in conjunction with the relatively minor differences between this study and regional statistics, as well as earlier literature \cite{Adalibieke2023Global1961}, indicate that our crop-specific fertilizer application rate dataset is comparatively reasonable across regions and years. 

\section*{Usage Notes}
In this study, we provide detailed estimates on global \nfer{}, \pfer{}, and \kfer{} fertilizer application rate based on the \ac{HGB} model output and compile a comprehensive dataset of these estimates by major crop groups between 1961 - 2019. Tabular data of the country- and crop-level predictions are made available as well as the 5-arcmin resolution gridded maps from our application, rendering an easy to use complete dataset. Subsequent analysis can be done both on the tabular data and the outputted maps, such as a trend analysis of fertilizer application rate or causal discovery to identify drivers of fertilizer application rate. Furthermore, our dataset can be leveraged as a source in other models where for example yield, ecological impact or fertilizer pricing can be seen as the output rather than use. 

Our results represent an improvement and advance in efforts to evaluate historical fertilizer consumption for different crop groups and fertilizers. However, as demonstrated during the validation process, this approach still has limitations that data source users should be made aware of. The limited amount of available data for some crops, nutrients, and regions can lead to biases, particularly in regions such as Africa, during certain years, especially in the 60s, and for certain nutrients, mainly \kfer{}. Hence, the \ac{ML} approach can be sensitive to outlying data points or noise and the limited data can make it prone to overfitting, which was mitigated as much as possible in the \ac{CV} setup. In addition, our model is trained on data provided by global datasets \cite{Martinez1982FertilizerProduction, AdolfoMartinez1990FertilizerT-37, FAO1992Fertilizer1, FAO1994Fertilizer2, FAO1996Fertilizer3, FAO1999Fertilizer4, FAO2002Fertilizer5, EFMA2002Fertilizer2001/02, EFMA2007Fertilizer2006/07, PattrickHeffer2009Asessment2007/08, FertilizerEurope2012Fertilizer2011/12, Heffer2017Assessment2014-2014/15, FertilizerEurope2015Fertilizer2014/15, Ludemann2022GlobalCountry}, which means that while our predictions may align closely with them, it is essential to acknowledge that they might diverge from national data mainly due to the difference between the two data sources as highlighted by the validation. This discrepancy between global and national databases such as the \ac{USDA} \cite{USDA2019} or \ac{DEFRA} \cite{DEFRA2023FertiliserUKdataset} databases highlights the complexity of accurately capturing historical fertilizer consumption trends across different regions and crop types. Moreover, the gridded cropland data provided by the HYDE 3.3 project \cite{klein2017anthropogenic}, is inconsistent with the one from satellite-derived land use (e.g., China and India \cite{liu2010china,tian2014history}) or data derived from a national census at regional scale  (e.g., \ac{USA} \cite{yu2018historical}), as stated by Adalibieke \textit{et al.} (2023) \cite{Adalibieke2023Global1961}. Furthermore, utilizing neighboring cells to allocate harvested areas across different crops, as well as leveraging the EARTHSTAT map \cite{monfreda2008farming}, implies some assumptions (see  \cref{eq:fer_sd:equation_maps_carea_rice,eq:fer_sd:equation_maps_carea_notrice,eq:fer_sd:equation_maps_carea__harea_rice,eq:fer_sd:equation_maps_carea__harea_notrice}). The main assumption is the suggestion that the distribution pattern of a specific cell mirrors that of its neighboring cells, which constrains potential changes in cropland over cells. The consistent use of the EARTHSTAT map \cite{monfreda2008farming} of the year 2000 also assumed that the crop group distribution of harvested area remains constant over time between 1961-2019. Finally, it is important to recognize that there are additional uncertainties stemming from the utilization of various data sources and methodological decisions within each data source, but these lie beyond the scope of our study. 

Nevertheless, our study extends the current literature by providing a more detailed historical geospatial distribution of fertilizer application rate by crop and using \ac{ML} to obtain detailed predictions with high precision. The detailed description and open-source code, in combination with the limited data sources used and ability to forecast, also make the method reproducible and easy to extend to forecast fertilizer application rate. In addition, our approach does not entail any assumptions, making it more flexible and robust than precious studies. Future research can build upon our study by expanding on more detailed specific fertilizer application rate. Considering the frequency of fertilizer application as well as the timing can be valuable for future research on the evaluation of fertilizer effectiveness and use. In addition, our study focuses on broad fertilizer applications, however, more detailed maps can be made for different types of specific fertilizers considered in our study (e.g., N fertilizer types). Furthermore, the time granularity of our maps can be improved. In addition, satellite data can be used to obtain even more fine-grained predictions, both in regions and more detailed time periods. Finally, a deeper investigation into the drivers of fertilizer application rate could enrich our understanding. While our focus has primarily been on the explainability of our model, exploring methodologies such as causal discovery or causal \ac{ML} within a temporal setting could unveil the drivers of fertilizer application rate over time, potentially providing valuable insights and facilitating more detailed predictions. 

\section*{Code availability}
Our Python (3.10.3) code, encompassing the model training, prediction generation, SHAP value computation, model validation and plot creation, as well as the R (4.2.2) scripts made for map generation are made available alongside the provided data map resources \cite{Janssens2024Code}. Open source packages used in the code are tabulated with their respective version in \cref{tab:fer_sd:code_packages}. Access to these resources is available at the designated location \cite{Coello2024Dataset,Janssens2024Code}.

\section*{Acknowledgements}
I.J. was supported by the European Commission: Horizon 2020 framework program for research and innovation under grant agreement No 964545, Bio-Accelerated Mineral weathering (BAM!). FC, JS and JP were supported by the Spanish Government grants PID2020115770RB-I, PID2022-140808NB-I00, and TED2021-132627 B–I00 funded by MCIN, AEI/10.13039/ 501100011033 European Union Next Generation EU/PRTR, the Fundación Ramón Areces grant CIVP20A6621, and the Catalan Government grants SGR 2021–1333 and AGAUR2023 CLIMA 00118.

\section*{Author contributions statement}
J.P., J.S., and T.V. designed the study. F.C. constructed the data. All authors analysed the data. I.J. and S.M. constructed the models and generated the \ac{SHAP} values. T.D. and F.C. created the spatial maps and model validation. I.J., F.C., S.M., and T.D. drafted the paper. All co-authors discuss the methods and results and reviewed and commented on the manuscript.

\section*{Competing interests}
The authors declare no competing interests.

\section{Tables}



\setlength\LTleft{-2cm}
\begin{longtable}[H]{p{5cm} p{1.5cm} p{1.5cm} p{1.5cm} p{5.5cm}} 
\caption{Fraction of \nfer{}, \pfer{}, and \kfer{} allocated for grasslands and fodder crops. The values given are the unique values or the range of values considered for the entire period. The mentioned sources give the information used to calculate these percentages, however, the specific country considerations are pointed throughout the Fertilizer use in other agricultural lands subsection}\\
\toprule
\textbf{Country} & \textbf{\nfer{} share} & \textbf{\pfer{} share} & \textbf{\kfer{} share} & \textbf{Source} \\
\midrule
Argentina & 0-9.8 & 0-28.0 & 0 & \cite{CEPAL1966, Martinez1982FertilizerProduction, AdolfoMartinez1990FertilizerT-37, FAO1992Fertilizer1, FAO1994Fertilizer2, FAO1996Fertilizer3, FAO1999Fertilizer4, FAO2002Fertilizer5, Argentina2006Grasslands, Argentina2011Grasslands, Argentina2012Grasslands, Argentina2013Grasslands, Argentina2014Grasslands, Argentina2015Grasslands, Argentina2016Grasslands, Argentina2017Grasslands, Argentina2018Grasslands}\\
Canada & 12.0 & 14.5 & 25.3 & \cite{Ludemann2022GlobalCountry, Pogue2023, Beaton1974} \\
Chile & 1.2-22.9 & 1.5-35.0 &  1.2-26.9 & \cite{CEPAL1966b, Martinez1982FertilizerProduction, AdolfoMartinez1990FertilizerT-37, FAO1992Fertilizer1, FAO1994Fertilizer2, FAO1996Fertilizer3, FAO1999Fertilizer4, FAO2002Fertilizer5, Heffer2017Assessment2014-2014/15, Ludemann2022GlobalCountry}\\
Dominican Republic & 0-3.1 & 0-3.0 & 0-2.5 & \cite{FAO1992Fertilizer1, FAO1996Fertilizer3, FAO1999Fertilizer4, Nuñez1999FertilizerDomRepublic}\\
Mexico & 0 & 2.6 & 0 & \cite{FAO1992Fertilizer1, FAO2002Fertilizer5}\\
United States of America & 6.6-16.6 & 4.0-17.2 & 6.8-19.1 & \cite{Nelson1968USFertilizerUse, Beaton1974, FAO1992Fertilizer1, FAO1996Fertilizer3, Heffer2017Assessment2014-2014/15} \\
Uruguay & 2.0-12.4 & 21.5-42.9 & 0 & \cite{Russell1974ForagesFertilizationTropical, AdolfoMartinez1990FertilizerT-37, FAO1992Fertilizer1, FAO2002Fertilizer5, Ludemann2022GlobalCountry}\\
\hline
Australia & 6.4 & 38.4 & 41.6 & \cite{ AdolfoMartinez1990FertilizerT-37, FAO1992Fertilizer1, FAO1994Fertilizer2, FAO1996Fertilizer3, FAO1999Fertilizer4, FAO2002Fertilizer5, Heffer2017Assessment2014-2014/15,  Ludemann2022GlobalCountry}\\
New Zealand & 91.1 & 93.0 & 88.8 & \cite{FAO1996Fertilizer3, FAO2002Fertilizer5, Heffer2017Assessment2014-2014/15, Ludemann2022GlobalCountry}\\
\hline
Austria & 20.8-31.4 & 27.1-30.3 & 19.5-21.8 & \cite{Einarsson2021, Martinez1982FertilizerProduction, AdolfoMartinez1990FertilizerT-37, FAO1992Fertilizer1, FAO1994Fertilizer2, FAO1996Fertilizer3, FAO1999Fertilizer4, FAO2002Fertilizer5}\\
Belgium and Luxembourg & 52.7-66.9 & 35.5-62.3 & 41.2-52.3 & \cite{Einarsson2021, Martinez1982FertilizerProduction, FAO1992Fertilizer1, FAO1994Fertilizer2, FAO1996Fertilizer3, FAO1999Fertilizer4, FAO2002Fertilizer5, EFMA2002Fertilizer2001/02, EFMA2007Fertilizer2006/07, FertilizerEurope2012Fertilizer2011/12, FertilizerEurope2015Fertilizer2014/15, Ludemann2022GlobalCountry} \\
Czech Republic & 16.7-19.7 & 13.6-16.0 & 13.9-16.3 & \cite{Einarsson2021, FAO2002Fertilizer5, FertilizerEurope2012Fertilizer2011/12, FertilizerEurope2015Fertilizer2014/15, Ludemann2022GlobalCountry}\\
Slovakia & 10.4-13.6 & 6.6-8.6 & 5.9-7.7 & \cite{Einarsson2021, FAO2002Fertilizer5, FertilizerEurope2012Fertilizer2011/12, FertilizerEurope2015Fertilizer2014/15, Ludemann2022GlobalCountry}\\
Czechoslovakia & 20.8-31.4 & 27.1-30.3 & 19.5-21.8 & \cite{Einarsson2021, FAO2002Fertilizer5, FertilizerEurope2012Fertilizer2011/12, FertilizerEurope2015Fertilizer2014/15, Ludemann2022GlobalCountry}\\
Denmark & 10.0-62.0 & 10.0-74.0 & 9.0-61.0 & \cite{Petersen1977DenmarkFertilizer, Martinez1982FertilizerProduction, FAO1992Fertilizer1, FAO1994Fertilizer2, FAO1996Fertilizer3, FAO1999Fertilizer4, FAO2002Fertilizer5, EFMA2002Fertilizer2001/02, EFMA2007Fertilizer2006/07, FertilizerEurope2012Fertilizer2011/12, FertilizerEurope2015Fertilizer2014/15, Ludemann2022GlobalCountry}\\
Finland & 37.0-49.0 & 22.0-37.0 & 21.0-64.0 & \cite{Martinez1982FertilizerProduction, FAO1992Fertilizer1, FAO1994Fertilizer2, FAO1996Fertilizer3, FAO1999Fertilizer4, FAO2002Fertilizer5, EFMA2002Fertilizer2001/02, EFMA2007Fertilizer2006/07, FertilizerEurope2012Fertilizer2011/12, FertilizerEurope2015Fertilizer2014/15, Ludemann2022GlobalCountry} \\
France & 7.0-39.0 & 9.0-48.0 & 12.0-52.0 & \cite{Einarsson2021, LeNoe2018FranceNutrients, SSP1984_EnqueteFrancePraire, SSP2001_EnqueteFrance, SSP2006_EnqueteFrance, SSP2020_EnqueteFrance, FAO1992Fertilizer1, FAO1994Fertilizer2, FAO1996Fertilizer3, FAO1999Fertilizer4, FAO2002Fertilizer5, EFMA2002Fertilizer2001/02, EFMA2007Fertilizer2006/07, FertilizerEurope2012Fertilizer2011/12, FertilizerEurope2015Fertilizer2014/15, Ludemann2022GlobalCountry, Beaton1974}\\
Germany & 11.0-43.0 & 10.0-42.0 & 9.0-39.0 & \cite{Einarsson2021, Martinez1982FertilizerProduction, Beaton1974, FAO1992Fertilizer1, FAO1994Fertilizer2, FAO1996Fertilizer3, FAO1999Fertilizer4, FAO2002Fertilizer5, EFMA2002Fertilizer2001/02, EFMA2007Fertilizer2006/07, FertilizerEurope2012Fertilizer2011/12, FertilizerEurope2015Fertilizer2014/15}\\
Greece & 0-10.0 & 0-13.0 & 0-10.0 & \cite{Einarsson2021, FAO1992Fertilizer1, FAO1994Fertilizer2, FAO1996Fertilizer3, FAO1999Fertilizer4, FAO2002Fertilizer5, EFMA2002Fertilizer2001/02, EFMA2007Fertilizer2006/07, FertilizerEurope2012Fertilizer2011/12, FertilizerEurope2015Fertilizer2014/15, Ludemann2022GlobalCountry}\\
Hungary & 1.0-20.0 & 1.0-18.0 & 1.0-20.0 & \cite{Einarsson2021, Martinez1982FertilizerProduction, FAO1992Fertilizer1, EFMA2007Fertilizer2006/07, FertilizerEurope2012Fertilizer2011/12, FertilizerEurope2015Fertilizer2014/15, Ludemann2022GlobalCountry}\\
Ireland & 24.0-90.0 & 20.0-82.0 & 19.0-83.0 & \cite{Walsh1957Ireland, IrelandCSOSurface1980,IrelandCSOSurface1990, IrelandCSOSurface2000, IrelandCSOSurface2010, IrelandCSOSurface2020, Murphy1978IrelandSurvey, Murphy1983IrelandSurvey, Murphy1987IrelandSurvey, Coulter2005IrelandSurvey, Lalor2010IrelandSurvey, Dillon2018IrelandSurvey, FAO1992Fertilizer1, FAO1994Fertilizer2, FAO1996Fertilizer3, FAO1999Fertilizer4, FAO2002Fertilizer5, EFMA2002Fertilizer2001/02, EFMA2007Fertilizer2006/07, FertilizerEurope2012Fertilizer2011/12, FertilizerEurope2015Fertilizer2014/15, Ludemann2022GlobalCountry}\\
Italy & 9.0-11.0 & 7.0-8.0 & 6.0-8.0 & \cite{Einarsson2021, Beaton1974, FAO1994Fertilizer2, FAO1996Fertilizer3, FAO1999Fertilizer4, FAO2002Fertilizer5, EFMA2002Fertilizer2001/02, EFMA2007Fertilizer2006/07, FertilizerEurope2012Fertilizer2011/12, FertilizerEurope2015Fertilizer2014/15, Ludemann2022GlobalCountry}\\
The Netherlands & 52.8-77.6 & 9.2-58.3 & 10.5-26.2 & \cite{Prins1983Netherlands, Einarsson2021, Martinez1982FertilizerProduction, FAO1996Fertilizer3, FAO1999Fertilizer4, FAO2002Fertilizer5, EFMA2002Fertilizer2001/02, EFMA2007Fertilizer2006/07, FertilizerEurope2012Fertilizer2011/12, FertilizerEurope2015Fertilizer2014/15, Ludemann2022GlobalCountry}\\
Poland & 1.0-43.0 & 1.0-40.0 & 1.0-33.0 & \cite{Einarsson2021, Kurek1971PolandSurvey, FAO1992Fertilizer1, FAO1999Fertilizer4, FAO2002Fertilizer5, EFMA2002Fertilizer2001/02, EFMA2007Fertilizer2006/07, FertilizerEurope2012Fertilizer2011/12, FertilizerEurope2015Fertilizer2014/15, Ludemann2022GlobalCountry}\\
Portugal & 2.0-23.0 & 3.0-23.0 & 2.0-29.0 & \cite{Einarsson2021, Martinez1982FertilizerProduction, FAO1992Fertilizer1, FAO1994Fertilizer2, FAO1999Fertilizer4, FAO2002Fertilizer5, EFMA2002Fertilizer2001/02, EFMA2007Fertilizer2006/07, FertilizerEurope2012Fertilizer2011/12, FertilizerEurope2015Fertilizer2014/15, Ludemann2022GlobalCountry} \\
Romania & 4.3-5.6 & 4.2-5.4 & 2.2-2.8 & \cite{Einarsson2021, FAO1992Fertilizer1, FAO1996Fertilizer3, FertilizerEurope2012Fertilizer2011/12, FertilizerEurope2015Fertilizer2014/15, EFMA2007Fertilizer2006/07}\\
Spain & 4.0-4.6 & 1.4-12.1 & 0-7.9 & \cite{Einarsson2021, GarciaSerranoSpain2010, Martinez1982FertilizerProduction, FAO1992Fertilizer1, FAO1994Fertilizer2, FAO1999Fertilizer4, FAO2002Fertilizer5, EFMA2002Fertilizer2001/02, EFMA2007Fertilizer2006/07, FertilizerEurope2012Fertilizer2011/12, FertilizerEurope2015Fertilizer2014/15, Ludemann2022GlobalCountry}\\
Sweden & 12.7-45.1 & 2.1-36.7 & 0-7.9 & \cite{Beaton1974, AdolfoMartinez1990FertilizerT-37, FAO1992Fertilizer1, FAO1994Fertilizer2, FAO1999Fertilizer4, FAO2002Fertilizer5, EFMA2002Fertilizer2001/02, EFMA2007Fertilizer2006/07, FertilizerEurope2012Fertilizer2011/12, FertilizerEurope2015Fertilizer2014/15, Ludemann2022GlobalCountry}\\
United Kingdom and Northern Ireland & 32.4-60.8 & 17.6-48.8 & 21.7-39.7 & \cite{FAO1992Fertilizer1, FAO1996Fertilizer3, EFMA2002Fertilizer2001/02, EFMA2007Fertilizer2006/07, FertilizerEurope2012Fertilizer2011/12, FertilizerEurope2015Fertilizer2014/15, Ludemann2022GlobalCountry,Church1971EnglandSurvey, DEFRA2023FertiliserUKdataset, DEFRA2024AreasEngland}\\
Iceland & 97.5 & 97.5 & 97.5 & \cite{Johannesson2010Iceland, Helgadottir2013Iceland}\\
Switzerland & 32.7-56.5 & 36.3-51.0 & 10.8-38.2 & \cite{Martinez1982FertilizerProduction, FAO1992Fertilizer1, FAO1994Fertilizer2, FAO1996Fertilizer3, FAO1999Fertilizer4, FAO2002Fertilizer5}\\
Norway & 64.0 & 50.0 & 66.0 & \cite{Beaton1974, FAO1994Fertilizer2, FAO1996Fertilizer3, FAO1999Fertilizer4, FAO2002Fertilizer5, EFMA2002Fertilizer2001/02, EFMA2007Fertilizer2006/07, FertilizerEurope2012Fertilizer2011/12, FertilizerEurope2015Fertilizer2014/15, Ludemann2022GlobalCountry}\\
Yugoslav SFR & 15.4 & 16.1 & 14.9 & \cite{Einarsson2021, FAO1994Fertilizer2, EFMA2002Fertilizer2001/02, Einarsson2021, YugoslaviaReport1995}\\
Croatia & 8.8-22.4 & 8.7-25.9 & 8.8-37.8 & \cite{Einarsson2021,FAO1994Fertilizer2, FAO2002Fertilizer5, EFMA2002Fertilizer2001/02, EFMA2007Fertilizer2006/07, FertilizerEurope2012Fertilizer2011/12, FertilizerEurope2015Fertilizer2014/15, Ludemann2022GlobalCountry}\\
Montenegro and North Macedonia & 10.2-25.1 & 10.5-28.2 & 10.4-36.0 & \cite{Einarsson2021,FAO1994Fertilizer2, FAO2002Fertilizer5, EFMA2002Fertilizer2001/02, EFMA2007Fertilizer2006/07, FertilizerEurope2012Fertilizer2011/12, FertilizerEurope2015Fertilizer2014/15, Ludemann2022GlobalCountry}\\
Serbia & 11.6-27.7 & 12.2-31.1 & 12.1-37.6 & \cite{Einarsson2021,FAO1994Fertilizer2, FAO2002Fertilizer5, EFMA2002Fertilizer2001/02, EFMA2007Fertilizer2006/07, FertilizerEurope2012Fertilizer2011/12, FertilizerEurope2015Fertilizer2014/15, Ludemann2022GlobalCountry,Lugic2010Serbia}\\
Slovenia & 45.6-70.9 & 43.2-77.9 & 35.1-76.5 & \cite{Einarsson2021,FAO1994Fertilizer2, FAO2002Fertilizer5, EFMA2002Fertilizer2001/02, EFMA2007Fertilizer2006/07, FertilizerEurope2012Fertilizer2011/12, FertilizerEurope2015Fertilizer2014/15, Ludemann2022GlobalCountry}\\
USSR & 0-34.0 & 0-34.0 & 0-32.0 & \cite{Loza1977USSR1975, Klatt1976USSR, FAO1992Fertilizer1, Shend1993USSRStats}\\
Armenia, Georgia and Azerbaijan & 4.0 & 7.0 & 9.0 & \cite{FAO2002Fertilizer5}\\
Kazakhstan, Kyrgyzstan, Tajikistan, Turkmenistan and Uzbekistan & 2.0 & 2.0 & 1.5 & \cite{Heffer2017Assessment2014-2014/15}\\
Estonia & 5.0-40.0 & 3.0-32.0 & 21.0-64.0 & \cite{FAO2002Fertilizer5, EFMA2002Fertilizer2001/02, EFMA2007Fertilizer2006/07, FertilizerEurope2012Fertilizer2011/12, FertilizerEurope2015Fertilizer2014/15, Ludemann2022GlobalCountry}\\
Latvia & 7.0-81.0 & 6.0-60.0 & 6.0-65.0 & \cite{FAO2002Fertilizer5, EFMA2002Fertilizer2001/02, EFMA2007Fertilizer2006/07, FertilizerEurope2012Fertilizer2011/12, FertilizerEurope2015Fertilizer2014/15, Ludemann2022GlobalCountry}\\
Lithuania & 18.0-59.0 & 16.0-45.0 & 16.0-63.0 & \cite{FAO2002Fertilizer5, EFMA2002Fertilizer2001/02, EFMA2007Fertilizer2006/07, FertilizerEurope2012Fertilizer2011/12, FertilizerEurope2015Fertilizer2014/15, Ludemann2022GlobalCountry}\\
Belarus & 27.0 & 14.0 & 26.0 & \cite{Heffer2017Assessment2014-2014/15, Ludemann2022GlobalCountry}\\
Republic of Moldova & 7.0 & 6.0 & 3.0 & \cite{FAO1996Fertilizer3}\\
Ukraine & 2.0 & 1.0 & 1.0 & \cite{Heffer2017Assessment2014-2014/15, Ludemann2022GlobalCountry}\\
Russian Federation & 6.5-43.4 & 1.8-19.4 & 2.8-33.4 & \cite{FAO1994Fertilizer2, FAO1996Fertilizer3, Heffer2017Assessment2014-2014/15}\\
\hline
Islamic Republic of Iran & 0-3.2 & 0-3.7 & 0-1.1 & \cite{FAO1992Fertilizer1, FAO1994Fertilizer2, Heffer2017Assessment2014-2014/15}\\
Japan and Republic of Korea & 17.3 & 16.9 & 15.6 & \cite{Martinez1982FertilizerProduction, FAO1992Fertilizer1, FAO1994Fertilizer2, FAO1996Fertilizer3, FAO1999Fertilizer4, FAO2002Fertilizer5, Heffer2017Assessment2014-2014/15, Ludemann2022GlobalCountry}\\
Turkey & 0-1.2 & 0-2.4 & 0-2.1 & \cite{Martinez1982FertilizerProduction, FAO1992Fertilizer1, FAO1994Fertilizer2, FAO1996Fertilizer3, FAO1999Fertilizer4, FAO2002Fertilizer5, Heffer2017Assessment2014-2014/15, Ludemann2022GlobalCountry}\\
\hline
Egypt & 4.0 & 8.6 & 1.0 & \cite{AdolfoMartinez1990FertilizerT-37, FAO1999Fertilizer4}\\
Morocco & 14.8 & 10.5 & 6.1 & \cite{FAO1992Fertilizer1, FAO1999Fertilizer4, FAO2006Morocco}\\
South Africa & 12.4 & 13.3 & 9.2 & \cite{FAO1992Fertilizer1, FAO1994Fertilizer2, FAO1996Fertilizer3, FAO1999Fertilizer4, FAO2002Fertilizer5, Heffer2017Assessment2014-2014/15, Ludemann2022GlobalCountry}\\
\bottomrule

\label{tab:fer_sd:grassland_share}
\end{longtable}

\newpage
\begin{longtable}[H]{p{0.2cm} p{3.55cm} p{8.3cm} p{1.7cm} p{0.8cm} p{1cm}} 
\caption{Environmental, agrological and socioeconomic features used in the prediction of the fertilizer application rates, accompanied by their description, unit and data source. The \textit{Model} column indicates whether the feature was an input for either the \nfer{}, \pfer{} or \kfer{} prediction, or for all 3 predictions.}\\

\toprule
& \textbf{Feature} & \textbf{Description} & \textbf{Unit} & \textbf{Model} & \textbf{Source} \\ 
\midrule
\endfirsthead
\bottomrule
\endlastfoot %

& Year & Year of the data && All &  \\ 
& Crop & Crop class && All &  \\ 
& Country & Code of the country or region in FAOSTAT && All & \\ 
& Country surface & Surface of the country & $km^2$ & All & \cite{UNArea} \\
& Region & World region && All & \cite{FAOSTAT2022Regions} \\ 

\cellcolor{env} & \ac{PET} & Annual potential evapotranspiration & mm/year & All & \cite{Harris2020VersionDataset} \\ 
\cellcolor{env} & \ac{MAP} & Annual precipitation & mm/year & All & \cite{Harris2020VersionDataset} \\
\cellcolor{env} & TMN & Average annual temperature & $^{\circ}$ C & All & \cite{Harris2020VersionDataset} \\
\cellcolor{env} & Aridity index & Aridity index && All & \cite{Harris2020VersionDataset} \\
\cellcolor{env} & Soil N & Average soil nitrogen content at 0-30 cm depth & cg/kg & All & \cite{Poggio2021SoilGridsUncertainty} \\
\cellcolor{env} & Soil \ac{OCS} & Average soil organic carbon stock at 0-30 cm depth & t/ha & All & \cite{Poggio2021SoilGridsUncertainty} \\
\cellcolor{env} & Soil sand & Average soil sand content at 0-30 cm depth & g/kg & All & \cite{Poggio2021SoilGridsUncertainty} \\
\cellcolor{env} & Soil silt & Average soil silt content at 0-30 cm depth & g/kg & All & \cite{Poggio2021SoilGridsUncertainty} \\
\cellcolor{env} & Soil clay & Average soil clay content at 0-30 cm depth & g/kg & All & \cite{Poggio2021SoilGridsUncertainty} \\
\cellcolor{env} & Soil pH & Average soil pH at 0-30 cm depth && All & \cite{Poggio2021SoilGridsUncertainty} \\
\cellcolor{env} \multirow{-12}{*}{\rotatebox[origin=c]{90}{\textcolor{white}{Environmental}}} & Soil \ac{CEC} & Average soil cation exchange capacity at pH 7 at 0-30 cm depth & mmol(c)/kg & All & \cite{Poggio2021SoilGridsUncertainty} \\

\cellcolor{agr}& Crop area & Harvested Area of the crop & ha & All & \cite{FAOSTAT2023AreaDataset} \\ 
\cellcolor{agr}& Crop area perc & Area of the crop over the total cropland area & \% & All & \cite{FAOSTAT2023AreaDataset} \\ 
\cellcolor{agr}& Country N per ha & Amount of \nfer{} fertilizer used per cropland area & t/ha & \nfer{} & \cite{FAOSTAT2023FertilizerDataset} \\
\cellcolor{agr}& Country \pfer{} per ha & Amount of \pfer{} fertilizer used per cropland area & t/ha & \pfer{} & \cite{FAOSTAT2023FertilizerDataset} \\
\cellcolor{agr}& Country \kfer{} per ha & Amount of \kfer{} fertilizer used per cropland area & t/ha & \kfer{} & \cite{FAOSTAT2023FertilizerDataset} \\
\cellcolor{agr}& Country N use & Amount of \nfer{} fertilizer used in the country & t & \nfer{} & \cite{FAOSTAT2023FertilizerDataset} \\ 
\cellcolor{agr}& Country \pfer{} use & Amount of \pfer{} fertilizer used in the country & t & \pfer{} & \cite{FAOSTAT2023FertilizerDataset} \\ 
\cellcolor{agr}& Country \kfer{} use & Amount of \kfer{} fertilizer used in the country & t & \kfer{} & \cite{FAOSTAT2023FertilizerDataset} \\
\cellcolor{agr}& Holding size stand & Standardized average size of farms for each country and year & ha & All & \cite{Zou2022} \\ 
\cellcolor{agr}& Crop N content & N content of the crop & kg/t & \nfer{} & \cite{Ludemann2023A19612020}\\
\cellcolor{agr}& Crop P content & P content of the crop & kg/t & \pfer{} & \cite{Ludemann2023A19612020}\\
\cellcolor{agr}& Crop K content & K content of the crop & kg/t & \kfer{} & \cite{Ludemann2023A19612020}\\
\cellcolor{agr}& Crop N removal per ha & Average N removal per ha for the crop, country and year & kg/ha & \nfer{} & \cite{Ludemann2023A19612020, FAOSTAT2023AreaDataset}\\ 
\cellcolor{agr}& Crop P removal per ha & Average P removal per ha for the crop, country and year & kg/ha & \pfer{} & \cite{Ludemann2023A19612020, FAOSTAT2023AreaDataset} \\ 
\cellcolor{agr}& Crop K removal per ha & Average K removal per ha for the crop, country and year & kg/ha & \kfer{} & \cite{Ludemann2023A19612020, FAOSTAT2023AreaDataset}\\ 
\cellcolor{agr}& Irrigation implementation & Share of agricultural land irrigated in the country & \% & All & \cite{WB2023Irri}\\
\cellcolor{agr}\multirow{-17}{*}{\rotatebox[origin=c]{90}{\textcolor{white}{Agrological}}} & Machinery use & Number of agriculture machinery per ha of arable land for the country and year & $ha^{-1}$ & All & \cite{WB2023Machinery, FAOSTAT2023AgArea} \\


\cellcolor{ec} & Global urea price & Current urea price per metric tonnes & \$ current & \nfer{} & \cite{WorldBank2022WorldSheet}\\
\cellcolor{ec}& Global P-rock price & Current P price per metric tonnes & \$ current & \pfer{} & \cite{WorldBank2022WorldSheet}\\
\cellcolor{ec}& Global \kfer{} price & Current \kfer{} price per metric tonnes & \$ current & \kfer{} & \cite{WorldBank2022WorldSheet}\\
\cellcolor{ec}& Global crop price & Real global crop price & \$ current & All & \cite{WorldBank2022WorldSheet}\\
\cellcolor{ec}& Education & Fraction of \ac{GDP} used for education & \% & All & \cite{WorldBank2023GovermentGDP}\\
\cellcolor{ec} & \ac{GDP} per capita & Current \ac{GDP} per capita & \$ current & All & \cite{UN2023GDPcap} \\ %
\cellcolor{ec}& \nfer{} cost from production & \nfer{} fertilizer cost from production &  & \nfer{} & \cite{McArthur2017FertilizingDevelopment}\\
\cellcolor{ec}& P cost from production & \pfer{} fertilizer cost from production & & \pfer{} & \cite{McArthur2017FertilizingDevelopment} \\
\cellcolor{ec}& K cost from production & \kfer{} fertilizer cost from production &  & \kfer{} & \cite{McArthur2017FertilizingDevelopment} \\
\cellcolor{ec}& Population pressure & Population per ha of agricultural land & persons/ha & All & \cite{FAOSTAT2023AgArea, FAOSTAT2023Population} \\

\cellcolor{ec}\multirow{-11}{*}{\rotatebox[origin=c]{90}{\textcolor{white}{Socioeconomic}}} & National crop price & Real price paid to farmer at the country-level & \$ current & All & \cite{FAOSTAT2023PodPrice, FAOSTAT2023PodPriceOld} \\ 

\label{tab:fer_sd:features}
\end{longtable}


\begin{longtable}[H]{p{2.3cm} p{0.6cm}p{2cm}p{11.15cm}} 
    
\caption{Crop Classification with FAOSTAT Item Codes}\\
\toprule
\textbf{Crop Class} & \textbf{Crop Code} & \textbf{Description} & \textbf{Crops FAOSTAT (FAOSTAT Item Code)} \\
\midrule
\endfirsthead
\bottomrule
\endlastfoot %
Wheat & 1\_1 & Wheat & Wheat (15) \\
Maize & 1\_2 & Maize, only for grain & Maize, corn (56) \\
Rice & 1\_3 & Rice & Rice (27) \\
Other Cereals & 1\_4 & Other cereals not mentioned above & Barley (44), Buckwheat (89), Canary seed (101), Fonio (94), Millet (79), Oats (75), Rye (71), Sorghum (83), Triticale (97), Quinoa (92), Cereal n.e.c (108) \\
Soybean & 2\_1 & Soybean & Soya beans (236) \\
Palm Oil fruit & 2\_2 & Palm oil fruit & Oil Palm fruit (254) \\
Other Oilseeds & 2\_3 & Other oilseed crops not soybean and palm oil fruit & Castor oil seeds (265), Coconut, in shell (249), Jojoba seeds (277), Linseed (333), Mustard seed (292), Olives (260), Poppy seeds (296), Rape and colza seed (270), Safflower (280), Sesame seed (289), Sunflower seed (267), Tallowtree seed (305), Tung nuts (275), Other oil seeds, n.e.c (339) \\
Vegetables & 3\_1 & Vegetables & Artichokes (366), Asparagus (367), Cabbages (358), Cauliflowers and broccoli (393), Chillies and peppers, green (401), Cucumber and gherkins (397), Eggplants (399), Green garlic (406), Leeks and alliaceous (407), Cantaloupes and other melons (568), Melonseed (299), Mushrooms and truffles (449), Okra (430), Onion and shallots, green (402), Onion and shallot, dry (403), Pumpkins, squash and gourds (394), Spinach (373), Tomatoes (388), Watermelons (567), Carrots and turnips (426), Lettuce and chickory (372), Cassava leaves (378), Green corn (446), Other vegetables fresh n.e.c (463) \\
Fruits & 3\_2 & Fruits & Apples (515), Apricots (526), Avocados (572), Bananas (486), Blueberries (552), Cherries (531), Sour cherries (530), Cranberries (554), Currants (550), Dates (577), Figs (569), Gooseberries (549), Pomelos and grapefruits (507), Grapes (560), Kiwi fruits (592), Lemos and limes (497), Oranges (490), Papayas (600), Peaches and nectarines (534), Pears (521), Perssimons (587), Pineapples (574), Plantains (489), Plums and sloes (536), Quinces (523), Raspberries (547), Strawberries (544), Tangerines, mandarins, clementines (495), other berries n.e.c (558), other citrus n.e.c. (512), other fruits n.e.c. (619), Other pome fruits n.e.c (542), Other stone fruits n.e.c (541), Other tropical fruits n.e.c (603) \\
Roots and tubers & 4 & Roots and tubers & Cassava (125), Potatoes (116), Sweet potatoes (122), Taro (136), Yams (137), Yautia (135), Edible roots and tubers n.e.c. (149) \\
Sugar crops & 5 & Sugar cane, sugar beet and only sugar crops & String beans (423), Sugar beet (157), Sugar cane (156), Locust beans (461), Other sugar Crops n.e.c. (161) \\
Fiber crops & 6 & Cotton and other fiber crops & Coir (813), True hemp (777), Hempseed (336), Jute (780), Kapok fruit (310), Kapok seed (311), Karite nuts (263), Abaca, manila, hemp (809), Ramie (788), Seed cotton (328), Sisal (789), Agave fibres (800), Flax (773), Kenak (782), Other fibre crops (821) \\
Other crops & 7 & Nuts, pulses, stimulants and aromatics, natural rubber, tobacco & Almonds (221), Areca nuts (226), Cashew nuts (217), Chestnuts (220), Hazelnuts (225), Pistachios (223), Walnut (222), Brazil nuts (216), Kola nuts (224), Other nuts (234),  Broad beans and horse beans, dry (181), Broad beans and horse beans, greens (420), Chick peas (191), Cow peas, dry (195), Lentils, dry (201), Lupins (210), Peas, dry (187), Peas, green (417), Pidgeon peas, dry (197), Bambara beans (203), Vetches (205), Other beans, green (414), Other pulses n.e.c (211), Coffee, green (656), Green tea (675), Tea leaves (667), Cocoa beans (661), Chickory roots (459), Mate leaves (671), Other stimulant, spice and aromatic n.e.c (723), Anise, badian, coriander, cumin, caraway, fennel and juniper (711), Cinnamon (693), Cloves (698), Ginger (720), Hop cones (677), Pepper (Piper spp.) (687), Nutmeg, mace, cardamoms (702), Vanilla (692), Chillies and peppers (689), Peppermint (748), Pyrethrum (754), Tobacco (826), Natural rubber (836), Balata, gutta-, percha-, chicle, and similar natural gums (839) \\
\label{tab:fer_sd:crop_classification}
\end{longtable}

\begin{table}[H]
\centering
\caption{\label{tab:fer_sd:hyperparameters_models}Overview of the explored hyperparameters for the Histogram-based Gradient Boosting (\ac{HGB}) and eXtreme Gradient Boosting (\ac{XGB}) regression models. }
\begin{tabular}{lll}
\toprule
\textbf{Method} & \textbf{Hyperparameter} & \textbf{Possible values} \\
\midrule
\multirow{4}{*}{\ac{HGB}} & max\_depth & 2, 5, 10, 20 \\
& max\_iter & 25, 50, 100, 200, 500 \\
& learning\_rate & 0.01, 0.1, 0.5, 1 \\
& min\_samples\_leaf & 5, 10, 20, 50 \\
\midrule
\multirow{5}{*}{\ac{XGB}} & max\_depth & 2, 3, 4, 5 \\
& n\_estimators & 25, 50, 100, 200, 300, 400 \\
& colsample\_by\_tree & 0.6, 0.7, 0.8, 0.9, 1.0 \\
& subsample & 0.6, 0.7, 0.8, 0.9, 1.0 \\
& min\_child\_weight & 3, 4, 5, 6, 8, 10 \\
\bottomrule
\end{tabular}
\end{table}

\begin{table}[H]
\centering
\caption{\label{tab:fer_sd:metrics}Performances of the eXtreme Gradient Boosting (XGB) and HistGradientBoosting (HGB) models on the test sets in a 2x5-fold nested cross validation grid search. The performance is quantified using the mean absolute error (MAE), root mean squared error (RMSE), mean squared error (MSE) and Pearson correlation coefficient ($R^2$). The naive performance of a model is defined as the performance of a model that uses the mean of all training samples as its prediction. It serves as a baseline value to compare the test performances of the models with. The best performances are indicated in boldface.}
\begin{tabular}{llllll}
\toprule
\textbf{Fertilizer} & \textbf{Model} & \textbf{\ac{MAE}} & \textbf{\ac{RMSE}} & \textbf{\ac{MSE}} & $\mathbf{R^2}$ \\
\midrule
\multirow{3}{*}{\nfer{}} & \ac{HGB} & $\mathbf{26.01 \pm 0.94}$ & $43.50 \pm 5.13$ & $\mathbf{1905 \pm 446}$ & $\mathbf{0.62 \pm 0.04}$ \\
& \ac{XGB} & $26.67 \pm 1.48$ & $\mathbf{43.35 \pm 7.12}$ & $\mathbf{1905 \pm 617}$ & $\mathbf{0.62 \pm 0.08}$ \\
& naive & $53.09 \pm 0.75$ & $70.13 \pm 4.19$ & $4927 \pm 588$ & $0.00 \pm 0.00$ \\
\midrule
\multirow{3}{*}{\pfer{}} & \ac{HGB} & $\mathbf{15.19 \pm 0.67}$ & $\mathbf{25.68 \pm 1.18}$ & $\mathbf{660 \pm 61}$ & $\mathbf{0.63 \pm 0.05}$ \\
& \ac{XGB} & $16.83 \pm 0.23$ & $26.40 \pm 0.74$ & $697 \pm 39$ & $0.61 \pm 0.04$ \\
& naive & $29.97 \pm 0.23$ & $42.12 \pm 1.02$ & $1774 \pm 86$ & $0.00 \pm 0.00$ \\
\midrule
\multirow{3}{*}{\kfer{}} & \ac{HGB} & $\mathbf{19.18 \pm 0.27}$ & $\mathbf{35.74 \pm 4.56}$ & $\mathbf{1287 \pm 326}$ & $\mathbf{0.65 \pm 0.08}$ \\
& \ac{XGB} & $19.99 \pm 0.20$ & $36.24 \pm 4.66$ & $1324 \pm 338$ & $0.64 \pm 0.09$ \\
& naive & $43.08 \pm 0.76$ & $60.25 \pm 0.52$ & $3631 \pm 63$ & $0.00 \pm 0.00$ \\
\bottomrule
\end{tabular}
\end{table}

\clearpage
\begin{table}[H] 
\centering
\caption{\label{tab:fer_sd:validation_model_mae} Validation of our model predictions of the average application rate per ha against national database information for certain countries and crops per fertilizer. The validation is quantified using the mean absolute error (MAE) and mean percentage error (MAPE) per fertilizer between the two data sources, expressed in the table as \ac{MAE} and MAPE respectively (fertilizer). The NPK stands for the sum of all fertilizers used in the country for certain crops, this is only discussed for Pakistan as more granular data is not available. Unavailable data points are expressed as NA in the table. The sample size of the comparison per country is indicated in parentheses.}
\hspace*{-2cm}
\begin{tabular}{p{2.5cm}lllllllll}
\toprule
\textbf{Country} & \textbf{Crop} & \multicolumn{2}{c}{\textbf{\nfer}} & \multicolumn{2}{c}{\textbf{\pfer}} & \multicolumn{2}{c}{\textbf{\kfer}} & \multicolumn{2}{c}{\textbf{NPK}} \\
\cline{3-10}
 &  & \textbf{\ac{MAE}} & \textbf{\ac{MAPE}} & \textbf{\ac{MAE}} & \textbf{\ac{MAPE}} & \textbf{\ac{MAE}} & \textbf{\ac{MAPE}} & \textbf{\ac{MAE}} & \textbf{\ac{MAPE}} \\
\midrule
\multirow{4}{*}{\shortstack[l]{United States\\of America}} & Soybean (45) & 16.25&	79.27	&13.00&	45.30&	13.08	&37.11 & NA & NA \\
& Maize (45) & 9.31	&7.03	&7.08	&12.84	&10.38	&19.72 & NA & NA \\
& Wheat (46) & 16.20	&26.38&	15.23&	41.01&	14.21&	62.69 & NA & NA \\
& Fiber crops (43) & 14.76&	16.55	&11.90	&27.51&	11.87	&32.06 & NA & NA \\
\midrule
\multirow{4}{*}{United Kingdom } & Other Oilseeds (22)& 37.20&	16.00	&3.66	&10.36&	16.29	&32.08 & NA & NA \\
& Wheat (22)& 47.87	&19.98	&2.21	&7.09&	4.07&	9.78 & NA & NA \\
& Sugar crops (22)& 72.83&	43.49&	11.99&	29.32&	24.85&	22.04 & NA & NA \\
& Roots and tubers (22)& 46.62	&23.20	&17.95	&17.03	&40.51&	21.44 & NA & NA \\
\midrule
\multirow{6}{*}{India} & Other Cereals (6)& 11.24&	30.43&	6.08	&34.93	&1.59	&38.14 & NA & NA \\
& Rice (8)& 11.43&	12.70	&3.54	&15.40	&2.86	&36.29 & NA & NA \\
& Maize (8)& 5.58&	12.95	&5.16	&35.56	&3.08	&59.67 & NA & NA \\
& Wheat (8)& 7.56&	15.47	&4.43	&21.64&	1.40&	36.68 & NA & NA \\
& Fiber crops (6)& 15.91	&15.97&	12.63&	49.14	&4.06	&91.06 & NA & NA \\
& Sugar crops (8)   & 13.59	&10.57	&13.36&	25.17	&12.34	&30.65 & NA & NA \\
\midrule
\multirow{3}{*}{Sweden} & Maize (4)& 176.04	&67.81	&50.75	&68.76	&52.41	&55.16 & NA & NA \\
& Wheat (4)& 54.80	&29.05	&11.15	&48.54	&24.17&	108.43 & NA & NA \\
& Sugar crops (4)& 66.60&	39.80&	22.16	&50.53	&51.88&	49.24 & NA & NA \\
\midrule
\multirow{2}{*}{Philippines} & Rice (13)& 8.34	&18.91	&5.01&	75.49 & NA & NA & NA & NA \\
& Maize (12)& 33.96&	89.01&	5.66	&64.65 & NA & NA & NA & NA \\
\midrule
\multirow{2}{*}{New Zealand} & Fruits (5)& 71.92	&61.17	&80.43&	81.74 & NA & NA & NA & NA \\
& Vegetables (5)& 105.43&	55.06	&141.94&	87.28 & NA & NA & NA & NA \\
\midrule
\multirow{3}{*}{Pakistan} & Rice (23)& NA & NA & NA & NA & NA & NA & 66.06&	41.12 \\
& Wheat (23)& NA & NA & NA & NA & NA & NA & 31.18	&18.34\\
& Fiber crops (23)& NA & NA & NA & NA & NA & NA & 71.25	&29.69  \\
\bottomrule
\end{tabular}
\hspace*{2cm}
\end{table}

\begin{table}[H] 
\centering
\caption{\label{tab:fer_sd:FAOSTAT_vs_NDB} Comparison of the data reported by global datasets \cite{Martinez1982FertilizerProduction, AdolfoMartinez1990FertilizerT-37, FAO1994Fertilizer2, FAO1996Fertilizer3, FAO1999Fertilizer4, FAO2002Fertilizer5, EFMA2002Fertilizer2001/02, EFMA2007Fertilizer2006/07, FertilizerEurope2012Fertilizer2011/12, FertilizerEurope2015Fertilizer2014/15, PattrickHeffer2009Asessment2007/08,Heffer2017Assessment2014-2014/15, Ludemann2022GlobalCountry} of the average application rate per fertilizer per ha against national database information for certain countries and crops per fertilizer. The validation is quantified using the mean absolute error (MAE) and mean percentage error (MAPE) per fertilizer between the two data sources, expressed in the table as \ac{MAE} and MAPE respectively (fertilizer). The NPK stands for the sum of all fertilizers used in the country for certain crops, this is only discussed for Pakistan as more granular data is not available. Unavailable data points are expressed as NA in the table.}
\hspace*{-2cm}
\begin{tabular}{llllllllll}
\toprule
\textbf{Country} & \textbf{Crop} & \multicolumn{2}{c}{\textbf{\nfer}} & \multicolumn{2}{c}{\textbf{\pfer}} & \multicolumn{2}{c}{\textbf{\kfer}} & \multicolumn{2}{c}{\textbf{NPK}} \\
\cline{3-10}
 &  & \textbf{\ac{MAE}} & \textbf{\ac{MAPE}} & \textbf{\ac{MAE}} & \textbf{\ac{MAPE}} & \textbf{\ac{MAE}} & \textbf{\ac{MAPE}} & \textbf{\ac{MAE}} & \textbf{\ac{MAPE}} \\
\midrule
\multirow{4}{*}{United States of America} & Soybean &  3.04&	25.37&	4.94&	17.33&7.78	&18.98 & NA & NA \\
& Maize & 10.45&	6.38&	8.41&	12.36	&13.69	&45.07 & NA & NA \\
& Wheat & 6.04&	8.96&	5.34&	18.20&	9.65&	66.89 & NA & NA \\
& Fiber crops & 21.51&	20.80&	8.74&	24.49&	11.76&	32.04 & NA & NA \\
\midrule
\multirow{4}{*}{United Kingdom} & Other Oilseeds & 12.42&	7.31&	3.41&	10.55&	3.01&	7.85 & NA & NA \\
& Wheat & 5.80	&3.04	&2.00	&6.67&	1.20&	3.32 & NA & NA \\
& Sugar crops & 3.80&	3.93&	3.20&	9.45&	9.20&	10.03 & NA & NA \\
& Roots and tubers & 11.20&	7.97	&6.60&	5.77&	12.40	&5.74 & NA & NA \\
\midrule
\multirow{6}{*}{India} & Other Cereals & 0.30&	1.03&	2.49&	18.89	&1.01	&36.54 & NA & NA \\
& Rice & 12.30	&13.05	&1.81	&5.73&	0.78	&4.31 & NA & NA \\
& Maize & 19.12&	43.84&	13.13&	124.63	&0.57&	19.34 & NA & NA \\
& Wheat & 3.71&	3.40	&1.05	&2.51	&0.29	&4.16 & NA & NA \\
& Fiber crops & 26.52&	29.83&	5.70	&12.94&	4.28&	33.55& NA & NA \\
& Sugar crops & 14.93&	9.11&	2.97&	5.01&	26.03	&46.83 & NA & NA \\
\midrule
\multirow{2}{*}{Philippines} & Rice & 8.30 & 16.67 & 6.23 & 76.49 & NA & NA & NA & NA \\
& Maize & 31.32	& 69.56 & 6.60 & 78.31  & NA & NA & NA & NA \\
\midrule
\multirow{3}{*}{Pakistan} & Rice & NA & NA & NA & NA & NA & NA & 44.64 & 27.54 \\
& Wheat & NA & NA & NA & NA & NA & NA & 29.97 &	18.68 \\
& Fiber crops & NA & NA & NA & NA & NA & NA & 54.30 & 23.83 \\
\bottomrule
\end{tabular}
\hspace*{2cm}
\end{table}

\begin{table}[H]
\centering
\caption{\label{tab:fer_sd:code_packages}Overview of the used open source packages and respective programming language in the code for both model training, \acrfull{SHAP} value computation and validation, as well as map building.}
\begin{tabular}{lll}
\toprule
\textbf{Programming language} & \textbf{Package} & \textbf{Version} \\
\midrule
Python & Python \cite{VanRossum2009PythonManual} & 3.10.3 \\
Python & numpy \cite{harris2020array} & 1.23.2 \\
Python & pandas \cite{mckinney_proc_scipy_2010} & 1.4.1 \\
Python & rasterio \cite{rasterio} & 1.3.9 \\
Python & scikit-learn \cite{Pedregosa2011Scikit-learn:Python} & 1.3.2 \\
Python & shap \cite{Lundberg2017APredictions} & 0.44.0 \\
Python & xgboost \cite{Chen2016XGBoost:System} & 2.0.3 \\
R & R \cite{R} & 4.2.2 \\
R & sf \cite{sf1,sf2} & 1.0-15 \\
R & ncdf4 \cite{pierce2019package} & 1.22 \\
R & exactextractr \cite{exactextractr} & 0.10.0 \\
R & readxl \cite{readxl} & 1.4.3 \\
R & stringr \cite{stringr} & 1.5.0 \\
R & dplyr \cite{dplyr} & 1.1.2 \\
R & readr \cite{readr} & 2.1.4 \\
R & ggplot2 \cite{ggplot2} & 3.4.2 \\
R & tidyverse \cite{tidyverse} & 2.0.0 \\
R & cshapes \cite{weidmann2010mapping} & 2.0 \\
R & terra \cite{hijmans2022package} & 1.7-65 \\
\bottomrule
\end{tabular}
\end{table}

\clearpage
\section{Figures}

\clearpage
\setlength{\marginparwidth}{50pt}
\setlength{\hoffset}{0pt}
\bibliographystyle{unsrt}
\bibliography{references}

\end{document}